%% file: main-file.tex
\documentclass[12pt]{scrartcl}

\usepackage[T1]{fontenc}
\usepackage[latin1]{inputenc}
\usepackage[english]{babel}
\usepackage{lmodern}

\usepackage{amsmath}
\allowdisplaybreaks
\usepackage{amssymb}
\usepackage{amsthm}
\theoremstyle{definition}  
\usepackage{bbm}

\usepackage{setspace}
\renewcommand{\baselinestretch}{1.5}
\usepackage[left=2.5cm,right=2.5cm,top=2.5cm,bottom=2.5cm]{geometry}
\usepackage{ragged2e}

\usepackage{graphicx}
\usepackage{caption}
\usepackage{subcaption}
\usepackage{float} 
\usepackage{rotating}

\usepackage{booktabs}
\usepackage{tabularx}
\usepackage{makecell}
\usepackage{multirow}
\usepackage{threeparttable}
\usepackage{dcolumn}
\newcolumntype{d}[1]{D{.}{\cdot}{#1} }
\usepackage{colortbl}
\DeclareCaptionLabelFormat{continued}{#1 #2 - continued}
\usepackage{arydshln}  
\usepackage{placeins}

\usepackage{indentfirst}
\usepackage{scrlayer-scrpage}
\setkomafont{section}{\Large\bfseries\rmfamily}
\setkomafont{subsection}{\large\bfseries\rmfamily}
\setkomafont{subsubsection}{\normalsize\rmfamily\mdseries\itshape}
\makeatletter
\renewcommand{\paragraph}{%
	\@startsection{paragraph}{4}{0pt}%
	{3.25ex plus 1ex minus .2ex} 
	{1.5ex plus .2ex} 
	{\normalsize\rmfamily\mdseries\itshape} 
}
\makeatother
\usepackage{appendix}

\usepackage[hang,flushmargin]{footmisc}
\usepackage{url}
\usepackage[unicode=true,pdfusetitle,
bookmarks=true,bookmarksnumbered=false,bookmarksopen=false,
breaklinks=true,pdfborder={0 0 0},pdfborderstyle={},backref=false,colorlinks=true]
{hyperref}
\hypersetup{
	linkcolor=magenta,citecolor=blue, urlcolor=black}
\addto\extrasenglish{

}

\usepackage{listings, xcolor}

\color{black} 
\usepackage{chngcntr}

\usepackage{titling} 

\usepackage{apacite}
\usepackage[longnamesfirst]{natbib}
\AtBeginDocument{%
}

\newtheorem{corr}{\noindent COROLLARY}
\newtheorem{lem}{\noindent LEMMA}
\newtheorem{thm}{\noindent THEOREM}
\newtheorem{prop}{\indent PROPOSITION}
\newtheoremstyle{custommydef} 
{10pt}       
{10pt}       
{\hangindent=20pt} 
{}          
{\bfseries} 
{.}         
{ }         
{}          
\theoremstyle{custommydef}
\newtheorem{mydef}{DEFINITION}

\newcommand{\thmref}[1]{Theorem~\ref{#1}}
\newcommand{\corref}[1]{Corollary~\ref{#1}}

\newcommand{\lemref}[1]{Lemma~\ref{#1}}
\newcommand{\propref}[1]{Proposition~\ref{#1}}
\newcommand{\defref}[1]{Definition~\ref{#1}}
\renewcommand{\proof}[1]{\vspace{5pt}
	{\noindent \textbf{Proof:}#1
		\renewcommand{\baselinestretch}{1}}
	$\blacksquare$ \vspace{10pt}}

\newcommand{\titleinfo}{\vspace{-1.5cm} \LARGE \textbf{Comprehensive Causal Machine Learning}}
\title{\titleinfo}
\def\authora{\large Michael Lechner}
\def\authorb{\large Jana Mareckova}
\def\emaila{\href{mailto:michael.lechner@unisg.ch}{michael.lech\-ner@unisg.ch}}
\def\emailb{\href{mailto:jana.mareckova@unisg.ch}{jana.mareckova@unisg.ch}}

\begin{document}
	
	\begin{titlepage}
		\title{\titleinfo \thanks{\scriptsize Results from two research projects that were part of the National Research Programmes ``Big Data'' (NRP 75, \url{www.nrp75.ch}, grant number 407540\_166999) and ``Digital Transformation'' (NRP 77, \url{www.nrp77.ch}, grant number 407740\_187301) of the Swiss National Science Foundation (SNSF) were the basis of this paper. The theoretical part on the mcf (contained in \ref{appendix-a}) is a revised version of the theoretical part of our unpublished ``Modified Causal Forest'' paper. We thank GPT-4 for some limited assistance in editing, and Phillip Heiler, Federica Mascolo, Fabian Muny, and Hannah Busshoff for helpful comments and suggestions. The paper was presented at research seminars at the Universities of Bern and the Collegio Carlo Alberto in Turino, at an invited session of the Annual Meeting of the German Economic Association in Berlin, at CompStat in Giessen, at American Causal Inference Conference in Seattle, at COMPIE in Amsterdam and at the Workshop on Causal Inference and Machine Learning in Groningen. We are grateful to participants for helpful remarks and interesting discussions.}}
		
		\author{\Large \authora\thanks{\scriptsize Michael Lechner is also affiliated with Örebro University, CEPR, London, CESIfo, Munich, IAB, Nuremberg, IZA, Bonn, and RWI, Essen.}  \\ \and \authorb}
		
		\renewcommand\maketitlehookc{\centering \vspace{-25pt} \includegraphics[width=0.3\textwidth]{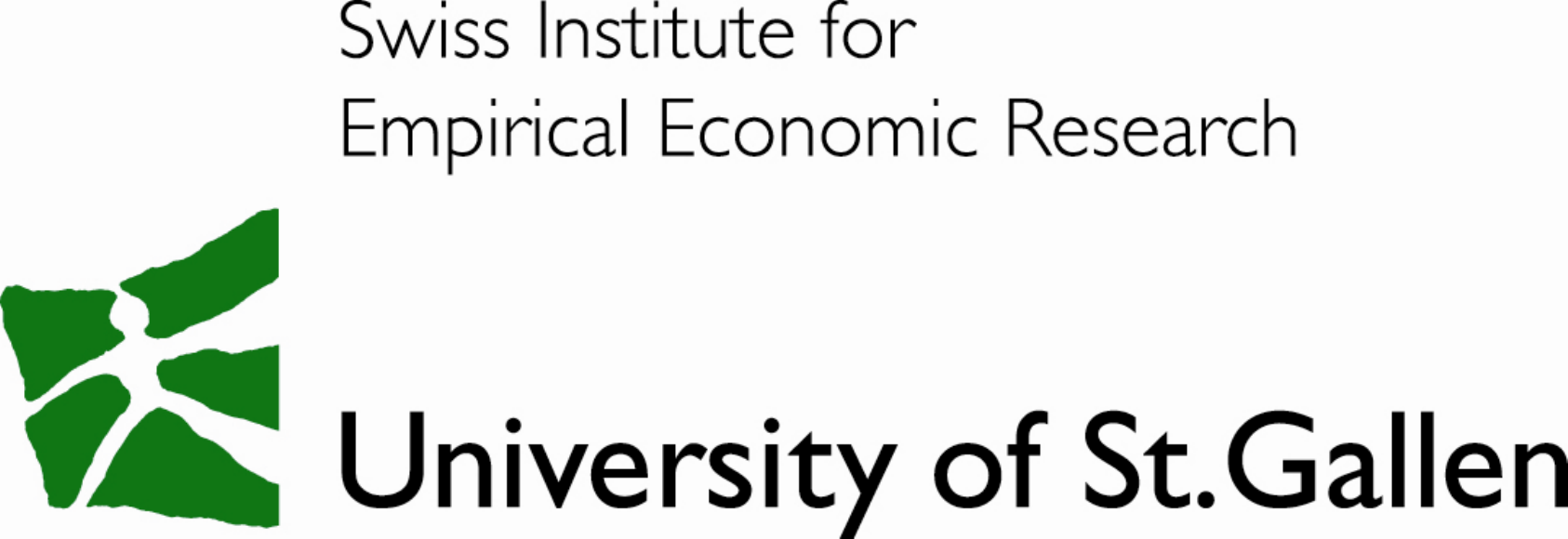}}
		
		\date{\large{{\today}}}
		
		\renewcommand\maketitlehookd{Comments are very welcome.}

		\maketitle
		
		\thispagestyle{empty}
		
		\begin{center}
		\footnotesize
		\textbf{Abstract}
		\end{center}
		\vspace{-25pt}
		\begin{abstract} \singlespacing	\footnotesize	
			Uncovering causal effects in multiple treatment
			setting at various levels of granularity provides substantial value to
			decision makers. Comprehensive machine learning approaches to causal
			effect estimation allow to use a single causal machine learning approach
			for estimation and inference of causal mean effects for all levels of
			granularity. Focusing on selection-on-observables, this paper compares
			three such approaches, the modified causal forest (\emph{mcf}), the
			generalized random forest (\emph{grf}), and double machine learning
			(\emph{dml}). It also compares the theoretical properties of the
			approaches and provides proven theoretical guarantees for the
			\emph{mcf.} The findings indicate that \emph{dml}-based methods excel
			for average treatment effects at the population level (ATE) and group
			level (GATE) with few groups, when selection into treatment is not too
			strong. However, for finer causal heterogeneity, explicitly
			outcome-centred forest-based approaches are superior. The \emph{mcf} has
			three additional benefits: (i) It is the most robust estimator in cases
			when \emph{dml}-based approaches underperform because of substantial
			selection into treatment; (ii) it is the best estimator for GATEs when
			the number of groups gets larger; and (iii), it is the only estimator
			that is internally consistent, in the sense that low-dimensional causal
			ATEs and GATEs are obtained as aggregates of finer-grained causal
			parameters.
		\end{abstract}
		\vfill
		\singlespacing \footnotesize	
		
		\vspace{5pt}
		
		\vspace{5pt}
		\noindent \textbf{Correspondence to}: Michael Lechner or Jana Mareckova,
		Professors of Econometrics, Swiss Institute for Empirical Eco­nomic
		Research (SEW), University of St.~Gallen, Switzerland,
		\emaila, \emailb, \url{www.sew.unisg.ch}.
	\end{titlepage}	

	\input{sections}

	\bibliography{ref.bib}
	\bibliographystyle{apacite}
	
	\newpage
	\begin{spacing}{1.5}	
	\appendix
	
	\setcounter{section}{0}  
	\setcounter{figure}{0}  
	\setcounter{table}{0}  
	\renewcommand{\thesection}{Appendix \Alph{section}}  
	\renewcommand{\thesubsection}{Appendix \Alph{section}.\arabic{subsection}}  
	\counterwithin{figure}{section}
	\renewcommand{\thefigure}{\Alph{section}.\arabic{figure}}
	\counterwithin{table}{section}
	\renewcommand{\thetable}{\Alph{section}.\arabic{table}}

\input{App}

	\end{spacing}
	
\end{document}

%% file: sections.tex
\newpage \pagenumbering{arabic}

\section{Introduction}\label{introduction}

Machine learning (ML) has paved its way into academia and industry,
impacting numerous fields from health care and finance to social
sciences. At the core of the ML revolution is the remarkable predictive
power of the methods. However, the growing debate around ML emphasizes
that prediction does not imply causation. Going beyond mere predictive
associations to identify cause-and-effect relationships is at the centre
of most questions concerning the \emph{effects} of policies, medical
treatments, marketing campaigns, business decisions, etc. \citep[see, e.g.,][]{athey2017beyond}. 
The different focus of causal modelling calls for ML
approaches that can reliably estimate causal effects, establishing
causal machine learning (CML).

CML integrates principles of ML and causal inference. The causality
literature provides conditions for identification and estimation of
\emph{effects}. Building on these conditions, causal problems can be
transformed into specific prediction problems \citep[see, e.g.,][]{imbens2009recent} 
for which ML methods are suitable \citep[see, e.g.,][for an overview on ML methods]{hastie2009}. The
flexibility of ML combined with careful use of data leads to versatile
causal effect estimators that are also able to uncover effect
heterogeneity \citep[for an overview, see][]{athey2017state}.

In recent years, researchers from different disciplines contributed to
method development in CML. Although such multidisciplinary work on a
common subject is an advantage from a scientific point of view, it
raises the question which of these many proposed methods to use in any
specific empirical study. In this paper, we focus on the relative
performance of a specific class of estimators for a special empirical
case, which is popular in the CML literature, namely (i) when causal
effects are plausibly identified under unconfoundedness (i.e.,
selection-on-observables) in a multiple treatment setting, and (ii) the
researcher is interested in aggregated average causal effects as well as
their possibly fine-grained heterogeneity.

A key condition of any causal effect estimator to be attractive in
applications is that it has acceptable proven statistical properties.
Such theoretical guarantees and the implied ability to conduct inference
are crucial, because, different to prediction settings, realisations
from the true effects, the so-called `ground truth', are unobservable.
Therefore, they cannot be used to evaluate the performance of the
specific estimation. Another important factor in practice is to keep the
estimation of the many different causal parameters sufficiently simple,
for example, by using the same, or the same type of causal machine
learner for all of them. This avoids the time-intensive task of tuning
and monitoring many different estimators. Additionally, methods that can
predict heterogeneous treatment effects on so far unseen data containing
covariates only are also an advantage in empirical settings. We will
call methods that (i) have available statistical inference, (ii) can
estimate average and fine-grained effects using one type of ML
framework, and (iii) can predict using covariates only, as Comprehensive
Causal Machine Learners (CCMLs). Finally, it is advantageous to have
\emph{internal consistency} of the possibly many estimated effects, in
the sense that effects at the higher aggregation levels are close to
appropriately aggregated lower-level effects.\footnote{A similar concept
	is known as \emph{coherence} in hierarchical time series forecasting,
	where it led to development of methods that ensure coherent forecasts
	across collections of time series formed through aggregation.
	\citet{athanasopoulos2024forecast} provide a
	review of such hierarchical time series methods.} Internal consistency
avoids scenarios in which conclusions across different aggregation
levels might be contradictory.

\defcitealias{chernozhukov2018dml}{Chernozhukov, Chetverikov, Demirer, Duflo, Hansen, Newey, and Robins, 2018}  

In the light of these arguably desirable properties, we analyse three
estimation principles that belong to the class of Comprehensive Causal
Machine Learning. The methods are Debiased/Double Machine Learning
(\emph{dml}; \citetalias{chernozhukov2018dml}), the Generalized Random Forest (\emph{grf}; \citealp{athey2019grf}), and the Modified Causal Forest
(\emph{mcf}; \citealp{lechner2018modified}). Only the latter fulfils the condition of
internal consistency by construction.\footnote{Within the classification
	of methods yielding coherent predictions for hierarchical time series,
	\emph{mcf} would be classified as a single-level, bottom-up approach.}

The intended contribution of this paper to the literature is two-fold:
The first contribution is a large scale-simulation exercise to evaluate
the finite-sample performance of all three CCMLs in many different
settings to better understand which estimator may have advantages or
disadvantages in certain situations and draw practical conclusions. The
simulation assesses the CCMLs at three different levels of treatment
effect granularity: (i) individual, (ii) group, and (iii) population
level. Crucially, these simulations also cover the inference procedures
that usually were absent from the small number of existing and far more
limited simulation studies. The second contribution are the theoretical
guarantees for the \emph{mcf} and a comparison of the theoretical
guarantees across the three different CCMLs. Additionally, similarities
and differences between the three approaches are documented.

The results yield several practical recommendations: First,
\emph{dml}-based methods predominantly excel in estimation of average
treatment effects (ATEs) or group treatment effects (GATEs) with fewer
groups. Second, for finer causal heterogeneity, explicitly
outcome-centred forest-based approaches are superior. Third, the
\emph{mcf} offers three additional benefits: (i) it is the most robust
estimator even for the ATE in cases when \emph{dml}-based approaches
underperform because of substantial selection into treatment that needs
to be corrected for; (ii) it outperforms both \emph{dml} and \emph{grf}
in GATE estimation when the number of groups gets larger; (iii) it is
the only estimator that is internally consistent.

The structure of the paper is as follows: \autoref{2-review} provides a review
with a particular focus on the broad contribution of CML methods to the
estimation of the ATE and more fine-grained conditional average
treatment effects (CATEs) under unconfoundedness. After a general
overview of CML methods, the section focuses on methods able to estimate
effects at all levels of granularity and identifies the CCML methods.
The section ends with an overview of CML simulation comparisons and
applications. \autoref{3-causal-framework} introduces the notation of the potential outcome
model, defines the parameters of interest, and states the identifying
assumptions imposed under unconfoundedness. \autoref{4-ccml} discusses the
three CCMLs. The large-scale simulation study is summarized in 
\autoref{5-mcs}. \autoref{6-conclusion} concludes. \ref{appendix-a} provides proven theoretical
guarantees for the \emph{mcf}, its inference procedure, a comparison to
the \emph{grf}, as well as some new results for the \emph{grf}.
Implementational details of the simulation study are collected in
\ref{appendix-b}, while \ref{appendix-c} holds its detailed results.

\section{Review of causal machine learning estimation}\label{2-review}

\subsection{Average treatment effect}\label{21-ate}

For causal inference under the unconfoundedness assumption \citep[see][]{rubin1974estimating}, 
the ATE represents one of the main parameters of interest
capturing the overall impact of a treatment or an intervention. For ATE
estimators, such as inverse probability weighting (IPW) and propensity
score matching \citep[see, e.g.,][]{imbens2004nonparametric}, propensity score estimation
plays a central role. Both estimators are consistent when a parametric
model for the propensity score is correctly specified. Consequently, one
of the first applications of ML methods in ATE estimation involved
flexible propensity score estimation. Various ML approaches, such as
Neural Networks, Support Vector Machines, Decision Trees \citep{westreich2010propensity}
or Boosting Models \citep{mccaffrey2004propensity,westreich2010propensity}, have been proposed to improve the
estimation of the propensity score.

However, even flexible ML methods may exhibit poor performance if they
prioritize fitting the propensity score well instead of explicitly
balancing the covariates across treatment groups, which is important to
reduce the bias of the estimator for the causal parameter of interest.
An estimator of the propensity score that explicitly aims to maximize
covariate balancing is introduced in \citet{imai2014covariate}.
Alternatively, \citet{graham2012inverse} improve IPW estimation by
covariate balancing within the empirical likelihood framework, without
explicit modelling of the propensity score.

\citet{cannas2019comparison} investigate the performance of matching and
weighting estimators based on propensity scores obtained from Logistic
Regression, Decision Tree, Bagging, Boosting, Random Forest, Neural
Networks, and Na\"{i}ve Bayes. Their simulation results indicate that Random
Forests consistently outperform other methods, particularly in the
context of IPW. \citet{goller2020does} compare
Random Forests, LASSO Logit Regression, and Probit estimation of the
propensity score for a matching estimator in an active labour market
policy setting, revealing that LASSO may yield more credible results
than conventional propensity score estimation methods, but overall
results were mixed.

Under the standard unconfoundedness assumptions, to be detailed in
\autoref{3-causal-framework}, ATE is alternatively identified as the expected difference
between conditional expectations of the outcome variable for the treated
and non-treated populations. This leads to regression-based estimation
of the two conditional expectations. Assuming a correct specification of
the outcome models, averaging their estimated differences yields a
consistent ATE estimator. In the ML literature, tree-based methods such
as Bayesian Additive Regression Trees (BART) \citep{hill2011bayesian} and Ensemble
Methods \citep{austin2012using} were introduced to obtain regression-based
estimators of ATEs. However, issues like regularization bias \citep{athey2018approximate}
and slow convergence of ML methods render the
outcome-based approach less popular.

Methods that combine propensity score-based and regression-based
estimators to increase robustness to misspecification have gained
popularity. Two related approaches prevail in the literature:
Double-robust (DR) estimators and Neyman-orthogonal scores. DR
estimators utilize two nuisance functions, i.e., the propensity score
and outcome models. They are consistent as long as at least one of the
nuisance functions is correctly specified. Certain DR estimators were
shown to be semi-parametrically efficient if both nuisance functions are
correctly specified. Ideas to combine propensity score and outcome
modelling in a double-robust way originally emerged in \citet{robins1994estimation, robins1995analysis},
leading to the Augmented Inverse
Probability Weighting (AIPW) estimator for the ATE. Nevertheless,
sensitivity to extreme propensity scores, as in the case of the IPW
estimator, remains an issue.\footnote{For potential benefits of trimming
	extreme propensity scores, see e.g., \citet{huber2013performance}.}

Approaches based on Neyman-orthogonality ensure that the moment
conditions identifying the target parameters are not locally affected by
small perturbances in the nuisance parameter estimates \citep{chernozhukov2018dml}.
Such an orthogonality property is leveraged, e.g., in \citet{belloni2014inference}
for ATE estimation with binary
treatment and in \citet{farrell2015robust} for multiple treatments. As
Neyman-orthogonal scores mitigate first order bias arising from ML
estimation of nuisance parameters \citep[p.~12]{bach2024dmlr},
Neyman-orthogonality emerged as a key element for
the Double Machine Learning (\emph{dml}) framework introduced in
\citet{chernozhukov2018dml}. \emph{dml} is a generic framework for
obtaining consistent estimators and valid inference for low-dimensional
parameters. This framework combines Neyman-orthogonal scores with
cross-fitting, a data partitioning technique that helps avoid further
biases arising from the ML estimation of nuisance parameters. \emph{dml}
is well-suited for high-dimensional settings, as it operates under
convergence rate conditions for nuisance parameter estimation that are
achievable by many ML methods. \citet{chernozhukov2018dml} apply the
\emph{dml} framework to ATE estimation using the DR score of 
\citet{robins1995semiparametric}\footnote{Note that the DR score of 
	\citet{robins1995semiparametric} is Neyman-orthogonal.} and derive 
its statistical properties.

Another approach for ATE estimation is Targeted Minimum Loss Estimation
(\emph{tmle}), a general method that combines ML methods with a targeted
updating step to reduce bias and variance of the estimated parameter of
interest in semi-parametric and non-parametric models \citep{van2006targeted}. 
Among various \emph{tmle} adaptations\footnote{See \citet{Laan2018}
	for a comprehensive overview.}, cross-validated
\emph{tmle}, \emph{cv-tmle}, proposed by \citet{Zheng2011}, shares 
considerable similarities with \emph{dml}. Both
approaches require estimation of nuisance parameters that are combined
in a way that yields a semi-parametrically efficient estimator of the
ATE that is robust to small errors in nuisance parameter estimation
under the same convergence rate conditions for nuisance parameter
estimation and use of data splitting. Meanwhile \emph{dml} utilizes the
nuisance parameters in a Neyman-orthogonal score to mitigate the bias,
\emph{cv-tmle} combines them in an iterative estimation procedure
addressing the bias and yielding quantities for each observation that
can be averaged into an ATE. Despite this difference, \emph{cv-tmle} and
\emph{dml} are asymptotically equivalent and can be expected to yield
similar results in large samples under the same choice of methods
estimating the nuisance parameters, as noted by \citet{knaus2021double}.

Automatic Debiased Machine Learning, introduced in \citet{chernozhukov2022automatic},
offers another alternative for ATE estimation that can
deal with bias stemming from using ML methods for estimation of nuisance
parameters. The key idea is based on a ``de-biased moment'' utilizing an
existence of a Riesz representer for parameters such as the ATE. The
debiasing is automatic in the sense that it does not require any
knowledge about the form of the Riesz representer. The Riesz representer
is directly estimated within the procedure. In case of the ATE, the
de-biased moment function corresponds to the DR moment function of
\citet{robins1995semiparametric}. The estimation of the Riesz representer and
nuisance parameters requires cross-fitting as in \emph{dml}. The method
is shown to yield a consistent and asymptotically normal estimator and
provides a consistent asymptotic variance estimator.

\subsection{Conditional average treatment effect}\label{22-cate}

\defcitealias{wei2023efficient}{Wei, Petersen, van der Laan, Zheng, Wu, and Wang (2023)}

CML methods extend beyond ATE estimation and help uncover heterogeneity
of the treatment effects through analysis of Conditional Average
Treatment Effects (CATEs). CATEs quantify how the treatment affects an
individual unit with specific low or large-dimensional characteristics.
Prominent methods estimating CATEs include Causal Forests
\citep{wager2018estimation,athey2019grf,lechner2018modified}, along with
\emph{dml}-based estimators yielding estimated components of the DR
efficient score that are further regressed on the covariates as first
proposed by \citet{Laan2006}. Alternative approaches include
Meta-learners, which use multiple machine learning algorithms as ``base
learners'' to estimate large-dimensional CATEs. The \emph{X-learner},
introduced by \citet{kunzel2019metalearners}, is effective in
cases with uneven treatment groups, while the \emph{R-learner},
developed by \citet{nie2021quasi}, leverages the \citet{robinson1988root}
transformation to estimate heterogeneous causal effects. When interested
in CATEs for specific subgroups, the above-mentioned quantities of the
\emph{tmle} iterative procedure can be averaged into a CATE, using the
corresponding part of the sample for the estimation.\footnote{\citetalias{wei2023efficient}
	note that the
	natural implementation of \emph{tmle} ``one-group-at-a-time'' for CATE
	estimation with discrete groups might be computationally costly. As a
	remedy, they propose a \emph{tmle} method that targets multi-subgroup
	treatment effects simultaneously.} In case of a CATE for the finest
granularity level, the \emph{tmle} estimand depends on an observed
outcome and cannot serve for a prediction on a new sample containing
covariates only.\footnote{Based on \citet{levy2021fundamental}
	who introduces a \emph{tmle} method for simultaneous
	estimation of ATE and variance of treatment effects, where estimated
	CATEs serve as nuisance parameters.}

The influential paper by \citet{chernozhukov2018dml}, averaging the
estimated components of individual DR efficient scores of \citet{robins1995semiparametric}
to estimate ATE in a \emph{dml} framework, spanned
further methodological contributions, particularly in discovering
heterogeneity along a chosen set of covariates, such as gender or age
group. \emph{dml}-based estimation of CATEs exploits the fact that the
conditional expectation of the DR efficient score given such covariates
identifies the respective CATEs. For low-dimensional sets of chosen
covariates, OLS, Series, or Kernel Regressions of the estimated DR score
components on the chosen set of covariates estimate the low-dimensional
CATE and standard statistical inference applies, as shown in 
\citet{semenova2021debiased}, \citet{zimmert2019nonparametric}, and \citet{fan2022estimation}. 
For a higher dimensionality of the set of the
chosen covariates, \citet{kennedy2023towards} introduces the DR-learner that takes
the components of a classic DR estimator of ATE and regresses them on
all covariates to estimate a large-dimensional CATE and derives an upper
bound on the DR-learner error relative to an oracle.

The foundations for heterogeneous treatment effect estimation by Causal
Forests are laid in a seminal paper by \citet{athey2016recursive}. This
work introduces splitting rules for treatment effect estimation and
honest data splitting, i.e.~using different parts of data to build the
tree and to estimate the parameters, to ensure valid inference for CATEs
through tree-based methods. Subsequently, \citet{wager2018estimation} develop
a nonparametric Causal Forest algorithm based on honest trees and
splitting rules maximizing heterogeneity in treatment effects across
final leaves. The last step of the estimator of the possibly
large-dimensional CATE is obtained by averaging individual tree
estimates. Concerning its asymptotic guarantees, the Causal Forest is
consistent and pointwise asymptotically normal under further regularity
conditions. In a distinct but related approach, \citet{athey2019grf}
introduce the Generalized Random Forests (\emph{grf}). This method is
based on an honest forest with a gradient-based approximation of an
estimator-specific splitting rule. The forest estimation determines
importance weights of each observation for the parameters of interest.
In other words, these weights determine a neighbourhood of observations
that will contribute to the CATE estimation. Instead of estimating the
CATE directly as an average across honest trees as in \citet{wager2018estimation},
the forest weights enter a set of local moment conditions
identifying the CATE. This estimator is consistent and pointwise
asymptotically Gaussian and generalizes beyond unconfoundedness.

Additionally, \citet{lechner2018modified} introduces a Modified Causal Forest
(\emph{mcf}) estimation procedure for ATEs and CATEs. In a multiple
treatment framework, he proposes a weights-based inference procedure
utilizing the weighted-outcome representation of forest estimator.
\emph{mcf} further differs from the other Causal Forests in using the
mean squared error of the CATE directly to find the best split.
Furthermore, it introduces a two-sample honesty, detailed in \autoref{43},
for building the forest and the estimation of the effects. A
simulation study in \citet{lechner2018modified} shows superior performance of
\emph{mcf} to the Causal Forest of \citet{wager2018estimation}. \citet{bodory2022high}
show that the \emph{mcf} works in
applications replicating results from several papers in different
fields.

\subsection{Comprehensive treatment effect estimation}\label{23-ccml}

Methods that (i) can estimate average and fine-grained effects using one
type of ML framework, (ii) have available statistical inference for all
the effects, and (iii) can predict using covariates only, are labelled
as Comprehensive Causal Machine Learners (CCMLs). Some of the
above-mentioned advances provide practitioners with methods able to
estimate treatment effects across all different levels of granularity.
Table \ref{tab1} provides an overview of commonly used CML approaches for
estimation of such treatment effects and evaluates them based on (i)
available statistical inference, (ii) their ability to predict
large-dimensional CATEs on new data (with covariate information only),
and (iii) internal consistency\footnote{An estimator is considered being
	internally consistent when effects at the higher aggregation levels
	are close to appropriately aggregated lower-level effects.} of
estimators. The overview reveals that \emph{tmle} and Meta-learners do
not have available statistical inference for all levels and that
\emph{tmle} cannot predict CATEs at the finest granularity level on a
new data set that contains covariates only. Thus, these two methods are
not considered to be CCMLs. On the other hand, \emph{dml} and Causal
Forests are approaches that fulfil the conditions for comprehensive
treatment effect evaluation with available statistical inference and
predictions for new data containing covariates only. For Causal Forests,
there is a Bayesian alternative which will not be pursued further
because its inference procedure is not based on repeated sampling
inference.

Focusing on estimation at different levels of granularity via
CCMLs, \emph{dml}-based estimation operates via the estimated components
of the DR efficient scores that are regressed on covariates to yield
lower and higher-level aggregates of average treatment effects. Causal
Forests, \emph{grf} and \emph{mcf}, provide in their first stage CATE
estimates at the highest level of granularity. Using direct aggregation
step of these CATEs, \emph{mcf} leverages the weighted-outcome
representation of forest predictions, while \emph{grf} uses a variant of
the AIPW score to yield higher level aggregations via a regression step.
Table \ref{tab1} summarizes the estimation at different levels of granularity for
CCMLs and other methods.

\renewcommand{\dashlinedash}{2pt}  
\renewcommand{\dashlinegap}{1.5pt} 

\begin{table}[h]
	\centering
	\sffamily
	\caption{Properties of commonly used CML methods for treatment effect estimation}
	\label{tab1}
	\small
	\begin{tabularx}{\textwidth}{lX>{\centering\arraybackslash}X>{\centering\arraybackslash}X>{\centering\arraybackslash}Xccc}
		\toprule
		\multicolumn{2}{l}{\textbf{Approach}} & \multicolumn{3}{c}{\textbf{Parameters}} & \textbf{Statistical} & \textbf{CATE pre-} & \textbf{Internal} \\
		& & & & & \textbf{inference} & \textbf{diction de-} & \textbf{con-} \\
		& & \multicolumn{2}{c}{\textit{Dimension of CATE}} & \emph{ATE} & & \textbf{pends only} & \textbf{sistency} \\
		& & large & low & & & \textbf{on covari-} & \\		
		& & & & & & \textbf{ates} & \\				
		\midrule
		\multirow{3}{*}{\rotatebox{90}{\textbf{CCML}}} & \textbf{\emph{dml}} & Regress & Regress & Regress & Yes & Yes & No \\
		& \textbf{\emph{grf}} & LocalGMM & Regress & Regress & Yes & Yes & No \\
		& \textbf{\emph{mcf}} & MCF & Aggreg & Aggreg & Yes & Yes & Yes \\
		\hdashline
		& \textbf{Meta-L} & Meta-L & Aggreg* & Aggreg* & No** & Yes & Yes \\
		& \textbf{\emph{tmle}} & TMLE & TMLE & TMLE & Yes (low-dimen.) & No & Yes \\
		& \textbf{\emph{bcf}} & BCF & BayAggr & BayAggr & Bayesian & Yes & Yes \\		
		\bottomrule
	\end{tabularx}
	\vspace{-8pt}
	\captionsetup{justification=justified,singlelinecheck=false}
	\caption*{\footnotesize \textsf{Note: \emph{dml}: Double / Debiased Machine Learning; \emph{LocalGMM}: Local General Method of Moments; \emph{mcf/MCF}: Modified Causal Forest; \emph{bcf/BCF}: Bayesian Causal Forest; \emph{tmle/TMLE}: Targeted Minimum Loss Estimation; \emph{Meta-L}: Meta-learner. \\ 
	\emph{Regress}: Regression-based estimation; \emph{Aggreg}: Aggregation of estimated IATEs; \emph{BayAggr}: Bayesian Aggregation. \\ 
	* Averaging of the finest large-dimensional CATEs for Meta-learners was implemented in, e.g., \citet{salditt2024tutorial}. \\
	** However, see recent developments in constructing valid confidence intervals for Meta-learners based on conformal prediction in \citet{alaa2024conformal}.}}
\end{table}

\subsection{Simulations and empirical work}\label{24}

\defcitealias{wendling2018comparing}{Wendling, Jung, Callahan, Schuler, Shah, and Gallego (2018)}  
\defcitealias{audrino2024does}{Audrino, Chassot, Huang, Knaus, Lechner, and Ortega, 2024}  
\defcitealias{langenberger2023exploring}{Langenberger, Steinbeck, Sch\"{o}ner, Busse, Pross, and Kuklinski, 2023}  

The literature on large scale simulation comparisons of finite sample
properties of CML methods is very limited. Among notable contributions,
\citet{knaus2021machine} look at the finite-sample
performance of selected machine learning estimators, covering \emph{grf}
and generic approaches that can be combined with any ML method, yielding
large-dimensional CATE estimates based on pseudo-outcomes or modified
covariates.\footnote{At the time the simulations of \citet{knaus2021machine}
	were performed, the \emph{mcf} was not yet
	available.} Higher level aggregates are simple averages of the
corresponding CATEs. Using the Empirical Monte Carlo Study (EMCS)
approach, many components of their simulations are based on real data.
The results show that best-performing estimators at the
large-dimensional CATE level produced most reliable estimates at higher
aggregation levels, such as for the ATE, in terms of their mean squared
error (MSE). In general, methods that utilized both outcome and
treatment equations are among the best-performing methods, including
\emph{grf} with local centering. \citet{caron2022estimating}
evaluate several Meta-learners in an EMCS based on health data. Their
results highlight variability in performance that depends on the
complexity of the data generating process. In a setting with a complex
CATE, multitask learners designed to estimate both outcome equations
jointly, such as Multitask Gaussian Process introduced in \citet{alaa2017bayesian},
perform best in terms of MSE. The \emph{X-learner}
performs the best in a setting with a simple CATE function and slightly
unbalanced treatment groups. Another health-data-based EMCS of \citetalias{wendling2018comparing}
concentrates on
large-dimensional CATE estimation in a common healthcare setting when
outcomes are binary and rare. The compared methods include BART,
\emph{grf} with local centering, Causal Boosting, and Causal
Multivariate Adaptive Regression Splines (MARS). The findings show that
BART and Causal Boosting perform better across the scenarios, as
evaluated by the root MSE in the conditional probability (risk)
difference estimates. Except for one scenario involving highly
heterogeneous treatment effects, the coverage rate for BART, \emph{grf},
and causal MARS was close to its nominal level.\footnote{To the best of
	our knowledge, this is the only study that also investigates the
	performance of the corresponding inference procedures. The simulation
	study in \citet{wendling2018comparing} differs from the simulation in \autoref{5-mcs}
	by using fixed covariates across replications. Despite this
	difference, their results are in line with the results in \autoref{5-mcs}
	confirming that high treatment heterogeneity negatively affects the
	coverage probability. Meanwhile DGPs with low treatment heterogeneity
	have coverage probability closer to nominal rates.}

Empirical applications using CML for treatment estimation span across
many fields and topics. The following non-exhaustive list illustrates
the variety of applications ranging from active labour market policy
\citep{davis2017using, bertrand2021workfare, Pytka2021Understanding,
knaus2022heterogeneous, cockx2023priority, burlat2024everybody},
education \citep{knaus2021double, farbmacher2021heterogeneous},
social experiments \citep{athey2019estimating, strittmatter2023value},
energy use \citep{knittel2021machine}, natural resource rents \citep{hodler2023institutions},
finance \citepalias{audrino2024does}, and
the dating market \citep{boller2021effect} to medicine \citepalias{langenberger2023exploring}.

Regarding the comprehensive nature of the \emph{dml} framework, \citet{knaus2022double}
demonstrates its flexibility in an active labour market programme
evaluation setting. \citet{rehill2024applied} documents and evaluates some practices
of using Causal Forests in applied work. Further illustrations of
potential benefits of CML compared to traditional methods in empirical
work is provided in \citet{baiardi2024value}. They revisit estimates of
average and heterogenous treatment effects from published observational
studies and randomized control trials. Their results illustrate the
potential of CML to capture non-linear confounding, its suitability in
scenarios with large number of covariates relative to the sample size
and its ability to discover treatment effect heterogeneity.

\section{Causal framework}\label{3-causal-framework}

\subsection{The potential outcome model}\label{31}

We use Rubin's potential outcome language \citep{rubin1974estimating} 
to describe a multiple
treatment model under unconfoundedness, also referred to as
selection-on-observables or conditional independence \citep{imbens2000role, Lechner2001}.
Let $D$ denote the treatment that may take a known
number of $M$ different integer values from 0 to $M-1$.
The (potential) outcome of interest that realises under treatment
$d$ is denoted by $Y^d$. For each
observation, we observe only the particular potential outcome
corresponding to the observed treatment status, i.e.,
${y_i} = \sum\limits_{d = 0}^{M - 1} {\underline 1 ({d_i} = d)y_i^d}$, 
where $\underline 1(\cdot)$ is the indicator function which equals one if its
argument is true.\footnote{If not obvious otherwise, capital letters
	denote random variables, and small letters their values. Small values
	subscripted by `$i$' denote the value of the respective variable
	of observation `$i$'.} There are two groups of variables to
condition on, $\tilde X$ and
$Z$. $\tilde X$ contains
those covariates needed to correct for selection bias (confounders),
while $Z$ contains variables that define (groups of) population
members for which an average causal effect estimate is desired.
$\tilde X$ and $Z$ may be
discrete, continuous, or both. They may overlap in any way. The union of
these two groups of variables is denoted by $X$, $X =
(\tilde X,Z)$, $dim(X)=p$.\footnote{In all CCMLs $p$ is assumed to be finite.}

Below, we investigate the following average causal effects:
\begin{align*}
& IATE(m,l;x) = E({Y^m} - {Y^l}| X = x), \\
& GATE(m,l;z) = E({Y^m} - {Y^l}| Z = z) = \int{IATE(m,l;\tilde x,z){f_{\tilde X| Z = z}}(\tilde x)}d\tilde x, \\
& ATE(m,l) = E({Y^m} - {Y^l}) = \int {IATE(m,l;x){f_X}(x)}dx.
\end{align*}

The \textbf{I}ndividualized \textbf{A}verage \textbf{T}reatment
\textbf{E}ffects (IATEs) measure the mean impact of treatment $m$
compared to treatment $l$ for units with covariates $x$. The
IATEs represent the causal parameters at the finest aggregation level of
the covariates. On the other extreme, the \textbf{A}verage
\textbf{T}reatment \textbf{E}ffects (ATEs) represent the population
averages. ATE is a classical parameter investigated in many empirical
studies. The \textbf{G}roup \textbf{A}verage \textbf{T}reatment
\textbf{E}ffect (GATE) parameters are in-between those two extremes with
respect to their aggregation levels.\footnote{We presume that the
	analyst selects the variables $Z$ prior to estimation. However,
	the estimated IATEs may be analysed by methods picking $Z$ in a
	data-driven way to describe their dependence on certain covariates.
	See Section 6 in \citet{lechner2018modified} for more details. Note that \cite{abrevaya2015estimating}
	and \citet{lee2017doubly} introduce
	similar aggregated parameters that depend on a reduced conditioning
	set and discuss inference in their specific settings.} IATEs and GATEs
are special cases of the already mentioned \textbf{C}onditional
\textbf{A}verage \textbf{T}reatment \textbf{E}ffects (CATEs).

\subsection{Identifying assumptions}\label{32}
\newcommand\rotpi{\rotatebox[origin=c]{180}{$\ \boldsymbol{\Pi} \ $}}

The following set of assumptions identifies the causal effects discussed
in the previous section \citep[see][]{imbens2000role, Lechner2001} (see Imbens, 2000, Lechner 2001):\footnote{To
	simplify the notation, we take the strongest form of these
	assumptions. Some parameters are identified under weaker conditions as
	well (for details see \citet{imbens2000role, imbens2004nonparametric}, or \citet{Lechner2001}.}
\begin{align*}
	&Y^d \rotpi D | X = x, & & & & \text{(CIA)} \\
	&0 < P(D=d|X=x) = p_d(x), & & \forall x \in \chi, \forall d \in \{0,\ldots, M-1\}; & & \text{(CS)} \\
	&Y = \sum\limits_{d = 0}^{M-1}\underline{1}(D=d)Y^d; & & & & \text{(Observation rule)}
\end{align*}

The conditional independence assumption (CIA) holds if there are no
covariates other than $X$ that jointly influence treatment and
potential outcomes (for the values of $X$ that are in the support
of interest, $\chi$). The
common support (CS) assumption stipulates that for each value in
$\chi$, there must be the
possibility to observe all treatments. The stable-unit-treatment-value
assumption (SUTVA, also referred to as observation rule or consistency
condition) implies that the observed value of the treatment and the
outcome does not depend on the treatment allocation of the other
population members (ruling out spillover and treatment scale effects).
Usually, to have an interesting interpretation of the effects, it is
required that $X$ is not influenced by the treatment (exogeneity).
In addition to these \emph{identifying} assumptions, assume that a large
random sample of size $N$ from the i.i.d.~random variables $Y,
	X, D$, $(y_{i}, x_{i}, d_{i})$,
$i=1, \ldots, N$, is available and that all necessary moments of
these random variables exist.

If these assumptions hold, all IATEs are identified in the sense that
they can be uniquely deduced from variables that have observable sample
realisations \citep[see][]{hurwicz1950}:
\small
\begin{align*}
IATE(m,l;x) & = E({Y^m} - {Y^l}| X = x) \\ 
& = E({Y^m}| X = x,D = m) - E({Y^l}| X = x,D = l)\\ 
& = E(Y| X = x,D = m) - E(Y| X = x,D = l) \quad {\forall x \in \chi ,\forall m \ne l \in \{0,...,M -1\}.}
\end{align*}
\normalsize

Since the distributions used for aggregation,
${f_{\tilde X| Z =
z}}(\tilde x)$ and
${f_X}(x)$, relate to observable
variables ($X$) only, they are identified as well (under standard
regularity conditions). This in turn implies that the GATE and ATE
parameters are identified.

\section{Comprehensive approaches for estimation and inference}\label{4-ccml}

In this section, three comprehensive approaches for treatment effect
estimation are presented. In the first two subsections, the underlying
principles as well as the concrete estimation and inference algorithms
for Double Machine Learning (\emph{dml)} and Generalized Random Forest
(\emph{grf)} are reviewed. The third subsection focuses on the Modified
Causal Forest (\emph{mcf)}. While the basic principles of the \emph{mcf}
are introduced in \citet{lechner2018modified}, that paper does not contain explicit
theoretical guarantees. Therefore, \ref{appendix-a1} provides the proofs of
consistency and asymptotic Gaussianity properties for IATEs, GATEs, and
ATEs.

\subsection{Double Machine Learning}\label{41}

\emph{dml} estimation \citep{chernozhukov2018dml} is based on moment
conditions with scores satisfying the identification condition as well
as Neyman-orthogonality, yielding estimators that are robust to small
estimation errors in nuisance parameters \citep[as in][]{neyman1959optimal}.
A Neyman-orthogonal score that identifies ATE, GATE, and IATE for
treatments $m$ and $l$ under assumptions outlined in
\autoref{32} is the DR score of \citet{robins1995semiparametric}:
\begin{align*}
& \psi_{m,l}^{dml}(O;{\theta_{m,l}},\eta_{m,l}^{dml}(X)) = \Gamma _{m,l}^{dml}(O;\eta_{m,l}^{dml}(X)) - {\theta_{m,l}}, \\
& \Gamma_{m,l}^{dml}(O;\eta_{m,l}^{dml}(X)) = \mu_m(X) - \mu_l(X) + \frac{\underline{1}(D=m)(Y-\mu_m(X))}{p_m(X)} - \frac{\underline{1}(D=l)(Y-\mu_l(X))}{p_l(X)},
\end{align*}
where $\mu_{d}(x)=E(Y| X=x, D=d)$, $\eta_{m,l}^{dml}(X)=(\mu_{m}(X),\mu_{l}(X), p_{m}(X), p_{l}(X))$
captures the nuisance parameters, and $\theta$ represents the treatment
effect of interest. $O$ represents all observable variables, i.e.,
$O = (X, Y, D)$. Noting that the difference of the third and
fourth term in this score has expectation zero conditional on $X$
at the true values of nuisance parameters, and letting
$\eta^{0}$ and $\theta^{0}$ denote the
true values of $\eta$ and $\theta$ respectively,
$\psi_{m,l}^{dml}(O;{\theta_{m,l}},\eta_{m,l}^{dml}(X))$
identifies the different treatment effects:
\begin{align*}
& E\left({\psi_{m,l}^{dml}(O;\theta_{m,l}^0,\eta _{m,l}^{dml,0})}\right) = 0,& & \theta _{m,l}^0 = ATE(m,l), \\
& E\left({\psi_{m,l}^{dml}(O;\theta_{m,l}^0(z),\eta_{m,l}^{dml,0})\left| {Z = z}\right.} \right) = 0, & & \theta _{m,l}^0(z) = GATE(m,l;z), \\
& E\left( {\psi_{m,l}^{dml}(O;\theta_{m,l}^0(x),\eta_{m,l}^{dml,0})\left| {X = x}\right.} \right) = 0,& & \theta _{m,l}^0(x) = IATE(m,l;x),
\text{ for all } x \in \chi.
\end{align*}

Effectively, $ATE(m,l)$ is estimated as an average of the estimated
component of the DR score, $\Gamma_{m,l}^{dml}\left({{o_i};\hat \eta _{m,l}^{dml, -k(i)}({x_i})} \right)$, across all
observations, i.e.,
$$\widehat {AT{E^{dml}}}(m,l) =
\hat \theta _{m,l}^{dml} =
\frac{1}{N}\sum\limits_{i
= 1}^N {\Gamma
_{m,l}^{dml}\left( {{o_i};\hat
\eta _{m,l}^{dml, - k(i)}({x_i})}
\right)}.$$
This process involves the estimation of nuisance parameters, symbolized
as $\hat \eta_{m,l}^{dml, - k(i)}({x_i})$, through
$K$-fold cross-fitting. The latter ensures that for each
observation $i$, the corresponding nuisance parameters are
estimated without using that specific observation in their training
data. Particularly, the elements of the vector
$\hat \eta_{m,l}^{dml, - k(i)}({x_i})$ are estimated on
$K-1$ folds not containing the observation $i$, which resides
in the left-out fold $k$. This is indicated by the superscript
$-k(i)$.

Neyman-orthogonal scores mitigate regularization bias in the estimation
of the ATE, provided that the product of the convergence rates of the
two nuisance parameter estimators within each treatment group is at
least $\sqrt N $. This
condition, which many ML methods achieve, allows for flexible estimation
of nuisance parameters leading to estimators robust to small estimation
errors in these nuisance parameters. The additional combination with
cross-fitting is important to avoid overfitting when estimating the
nuisance parameters and to subsequently guarantee
$\sqrt N - $consistent
estimation of the main parameters of interest. For technical details,
refer to \citet{chernozhukov2018dml}. The variance estimator of the
\emph{dml} can be computed as the sample variance of the estimated DR
scores.\footnote{See Theorem 3.2 in \citet{chernozhukov2018dml}, and
	Theorem 1 and Remark 1 in \citet{bach2024dmlr}.} The
\emph{dml} estimator asymptotically reaches the semiparametric
efficiency bound of \citet{hahn1998role}.\footnote{See Theorem 5.1 in
	\citet{chernozhukov2018dml}.}

Building on the identification result, $GATE(m,l;z)$ can be
estimated by regressing $\Gamma_{m,l}^{dml}\left({{o_i};\hat \eta _{m,l}^{dml, -k(i)}({x_i})} \right)$
on a
low-dimensional vector $Z$ of pre-specified variables. \citet{semenova2021debiased}
provide statistical inference for the best linear
predictor in this case, a method that is implemented in \autoref{5-mcs}.
Standard errors are estimated via heteroscedasticity-robust standard
errors for asymptotically valid pointwise and uniform confidence bands.
In general, the results in \citet{semenova2021debiased} apply to
linear projections onto a vector of basis functions. Under sufficient
smoothness and richness of the basis functions, inference targets the
$GATE(m,l;z)$ function. When basis functions are group indicators,
it targets the corresponding group parameters. For continuous
regressors, kernel regression offers a viable alternative, enabling the
estimation of $GATE(m,l;z)$ and facilitating statistical inference,
as demonstrated in \citet{zimmert2019nonparametric} and \citet{fan2022estimation}.

In a similar fashion, a natural approach to obtain an estimator of
$IATE(m,l;x)$ is by regressing $\Gamma_{m,l}^{dml}\left({{o_i};\hat \eta _{m,l}^{dml, -k(i)}({x_i})} \right)$ on the
covariates.\footnote{In \autoref{44} in Table \ref{tab2}, this step is called a
	smoothing step.} \citet{foster2023orthogonal} and \citet{kennedy2023towards} derive
error bounds for this two-step procedure, referred to as DR-learner in
 \citet{kennedy2023towards}. Particularly, the findings in  \citet{kennedy2023towards} provide a
theoretical foundation for the validity of inferential methods used in
this context for linear smoothers satisfying the required stability
condition and under further small-bias condition on
$\Gamma
_{m,l}^{dml}\left( {{o_i};\hat
\eta _{m,l}^{dml, - k(i)}({x_i})}
\right)$. In practical applications,
different techniques have been employed: \citet{goller2023analysing} utilizes linear
regression, while \citet{knaus2022double} opts for a Random Forest approach.

\subsection{Generalized Random Forests}\label{42}
\defcitealias{athey2019grf}{ATW19}

\emph{grf} (\citealp{athey2019grf}, abbreviated as ATW19 in this section)
extends the Random Forest into a nonparametric method estimating
parameters identified by local moment conditions:  
$$E\left(\psi^{grf}(O;{\theta^0}(x),{\eta^{grf,0}}(x))\left| {X = x}\right. \right) = 0 \quad  \text{for all } x \in \chi,$$
where ${\psi ^{grf}}(\cdot )$ is a score function identifying the
true values of $\theta(x)$. As before, $\eta^{grf}(x)$
captures nuisance parameters, $\theta^{0}(x)$ and
$\eta^{grf,0}(x)$ are the true values of $\theta(x)$
and $\eta^{grf}(x)$, and $O$ represents all
observable variables. \emph{grf} estimation of heterogeneous treatment
effects is based on $\psi^{grf}(Y,\tilde D,X;\theta(x),\eta_{IATE}^{grf}(x)) = \left({Y - \tilde D'\theta (x) -
\eta _{IATE}^{grf}(x)}\right) \left(1 \ \ \tilde D'\right)',$ where $\tilde D$ is an
$(M-1) \times 1$ vector containing dummy variables indicating whether
treatment $d \in
\left\{ {1,...,M - 1}
\right\}$ was received,
$\theta(x)$ is a parameter vector representing corresponding treatment
effects (all treatment $d$ vs 0 pairs) for point $x$, and
$\eta^{grf}_{IATE}(x)$ captures the
intercept and all confounders in a single nuisance parameter.\footnote{This
	part is coined as an \emph{intercept} term $c(x)$ in \citetalias{athey2019grf}.} Note
that the score function in \emph{dml} is based on a fully nonparametric
DR score identifying treatment effects for all pairwise combinations of
treatments, while the score function in \emph{grf} stems from a partial
linear model in which only treatment effects with respect to a reference
treatment are identified. The reference treatment is the one left out
from the vector $\tilde D$,
here the control group with $d = 0$. This is not too restrictive if
effects that are conditioned on the treatment status are not of interest
(as for IATE estimation), because the point estimates of the other
treatment combinations can be obtained as
$\theta_{m,l}(x)=\theta_{m,0}(x)-\theta_{l,0}(x)$.

Building on the \emph{local} generalized method of moments, \emph{grf}
estimates IATEs, $\theta(x)$, as
$$\left( {{{\hat
\theta }^{grf}}(x),\hat
\eta _{IATE}^{grf}(x)} \right)
\in \mathop {\arg
\min }\limits_{\theta
(x),\eta (x)}
\;{\left\|
{\sum\limits_{i = 1}^N
{w_i^{grf}(x){\psi
^{grf}}({o_i};\theta (x),\eta
_{IATE}^{grf}(x))} }
\right\|_2},$$
where the weights, $w_i^{grf}(x)$, are
obtained from an honest Random Forest whose gradient-based splitting
rule is designed to maximize treatment heterogeneity in the daughter
leaves. The required \emph{Honesty} assumption involves a data
partitioning strategy to prevent overfitting. The subsample drawn for
each tree is further split into two halves: one for building the tree
structure and the other one for estimating the parameters of interest --
in this case, the tree weights which are subsequently aggregated into
forest weights. As all observations alter between the two halves across
trees when subsampled, we call this procedure `one-sample honesty'. The
forest weights, summing up to 1, can be seen as a measure of relevance
of the observation $i$ for the estimation of $\theta(x)$. Given the
forest weights $w_i^{grf}(x)$, the
solution to the given optimization problem is:
\begin{align*}
\hat \theta^{grf}(x) & = \widehat {IATE^{grf}}(x) \\
& =\left(\sum\limits_{i =1}^N w_i^{grf}(x)\left(\tilde d_i - \bar{\tilde{d}}_w\right)\left(\tilde d_i - \bar{\tilde{d}}_w\right)'\right)^{-1} \left(\sum\limits_{i = 1}^N	w_i^{grf}(x)\left(\tilde d_i - \bar{\tilde{d}}_w\right)\left(y_i - \bar y_w \right) \right),
\end{align*}
where ${\bar{\tilde{d}}_w} =
\sum\limits_{i = 1}^N
{w_i^{grf}(x){{\tilde d}_i}} $
and ${\bar y_w} =
\sum\limits_{i = 1}^N
{w_i^{grf}(x){y_i}} $.\footnote{Compare with
	eq.~(19) in \citetalias{athey2019grf}.}

As the splits are chosen to maximize treatment heterogeneity across
daughter leaves, to mitigate confounding effects at early splits, \citetalias{athey2019grf}
recommend to partial out effects of the covariates $X$ on the
outcome $Y$ and treatment assignment $D$ as in \citet{robinson1988root}.
This means building the forest using locally centred values
$y_i^{cent} = {y_i} - {\hat
\mu ^{( - i)}}({x_i})$ and
$\tilde d_i^{cent} =
{\tilde d_i} - \hat p_d^{( -
i)}({x_i})$ for all observations, where
$\mu(x)=E(Y| X=x)$ and the superscript $(-i)$ denotes
that the prediction did not use observation $i$. The best practice
is to estimate $\mu(x)$ and $p_{d}(x)$, e.g., by
$K$-fold cross-fitting as mentioned in \citetalias{athey2019grf}. However, the
available \emph{grf} package performs local centring by subtracting
out-of-bag predictions\footnote{In \citetalias{athey2019grf} out-of-bag predictions are
	called leave-one-out predictions as the final prediction for
	observation $i$ averages predictions that are obtained from trees
	that did not use observation $i$ for splitting.} obtained from
regression forests specifically trained to predict the outcome variable
$Y$ and the probabilities of treatment assignment, given the
covariates $X$. Despite its computational attractiveness, this
implementation violates the requirement of complete separation of
training and prediction data consistent with the theoretical results.
Also, the simulation results in \ref{appendix-c1} reveal the necessity of
$K$-fold cross-fitting over out-of-bag predictions for local
centring to remove bias.

The asymptotic results for ${\hat
\theta ^{grf}}(x)$ are based on their
linear approximation motivated by the method of influence functions.
\citetalias{athey2019grf} prove that the two are coupled and asymptotic properties of one
apply to the other. Since the linear approximation can be interpreted as
an honest Causal Forest estimator introduced in \citet{wager2018estimation},
their asymptotic result can be applied to establish asymptotic
Gaussianity of ${\hat \theta
^{grf}}(x)$, given that $\theta(x)$ and
$\eta^{grf}_{IATE}(x)$ are consistently
estimated.\footnote{This can be achieved, e.g., via honest forests as
	outlined in \citetalias{athey2019grf} in Theorem 3.} The linear approximation is also
leveraged to derive an estimator of the pointwise standard errors for
${\hat \theta
^{grf}}(x)$. Derivation of the variance of the
linear approximation yields two terms. One term can be in general
consistently estimated by means of regressions and is a constant in case
of $IATE(x)$. The other can be seen as an output of a regression
forest with weights $w_i^{grf}(x)$ and
the score function as outcome variable. \citetalias{athey2019grf} propose to estimate this
term by a variant of a so-called bootstrap of little bags. This method
was introduced in \cite{sexton2009standard}. \citetalias{athey2019grf} motivate it by the
observation that ``\textit{building confidence intervals via half-sampling
	-- whereby evaluating an estimator on random halves of the training data
	to estimate its sampling error -- is closely related to the bootstrap
	\citet{efron1982}}''. A computationally efficient implementation to estimate
within one forest, both the forest weights and the element of the
variance by half-sampling, requires drawing for each small bag of trees
a random half of the sample that is available for further subsampling.
Honesty additionally requires the trees to be built on half of the
subsample, and the weights
$w_i^{grf}(x)$ to be computed on the
half that was not used to build the tree. Thus, each
$w_i^{grf}(x)$ is estimated on
approximately a quarter of all the trees in the forest when the
subsampling rate is close to 1.\footnote{Figure \ref{fig:a1} in \ref{appendix-a3}
	captures the \emph{grf} procedure graphically.} Consistent estimation
of the variance term yields asymptotically valid confidence intervals
for the IATE at point $x$.

Like \emph{dml}, \emph{grf} offers comprehensive estimation of treatment
effects at all levels of granularity. However, unlike \emph{dml}, which
estimates all effects in two steps by estimating the
$\Gamma
_{m,l}^{dml}\left( {{o_i};\hat
\eta _{m,l}^{dml, - k(i)}({x_i})}
\right)$ first, and subsequently
regressing them on a constant, group indicators or regressors to obtain
ATE, GATE and IATE estimators, \emph{grf} takes a more direct approach.
Specifically, \emph{grf} estimates IATEs ``directly'', while ATE and
GATEs are estimated through additional regressions. The \emph{grf}
package implements ATE estimation (treatment $m$ vs no treatment)
by a variant of the AIPW estimator, plugging in the estimates from IATE
estimation for the nuisance parameters:\footnote{The equations for
	estimating the ATE are reconstructed from the \emph{grf} R package
	(version 2.3.1). The derivation involves code implemented in the
	functions ``get\_scores.R'' and ``average\_treatment\_effect.R''.}
\begin{align*}
 \widehat {AT{E^{grf}}}(m,0) = \hat \theta _{m,0}^{grf} = & \ \frac{1}{N}\sum\limits_{i= 1}^N {\Gamma_{m,0}^{grf}} ({o_i};\hat \eta_{ATE(m,0)}^{grf}({x_i})), \\
 {\Gamma_{m,0}^{grf}} ({o_i};\hat \eta_{ATE(m,0)}^{grf}({x_i}))  = & \ \hat{\theta}_{m,0} + \frac{\underline{1}(d_i = m)}{\hat{p}_m(x_i)} (y_i - \hat{\mu}_m(x_i)) - \frac{\underline{1}(d_i = 0)}{\hat{p}_0(x_i)} (y_i - \hat{\mu}_0(x_i)), \\
	&\text{where} \quad \hat{\mu}_0(x) = \hat{\mu}(x_i) - \sum_{d>0} \hat{p}_d(x_i) \hat{\theta}_{d,0}(x_i) \\
	&\text{and} \quad \hat{\mu}_m(x_i) = \hat{\mu}_0(x_i) + \hat{\theta}_{d,0}(x_i).
\end{align*}

As default, the \emph{grf} package estimates the nuisance parameters
$\mu(x), p_{1}(x), \ldots, p_{M-1}(x)$ by
out-of-bag predictions from Random Forests, and
$\theta_{1,0}(x), \ldots, \theta_{M-1,0}(x)$ by a
weighted local moment condition with forest-based weights. Derivations
in \ref{appendix-a4} show the equivalence of the \emph{grf} score
$\psi
_{m,0}^{grf}(O;{\theta
_{m,0}},\eta _{ATE(m,0)}^{grf}(X)) =
\Gamma _{m,0}^{grf}(O;\eta
_{ATE(m,0)}^{grf}(X)) - {\theta
_{m,0}}$ to the DR score of \citet{robins1995semiparametric}
for the ATE. It can therefore be expected that the \emph{grf}
estimator of the ATE is asymptotically efficient when all nuisance
parameters are consistently estimated. Due to a different set of
nuisance parameters, consistent estimation of the ATE is robust only to
misspecifications of $\mu(x)$ and $\theta_{1,0}(x),
	\ldots, \theta_{M-1,0}(x)$ (see \ref{appendix-a4}). Consistent
estimation of the ATE based on the above \emph{grf} score therefore
requires consistent estimation of the propensity scores. \ref{appendix-a4}
also shows Neyman-orthogonality of the \emph{grf} score for the ATE,
guaranteeing that the moment function is locally insensitive to errors
in the nuisance parameter estimates. To obtain the GATE estimators,
$\Gamma
_{m,0}^{grf}({o_i};\hat \eta
_{ATE}^{grf}({x_i}))$ are regressed on
$Z$, as in \emph{dml}. Inference for the ATE and GATEs is conducted
as in \emph{dml}.

\subsection{Modified Causal Forest}\label{43}
\defcitealias{wager2018estimation}{WA18}  
\defcitealias{lechner2018modified}{L18}  

\citet[abbreviated as L18 in this section]{lechner2018modified} builds 
on \citet[abbreviated as WA18]{wager2018estimation} and introduces the Modified Causal
Forest estimator, \emph{mcf}. One of the procedures proposed by \citetalias{wager2018estimation}
builds trees by taking the outcome $Y$ as dependent variable and
finding splits that maximize treatment effect heterogeneity. The tree
estimator of an IATE at point $x$ is obtained from the final leaf
containing $x$ by differencing the average outcomes of the
treatment groups evaluated on the subsample that was not used to build
the tree. The final forest estimator of the IATE at point $x$ is
the average of all tree estimates. This procedure works well in an
experimental design with heterogeneous treatment effects but may do
poorly in the presence of confounding (see \citetalias{wager2018estimation}). This observation led
to further extensions addressing the confounding issue.

\emph{grf} introduces local centring inside its Random Forest to remove
the confounding effects by transforming the data when searching for
splits that maximize effect heterogeneity. \emph{mcf} takes a different
approach. Motivated by the fact that in the presence of selection bias
the difference of the outcome means of treated and controls within all
leaves will not correspond to the means of the true effects, \emph{mcf}
proposes a splitting criterion targeting the minimisation of the MSE of
IATE directly. In addition, a penalty is added penalizing squared
differences between propensity scores. Taken together, the chosen split
is predictive for $Y$, the differences of $Y$ across treatment
groups, and the conditional treatment assignment probabilities.

The IATE is identified as the difference of two outcome regressions
$\mu_{d}(x)$ in the different treatment groups. The
splitting criterion of the \emph{mcf} is therefore based on minimizing
the sum of the expected MSEs at a given point $x$ over all unique
treatment combinations, i.e., all IATEs relating to different treatment
pairs are of equal importance:
\begin{align*}
&{\overline{MSE}_x} =
\sum\limits_{m = 1}^{M - 1}
{\sum\limits_{l = 0}^{m - 1}
	{MSE\left( {\widehat {IATE}(m,l;x)}
		\right)} } , \\
&MSE\left( {\widehat
	{IATE}(m,l;x)} \right) = MS{E^m}(x) +
MS{E^l}(x) - 2MC{E^{m,l}}(x),
\end{align*}
where $MS{E^d}(x) = E\left(
{{{\hat \mu }_d}(x) -
\mu _d^0(x)} \right)$ and $MC{E^{m,l}}(x) = E\left({{{\hat \mu }_m}(x) -\mu _m^0(x)}\right)\left( {{\hat\mu }_l}(x)\right. \allowbreak - \allowbreak \left. \mu _l^0(x)\right)$.

For finding splits based on this criterion, the
$MSE^{d}$ and $MCE^{m,l}$
elements need to be estimated in the daughter leaves. Let
$N_{S(x)}^d$ denote the number of
observations with treatment value $d$ in a certain stratum (leaf)
$S(x)$, defined by the values of the covariates $x$. Then a
`natural' choice to estimate the $MSE^{d}$'s in leaf
$S(x)$ is:
\begin{equation*}
	\widehat{MSE_{S(x)}^d} =
	\frac{1}{N_{S(x)}^d} 
	\sum_{i=1}^{N} \underline{1}(x_i \in S(x)) \underline{1}(d_i = d)
	\left( \hat{y}_{S(x)}^d - y_i \right)^2,
\end{equation*}
where $\hat y_{S(x)}^d$ is an average of the observed outcomes
in treatment $d$ in leaf $S(x)$. To compute the $MCE$,
the \emph{mcf} uses the closest neighbour (in terms of similarity w.r.t.~$x$) 
available in the other treatment (which is denoted by
$\tilde y_{(i,d)}$ below).\footnote{Implementational details
	on finding the closest neighbours for the MCE estimation are in
	\ref{appendix-b31}.}
\begin{multline*}
	\widehat{MCE_{S(x)}^{m,l}} = \\
	\frac{1}{N_{S(x)}^m + N_{S(x)}^l} 
	\sum_{i=1}^{N} \underline{1}(x_i \in S(x))
	\left( \underline{1}(d_i = m) + \underline{1}(d_i = l) \right)
	\left( \hat{y}_{S(x)}^m - \tilde{y}_{(i,m)} \right)
	\left( \hat{y}_{S(x)}^l - \tilde{y}_{(i,l)} \right), \\
		\tilde{y}_{(i,d)} =
	\begin{cases}
		y_i, & \text{if } d_i = d, \\
		\tilde{y}_{(i,d)}, & \text{if } d_i \neq d.
	\end{cases}
\end{multline*}

Analysing the estimator of $\overline{MSE}_x$ reveals that its minimization favours
splits maximizing the differences of all
$\hat y_{S( \cdot
)}^d$ between the daughter leaves. Additionally, the
splits favour large differences between
$\hat y_{S( \cdot
)}^m$ and $\hat y_{S(
\cdot )}^l$ in the same leaf. In case of
a binary treatment, this is equivalent to maximising treatment effect
heterogeneity, as in \citetalias{wager2018estimation} and \citetalias{athey2019grf},
under no selection
into treatment. However, for multiple treatments, the \emph{mcf}
asymptotically targets treatment effect heterogeneity across daughter
leaves while simultaneously increasing treatment outcome variability
within leaves. Overlooking this dual focus may lead to splits that,
while maximizing treatment effect heterogeneity, fail to minimise the
sample MSE of parameters within leaves. Thus, \emph{mcf} utilizes a
splitting strategy that jointly considers all treatment combinations.

Although asymptotically confounding should be taken care of by the
splitting rule derived above, the \emph{mcf} adds a penalty term to the
splitting criterion as an additional safeguard. As before, let denote 
$S(x')$ and $S(x'')$ denote the values of the covariates in the
daughter leaves resulting from splitting some parent leaf. \citetalias{lechner2018modified} proposes
to add the following penalty to a combination of the two `final' MSEs in
the daughter leaves:
$$penalty(x',x'')= \lambda \left\{ {1 -
\frac{1}{M}\sum\limits_{d
= 0}^{M - 1} {{{\left( {P(D = d| X
\in S(x')) - P(D = d| X
\in S(x''))}
\right)}^2}} }
\right\}, \lambda >0.$$
The probabilities, local estimates of the propensity score, are computed
as relative shares of the respective treatments in the potential
daughter leaves. The penalty term is zero if the split leads to a
perfect treatment prediction of the treatment probabilities. It reaches
its maximum value, $\lambda$, when all probabilities are equal. Thus, the
algorithm prefers a split that is also predictive for $P(D
= d\left| {X = x} \right.)$. Of course, the choice of
the exact form of this penalty function is arbitrary. Furthermore, there
is the issue of how to choose $\lambda$ (\emph{without} expensive additional 
computations) which is taken up again in \autoref{522}.

Each tree is built on a subsample drawn from one half of the data, while
the other half serves for the estimation of the IATE at point $x$.
We name this procedure `two-sample honesty' as data points used to place
the splits (``split-construction data'') will not be used to estimate
the treatment effects (``leaf-populating data'') and vice
versa.\footnote{See Figure \ref{fig:a2} in \ref{appendix-a3} for a graphical
	representation.} The final estimator based on $B$ trees takes a
form:
$$\widehat {IAT{E^{mcf}}}(m,l;x) =
\hat \theta _{m,l}^{mcf}(x) =
\frac{1}{B}\sum\limits_{b
= 1}^B {\sum\limits_{i =
1}^{N/2} {w_{i,b}^{mcf}({d_i},{x_i};x,m,l){y_i}
= } \sum\limits_{i = 1}^{N/2}
{w_i^{mcf}({d_i},{x_i};x,m,l){y_i}} }, $$
where $w_{i,b}^{mcf}(d_i, x_i; x, m, l) =
\left( \frac{\underline{1}(d_i = m)}{N_{S_b(x)}^m} - \frac{\underline{1}(d_i = l)}{N_{S_b(x)}^l} \right)
\underline{1}(x_i \in S_b(x))$ represent the weights for the $IATE(x)$ estimate in tree
$b$. This estimator is differencing the average outcomes of the
treatment groups evaluated on the ``leaf-populating'' subsample in each
tree leaf containing point $x$,
${S_b}(x)$, and subsequently averages
these differences across trees. As in the \emph{grf}, the \emph{mcf}
forest weight
$w_{i}^{mcf}(d_{i},x_{i};x,m,l)$
captures how important observation $i$ is for estimation of the
parameter of interest at point $x$. While the \emph{mcf} weights
weigh the observed outcome directly, \emph{grf} applies the forest
weights in a locally weighted moment function. There is also a
difference in the number of weights estimated in each method due to
different honesty concepts. For a given $B$ and $N$,
\emph{grf} estimates $N$ weights on approximately $B/4$ trees,
while \emph{mcf} estimates $N/2$ weights on $B$ trees as
illustrated in \ref{appendix-a3} in Figures \ref{fig:a1} and \ref{fig:a2}. Thus, \emph{grf} has
an advantage in smoothing over more observations, while \emph{mcf} has
an advantage in the precision of the weights, potentially influencing
the finite sample properties of both methods.

The \emph{mcf} estimators of GATE and ATE are averages of corresponding
group or sample $\widehat
{IAT{E^{mcf}}}(m,l;{x_j})$'s. By construction,
they take the form of weighted observed outcomes. The \emph{mcf} GATE
estimator incorporates into the weighted average not only observed
outcomes of the observations in the pre-specified group but potentially
also observed outcomes of close neighbouring groups \footnote{This in
	general depends on whether variables $Z$ are among the splitting
	variables, type of $Z$, and the sample size. E.g., in settings with discrete $Z$ and when
	some trees do not (yet) split on all groups defined by a single value
	of $Z$, the observed outcomes from other groups that populate the
	same final leaves as $x_{j}$'s from group $z$
	are incorporated into the weighted average. When all
	trees split on all groups, only observed outcomes from group $z$
	are incorporated into the weighted average.}, defined by the non-zero
forest weights in the IATEs.

\ref{appendix-a1} contains the (so-far missing) theoretical guarantees for
the \emph{mcf}. Under similar assumptions as in \citetalias{wager2018estimation}, the \emph{mcf}
estimators of the IATEs, GATEs, and ATEs are consistent and pointwise
asymptotically normal. The \emph{mcf} inference procedures exploit the
fact that \emph{mcf} predictions, like for many Random Forest based
estimators, can be computed as weighted means of $Y$. The details
of this weights-based inference are presented in \ref{appendix-a2}, while
\ref{appendix-a3} explains some of the difference to the \emph{grf} at the
implementation level.

\subsection{Overview}\label{44}

This section summarizes the statistical guarantees of the three CCMLs
for estimation of ATE, GATE, and IATE. For the ATE, \emph{dml} and
\emph{grf} provide the strongest guarantees, as they are based on the DR
efficient score, and are expected to perform well in terms of bias, MSE
and inference in the simulations.\footnote{Assuming a certain quality of
	the nuisance parameter estimators as mentioned in \autoref{41} and \autoref{42}} 
The \emph{mcf} is also expected to perform well but with a
somewhat larger MSE as it does not necessarily reach the efficiency
bound.

Regarding GATE, all methods guarantee consistency and asymptotic
normality. \emph{dml} and \emph{grf} differ from the \emph{mcf} in the
utilization of the observed data points. While \emph{dml} and \emph{grf}
compute the average of estimated components of the DR score for
observations in the corresponding group of interest, \emph{mcf} builds a
weighted average of $Y$'s, potentially incorporating information
from groups close to the group of interest as determined by the forest
structure, as mentioned in \autoref{43}. This will affect the efficiency
of the GATE estimators in different scenarios as captured in the
following proposition.

\begin{prop} Let $J$ denote the number of groups for
	which GATEs are to be estimated and let $\mathit{\Pi}$ be a $\mathit{J \times 1}$
	vector containing the non-zero population shares of observations in the
	corresponding groups. Suppose that the total number of groups increases from $J$ to $J'$ and
	the population share of group $j$ under $J$ groups, $\pi_j$, decreases to $\pi_{j'}$ under $J'$ groups.
	Then, the asymptotic variances of the efficient \emph{dml} and \emph{grf} GATE
	estimators increase for each group $j$ within $J$
	proportionally to $\pi_j/\pi_{j'}$, while
	the variance of \emph{mcf} estimator of the GATE does not depend on the
	number of groups and remains stable. This implies $Var\left( {\widehat {GAT{E^{mcf}}}\left( {m,l;{z_j}} \right)} \right) \le Var\left( {\widehat {GAT{E^{dml/grf}}}\left( {m,l;{z_j}} \right)} \right)$
	for large $J$.
	\label{prop1-main}
\end{prop}

For the IATE, there are no specific statistical guarantees for
\emph{dml}. There is an upper bound on the error of the DR-learner
providing a general guarantee. \emph{grf} and \emph{mcf} guarantee
pointwise consistency and asymptotic normality and are both expected to
yield lower MSEs in the simulations than \emph{dml}-based DR-learner.
For \emph{grf}, the convergence rate is known and correspondingly slower
due to the non-parametric nature of the estimator. Its speed depends on
the subsampling rate which is implicitly restricted by the forest
parameters (see Theorem 5 in \citetalias{athey2019grf}). All the above-mentioned conjectures
are captured in Table \ref{tab2}.

\begin{table}[h]
	\centering
	\caption{Selected statistical properties of the CCMLs}
	\label{tab2}
	\begin{tabularx}{\textwidth}{X>{\RaggedRight\arraybackslash}X>{\RaggedRight\arraybackslash}X>{\RaggedRight\arraybackslash}X}
		\toprule
		\textbf{Approach} & \emph{ATE} & \emph{GATE} & \emph{IATE} \\
		\midrule
		\textbf{\emph{dml}} & 
		Consistent, asymptotically normal, asymptotically efficient & 
		Consistent, asymptotically normal, uniform convergence & 
		Depends on implementation of the smoothing (regression) step \\
		\midrule
		\textbf{\emph{grf}} & 
		Consistent, asymptotically normal, asymptotically efficient & 
		Consistent, asymptotically normal, uniform convergence & 
		Consistent, asymptotically normal, pointwise convergence known (slow) convergence rate \\
		\midrule
		\textbf{\emph{mcf}} & 
		Consistent, asymptotically normal & 
		Consistent, asymptotically normal, pointwise convergence & 
		Consistent, asymptotically normal, pointwise convergence \\
		& & & \\
		& & More efficient for large number of groups (compared to \emph{dml} and \emph{grf}) &  \\
		\bottomrule
	\end{tabularx}
\end{table}

\section{Monte Carlo study}\label{5-mcs}

\subsection{Concept}\label{51}

In this section, we compare the finite-sample performance of CCMLs in
estimating average effects across three specified aggregation levels in
a Monte Carlo simulation. This comparison includes evaluating both the
precision of point estimates and the robustness of inference procedures.

In executing this Monte Carlo approach, we meticulously construct
artificial datasets, varying numerous elements of the data generating
process (DGP). These elements include the degree of selection into
treatment (referred to as selectivity henceforth), the type and quantity
of covariates, sample size, diverse functional forms, varying degrees of
effect heterogeneity, the influence of covariates on outcomes and
heterogeneity, the share of treated within the sample, and the number of
treatments resulting in 77 different DGPs.\footnote{Table \ref{tab:b2} in
	\ref{appendix-b25} gives the list of the scenarios investigated. \ref{appendix-c} 
	collects the tables that contain the corresponding results.} Despite
its significant computational demands, this approach offers a
comprehensive exploration of various scenarios, surpassing the more
limited scope of Monte Carlo studies typically employed in papers
introducing a new methodology.

The subsequent subsection provides a concise overview of the simulation
designs' components. The procedure involves initially generating random
data, applying the different estimators to this training dataset,
predicting effects on a separate dataset of equivalent size derived from
the same DGP, saving the results, and repeating these steps $R$
times. Subsequently, we calculate a range of performance metrics that
capture different dimensions of the accuracy and reliability of both
estimation and inference processes.

\subsection{Key features of the simulation study}\label{52}

\subsubsection{Data Generating Process}\label{521}

The simulated (i.i.d.) data consist of the covariates, the treatment,
and the potential outcomes. The simulation of these components will be
discussed in turn (for additional details see \ref{appendix-b}).

The simulation involves generating a range of 10 to 50 independent
covariates ($p=10$, 20, 50). These covariates are normally,
uniformly, or binomially (dummy variables) distributed, or as
combinations of the three types. Among these $p$ covariates, the
first $k$ covariates influence both the selection into treatment
and the potential outcomes. The effect of these $k$ covariates is
modelled to decrease linearly, ranging from 1 to $1/k$. The base
specification consists of 20 uniformly and normally distributed
covariates, half of them relevant.\footnote{When covariates are
	simulated from both, the normal and uniform distributions, the
	1$^{\text{st}}$, 3$^{\text{rd}}$, 5$^{\text{th}}$,
	... covariates are drawn from the uniform distribution.}

\begin{figure}[h!]
	\caption{Distribution of treatment probabilities in treated and non-treated population}
	\label{fig1}
	\includegraphics[width = \textwidth]{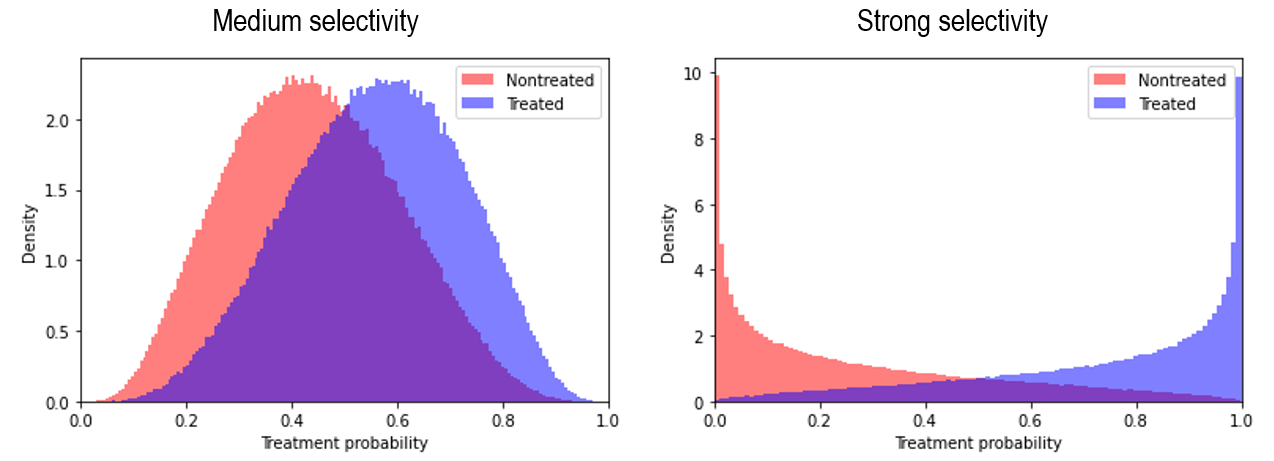}
	\caption*{\footnotesize \textsf{Note: Two treatments and 50\% treatment share (base specification). Figures are based on 1'000'000 observations.}}
\end{figure}

The selection process is based on a linear index function of the
$k$ relevant covariates plus noise. The quantiles of this index
function are used to generate the treatments. The base specification
considers cases with random selectivity (experiment), medium selectivity
(true $R^2$ of about 10\%), and strong selectivity
(true $R^2$ of about 42\%). Cases of two and four
treatments with equal as well as asymmetric treatment shares are
considered. The base specification consists of 2 treatments with equal
treatment shares. Figure \ref{fig1} shows the distribution of the true treatment
probabilities for the strong and medium selectivity case with two
treatments and 50\% treatment share.

\begin{figure}[h!]
	\caption{Shape of potential outcomes for different shapes of IATEs (base specification)}
	\label{fig2}
	\includegraphics[width = \textwidth]{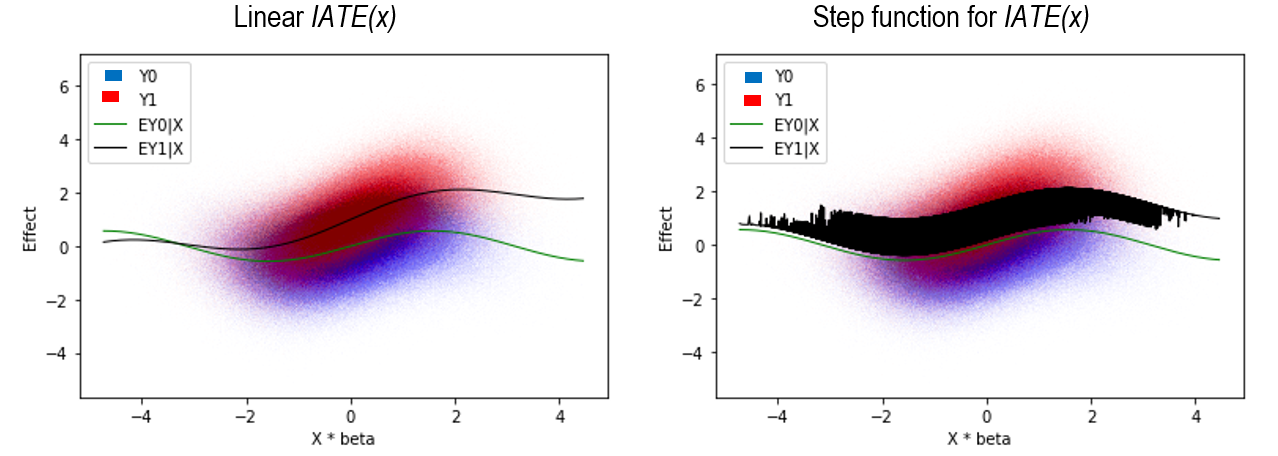}
	\caption*{\raggedright \footnotesize \textsf{Note: Figures are based on 1'000'000 observations.}}
\end{figure}

The non-treatment potential outcomes are obtained by simulating the
expected non-treatment potential outcome as a sine-function of the
linear index of the covariates plus noise. The relevance of the
sine-function relative to the noise level is varied in the simulations
from cases with true $R^2$ of 0 to 45\% (base
specification: 10\%). The potential treatment outcomes are obtained by
adding simulated IATEs plus noise to the expected non-treatment
potential outcomes.\footnote{The 2 (2 treatments) or 4 noise terms (4
	treatments) used to simulate potential outcomes are independent of
	each other.} The IATEs are generated as functions of the linear index
or as a smooth function of the first two, most important, covariates. In
the former case, a linear function, a logistic function, and a quadratic
function is specified. Finally, the smooth function approach follows
closely the specification of \citetalias{wager2018estimation}. In all these cases the ATE is close
to one. The case of zero IATEs, and thus zero ATE, is considered as
well. The specification of the IATE as a smooth function is used in the
base specification presented in \autoref{53}.\footnote{The base
	specifications with different degrees of selectivity and types of
	IATEs are captured in rows 1-15 in Table \ref{tab:b2} in \ref{appendix-b25}.} The
smooth function, when plotted as a function of the linear index,
exhibits a pattern that resembles a step function. Figure \ref{fig2} shows the
expected and realised potential outcomes and their relation to the
linear index. Thus, it summarizes main properties of the base
specification following \citetalias{wager2018estimation}, as well as of the linear specification of
the IATE. Similar plots for the other specifications can be found in
\ref{appendix-b24}.

The GATEs depend only on the first covariate, which is uniformly
distributed if there are uniformly as well as normally distributed
covariates in the DGP. This continuous covariate is split in groups with
the same expected sizes and corresponding indicator variables are
created. Thus, these dummy variables have no direct effect on the DGP.
They are passed to the GATE estimators.

A final aspect of the DGP is the sample size. We mainly consider sample
sizes of 2'500 and 10'000 as compromise between computational costs and
practical relevance.\footnote{$N$=40'000 is also considered, but,
	for computational reasons, only for one specification.} To keep the
noise from the simulations on the performance measures stable (at least
for the estimators of ATE and GATEs, which can be expected to show
$\sqrt N - $convergence), the
number of replications, $R$, declines at the same rate as the
sample increases. Thus, the results for $N$=2'500 are based on
$R$=1'000, while $R$ declines to 250 for $N$=10'000.

In our simulation study, it's crucial to highlight that
we adopt a methodology designed to replicate repeated sampling inference
more closely, especially in scenarios involving stochastic covariates.
To achieve this, the samples used for computing the
out-of-training-sample effects are freshly drawn from the same Data
Generating Process (DGP) for each replication. Additionally, these
samples are matched in size to the training sample. This approach
contrasts with the method used in \citet{knaus2021machine}, where the sample
for calculating effects is drawn once prior to the initial replication
and remains constant across subsequent replications, even when similar
DGPs are employed.

The methodology chosen for our study, however, comes with certain
trade-offs. Notably, for each IATE as a function of covariate values
($IATE(x)$), there exists only a single true and estimated value
across all replications. This uniqueness arises because all or at least
some covariates are continuous, leading to a scenario where each
specific covariate value appears only once in the simulations.
Consequently, for the IATEs this aspect of our simulation methodology
precludes the possibility of estimating moments other than the
expectation. This is an important consideration to bear in mind when
interpreting the results and the applicability of our findings.

\subsubsection{Estimators}\label{522}

Below, we present the results of the Modified Causal Forest and the
Generalized Random Forest, both with the outcome variable centred prior
to estimation, \emph{mcf-cent} and \emph{grf-cent}, as well as of
Double/debiased Machine Learning with normalized weights,
\emph{dml-norm}. For IATE estimation, we also include a more efficient
version of the centred \emph{mcf} for which no inference is available
(see below), \emph{mcf-cent-eff}. The tables in Appendix C contain
additional results for the uncentred \emph{mcf} and \emph{grf}, standard
(non-normalized) \emph{dml}, and \emph{OLS}. We describe the
implementation of these estimators briefly in this section and refer the
interested reader to \ref{appendix-b3} for details.

While \ref{appendix-b31} details the implementation of the \emph{mcf}
further, at least 2 points merit some more discussion. The first point
concerns the penalty parameter, $\lambda$. In the simulations, $\lambda$ is set equal to
$Var(Y)$. $Var(Y)$ corresponds to the MSE when the effects are
estimated by the sample mean without any splits. Thus, it provides some
ad-hoc benchmark for plausible values of $\lambda$. Generally, decreasing the
penalty increases biases and reduces variances, et vice
versa.\footnote{The sensitivity of the centred version of the \emph{mcf}
	to the choice of the penalty parameter seems to be generally very low.}
The simulations below will show that biases are more likely to occur
when selectivity is strong. Thus, if a priori knowledge about the
importance of selectivity is available, then the researcher might adjust
the penalty term accordingly.

Secondly, if inference is not a priority, like when using the IATEs as
inputs into the training of an optimal assignment algorithm, the
efficiency loss inherent in the \emph{mcf}'s two-sample honesty approach
can be avoided by cross-fitting, i.e., by repeating the estimation with
exchanged roles of the two samples and averaging the two (or more)
estimates. However, in such a case it is unclear how to compute the
weights-based inference for the averaged estimator \emph{mcf-cent-eff}
as the two components of this average are correlated.\footnote{For such
	a cross-fitted estimator (e.g., computed as mean of the single
	estimators), conservative inference could be obtained by basing
	inference on normality with a variance taken as average over the
	variances of the single estimations.} A similar efficiency loss can be
avoided in the \emph{grf} procedure by switching off the half-sampling
when inference is not relevant. However, the current implementation in
the \emph{grf} package does not support this option.

As mentioned in \autoref{42}, \emph{grf} performs local centring inside
the Random Forest to remove confounding bias. The default \emph{grf}
local centring, labelled here as \emph{uncentred}, refers to local
centring that uses out-of-bag predictions of the outcome and treatment
assignment from Random Forests trained on the full sample, to calculate
the residuals. The simulation study includes an additional local
centring strategy, labelled as \emph{centred}. The difference between
the centred and the uncentred \emph{mcf} and \emph{grf} is that the
former uses a transformed outcome variable. This transformation
subtracts a Random Forest prediction of $E(Y| X)$ from the
observed $Y$ (obtained with 5-fold-cross-fitting), and thus purges
them from much of the influence of $X$.\footnote{Note that the
	\emph{grf} package does not provide centring of this kind. We added
	this estimator as the original version used in the \emph{grf} package
	performed poorly in many of our DGPs for reasons discussed in \autoref{42}.}
For \emph{grf}, GATEs and ATE are estimated via linear
regression in which a variant of a local AIPW estimator is regressed on
group indicators or a constant.

Compared to the standard \emph{dml}, the normalized \emph{dml} is more
robust to extreme values of the estimated propensity score. It is
obtained by normalizing the weights that are implicit in the \emph{dml}
scores. While it is straightforward to use \emph{dml} for ATE, it is
less straightforward to use them for heterogeneity estimation. Here,
GATEs and IATEs are obtained by regression-type approaches in which the
estimated components of the \emph{dml} scores serve as dependent
variable. The GATEs are obtained as OLS-coefficients of a saturated
regression model with the indicators for the groups defined by the
discrete variable $Z$ as independent variables. IATEs are computed
by using $X$ as independent variables either in a regression Random
Forest or in an OLS regression. When OLS is used, inference is based on
the heteroscedasticity-robust covariance matrix of the corresponding
coefficients. No inference is obtained for the Random Forest based
IATEs.

All estimators are not tuned as computational costs would be prohibitive
given the already extensive simulations. Instead, the default values
provided in the respective software packages are used.

\subsubsection{Performance measures}\label{523}

The main performance measures are the biases of the effects and their
standard errors, the standard deviation of the effects, the mean
absolute error, and the root mean squared error (RMSE) of the effects.
The coverage probability (CovP) for the 95\% confidence interval is
reported to gauge the quality of the inference.\footnote{In addition to
	the CovP for the 80\% interval, the tables in the \ref{appendix-c} also
	report the skewness and excess kurtosis of the estimators, which are,
	however, in a large majority of cases in the `normal', unproblematic
	ranges.}

Note that, as mentioned above, the standard deviation, and thus the bias
of the estimated standard error, cannot be computed for the IATE due to
the simulation design. Whenever several parameters are involved (as for
the GATEs and IATEs), the performance measures are computed for each
parameter and then averaged.

\subsection{Results}\label{53}

In this subsection, we analyse \emph{dml-norm}, \emph{grf-cent},
and \emph{mcf-cent} for the \citetalias{wager2018estimation} base specification Data Generating
Processes (DGPs) and sample sizes of $N$=2,500 and $N$=10,000
(when there is no confusion, in this subsection, we will drop the
\emph{-norm} and \emph{-cent} ending in the text). Additionally, the
degree of selectivity within the data is varied.

Since it turned out that the relative performance of the estimators does
not only depend on the strength of selection into treatment, but also on
the specific parameter to be estimated, the results for the ATE, the
GATEs, and the IATEs are discussed in turn.

\subsubsection{Average treatment effects}\label{531}

Table \ref{tab3} shows the results for the ATE. If there is no selectivity, like
in an experiment, we obtain the expected results: the efficient
estimators, \emph{dml} and \emph{grf}, are similar and outperform the
\emph{mcf}, at least when selectivity is not too strong. All estimators
are essentially unbiased, and empirical coverage is close to the nominal
level. Furthermore, when the sample size quadruples, the standard
deviation and RMSE halve, which is indicative of
$\sqrt N - $convergence.

The analysis reveals a notable trend with increasing selectivity: the
performance of the normalized double/debiased machine learning
(\emph{dml-norm}) estimator deteriorates, particularly in terms of bias.
This leads to a large increase in the RMSE.\footnote{As can be seen from
	Table \ref{tab:c15} in \ref{appendix-c1}, this behaviour appears for the not
	normalized \emph{dml} as well and is driven mainly by the bias of the
	point estimator.} One plausible explanation is that (despite the
normalisation) the double-robust score becomes more problematic when
propensity score values become more extreme. A similar increase in bias
is visible for the \emph{grf}, although it is not as extreme as for
\emph{dml}. The resulting bias impacts also their coverage rates, which
fall to very low levels. These issues appear to a much lesser extent for
the \emph{mcf}.

One summary coming from this table is that \emph{mcf} appears to be more
robust to stronger selectivity at the cost of some additional RMSE when
selectivity does not matter much. The results in \ref{appendix-c1} show that
these performance patterns with respect to the degree of selectivity
also appear for linear IATEs and non-linear IATEs. However, in the case
of quadratic IATEs it turns out that the uncentred (!) \emph{grf}
outperforms the centred \emph{grf} in terms of bias even for strong
selectivity. One possible explanation for this surprising behaviour is
that the Random Forest estimator used to do the centring does not
capture the shape of $E(Y| X=x)$ well.

Fixing selectivity to medium levels and varying the other parameters of
the DGP (\ref{appendix-c2}), suggests that the patterns observed in Table \ref{tab3}
qualitatively appear almost in all other DGPs as well. The most
remarkable case is when covariates are more important for the
non-treatment potential outcome (Table \ref{tab:c17}, Table \ref{tab:c21}): all
estimators become biased, which increases the RMSE and leads to coverage
far below nominal levels.

\begin{table}[h!]
	\sffamily
	\small
	\caption{Simulation results for average treatment effect (ATE)}
	\label{tab3}
	\resizebox{\textwidth}{!}{
		\begin{tabular}{lcccccccc}
			\toprule
			\multicolumn{3}{c}{} & \multicolumn{4}{c}{\textbf{Estimation of effects}} & \multicolumn{2}{c}{\textbf{Inference}} \\
			\midrule
			\textbf{Estimator} & Selec- & Sample & Bias & Mean & Std. dev & RMSE& Bias (SE) & CovP  \\
			& tivity & size & & absolute & & & (SE) & (95) in \\
			& & & & error& & &  & \% \\
			\midrule
			\multicolumn{1}{c}{(1)} & (2) & (3) & (4) & (5) & (6) & (7) & (8) & (9) \\
			\midrule
			\textbf{dml-norm} & None   & 2'500  & -0.002 & 0.034 & 0.043 & 0.043 & 0.005  & 97 \\
			\textbf{grf-cent} &        &        & 0.003  & 0.033 & 0.042 & 0.042 & 0.000  & 95 \\
			\textbf{mcf-cent} &        &        & 0.003  & 0.048 & 0.061 & 0.061 & -0.001 & 94 \\ \hdashline
			\textbf{dml-norm} & Med-   & 2'500  & 0.020  & 0.040 & 0.045 & 0.050 & 0.004  & 95 \\
			\textbf{grf-cent} & ium    &        & 0.037  & 0.046 & 0.042 & 0.056 & -0.000 & 85 \\
			\textbf{mcf-cent} &        &        & 0.037  & 0.057 & 0.061 & 0.072 & 0.000  & 91 \\ \hdashline
			\textbf{dml-norm} & Strong & 2'500  & 0.140  & 0.140 & 0.056 & 0.150 & 0.002  & 23 \\
			\textbf{grf-cent} &        &        & 0.090  & 0.091 & 0.046 & 0.101 & -0.006 & 41 \\
			\textbf{mcf-cent} &        &        & 0.046  & 0.060 & 0.058 & 0.074 & 0.008  & 92 \\ \midrule
			\textbf{dml-norm} & None   & 10'000 & 0.000  & 0.016 & 0.020 & 0.020 & 0.003  & 98 \\
			\textbf{grf-cent} &        &        & 0.002  & 0.017 & 0.021 & 0.021 & 0.000  & 95 \\
			\textbf{mcf-cent} &        &        & 0.004  & 0.022 & 0.028 & 0.028 & 0.002  & 97 \\ \hdashline
			\textbf{dml-norm} & Med-   & 10'000 & 0.010  & 0.019 & 0.022 & 0.024 & 0.003  & 94 \\
			\textbf{grf-cent} & ium    &        & 0.015  & 0.022 & 0.022 & 0.027 & -0.001 & 92 \\
			\textbf{mcf-cent} &        &        & 0.014  & 0.025 & 0.027 & 0.030 & 0.003  & 96 \\ \hdashline
			\textbf{dml-norm} & Strong & 10'000 & 0.079  & 0.079 & 0.032 & 0.085 & 0.000  & 28 \\
			\textbf{grf-cent} &        &        & 0.042  & 0.044 & 0.027 & 0.050 & -0.006 & 46 \\
			\textbf{mcf-cent} &        &        & -0.013 & 0.025 & 0.028 & 0.031 & 0.007  & 96 \\ \bottomrule
		\end{tabular}
	}
\vspace{-8pt}
\captionsetup{justification=justified,singlelinecheck=false}
\caption*{\footnotesize \textsf{Note: RMSE abbreviates the Root Mean Squared Error. \emph{CovP (95)} denotes the (average) probability that the true value is part of the estimated 95\% confidence interval. 1'000 / 250 replications are used for 2'500~/ 10'000 observations.}}
\end{table}

\subsubsection{Conditional average treatment effects with a small number of groups (GATEs)}\label{532}

Table \ref{tab4} shows that the relative performance of the different estimators
for the GATEs depends not only on the strength of selectivity, but also
on the number of groups for which a GATE is computed for as captured in \propref{prop1-main}.
Table \ref{tab4} shows the results for the cases of 5 and 40
groups, while Tables \ref{tab:c28} to \ref{tab:c30} in \ref{appendix-c2} show the intermediate
cases of 10 and 20 groups as well.

\begin{table}[h!]
	\sffamily
	\scriptsize
	\caption{Simulation results for group average treatment effects (GATEs)}
	\label{tab4}
	\resizebox{\textwidth}{!}{
		\begin{tabular}{lcccccccc}
			\toprule
			\multicolumn{3}{c}{} & \multicolumn{4}{c}{\textbf{Estimation of effects}} & \multicolumn{2}{c}{\textbf{Inference}} \\
			\midrule
			\textbf{Estimator} & Selec- & Sample & Bias & Mean & Std. dev & RMSE& Bias (SE) & CovP  \\
			& tivity & size & & absolute & & & (SE) & (95) in \\
			& & & & error& & &  & \% \\
			\midrule
			\multicolumn{1}{c}{(1)} & (2) & (3) & (4) & (5) & (6) & (7) & (8) & (9) \\
			\midrule
			\multicolumn{9}{c}{5 Groups} \\
			\midrule
			\textbf{dml-norm} & None   & 2'500  & -0.002 & 0.075 & 0.094 & 0.094 & 0.002  & 95 \\
			\textbf{grf-cent} &        &        & 0.004  & 0.075 & 0.094 & 0.094 & 0.000  & 95 \\
			\textbf{mcf-cent} &        &        & 0.002  & 0.090 & 0.091 & 0.114 & -0.010 & 83 \\ \hdashline
			\textbf{dml-norm} & Med-   & 2'500  & 0.020  & 0.081 & 0.081 & 0.101 & 0.001  & 95 \\
			\textbf{grf-cent} & ium    &        & 0.037  & 0.081 & 0.093 & 0.102 & -0.000 & 93 \\
			\textbf{mcf-cent} &        &        & 0.037  & 0.094 & 0.090 & 0.121 & -0.006 & 84 \\ \hdashline
			\textbf{dml-norm} & Strong & 2'500  & 0.140  & 0.157 & 0.123 & 0.187 & -0.005 & 74 \\
			\textbf{grf-cent} &        &        & 0.090  & 0.119 & 0.096 & 0.148 & -0.007 & 75 \\
			\textbf{mcf-cent} &        &        & 0.046  & 0.112 & 0.085 & 0.146 & 0.009  & 81 \\ \midrule
			\textbf{dml-norm} & None   & 10'000 & -0.001 & 0.037 & 0.046 & 0.046 & 0.001  & 95 \\
			\textbf{grf-cent} &        &        & 0.002  & 0.036 & 0.046 & 0.046 & 0.001  & 95 \\
			\textbf{mcf-cent} &        &        & 0.004  & 0.045 & 0.045 & 0.058 & -0.002 & 87 \\ \hdashline
			\textbf{dml-norm} & Med-   & 10'000 & 0.010  & 0.041 & 0.050 & 0.051 & 0.000  & 94 \\
			\textbf{grf-cent} & ium    &        & 0.016  & 0.041 & 0.048 & 0.051 & -0.001 & 92 \\
			\textbf{mcf-cent} &        &        & 0.013  & 0.049 & 0.045 & 0.063 & 0.004  & 86 \\ \hdashline
			\textbf{dml-norm} & Strong & 10'000 & 0.078  & 0.090 & 0.073 & 0.108 & -0.007 & 71 \\
			\textbf{grf-cent} &        &        & 0.043  & 0.063 & 0.051 & 0.080 & -0.006 & 75 \\
			\textbf{mcf-cent} &        &        & -0.014 & 0.085 & 0.045 & 0.100 & 0.007  & 65 \\ \midrule
			\multicolumn{9}{c}{40 Groups} \\
			\midrule
			\textbf{dml-norm} & None   & 2'500  & -0.002 & 0.214 & 0.268 & 0.269 & -0.003 & 94 \\
			\textbf{grf-cent} &        &        & 0.004  & 0.213 & 0.269 & 0.269 & -0.001 & 95 \\
			\textbf{mcf-cent} &        &        & 0.002  & 0.096 & 0.096 & 0.120 & -0.013 & 82 \\ \hdashline
			\textbf{dml-norm} & Med-   & 2'500  & 0.020  & 0.226 & 0.283 & 0.284 & -0.005 & 94 \\
			\textbf{grf-cent} & ium    &        & 0.037  & 0.213 & 0.265 & 0.268 & -0.001 & 95 \\
			\textbf{mcf-cent} &        &        & 0.036  & 0.099 & 0.095 & 0.128 & -0.008 & 83 \\ \hdashline
			\textbf{dml-norm} & Strong & 2'500  & 0.140  & 0.293 & 0.338 & 0.367 & -0.020 & 89 \\
			\textbf{grf-cent} &        &        & 0.090  & 0.223 & 0.255 & 0.280 & -0.003 & 92 \\
			\textbf{mcf-cent} &        &        & 0.043  & 0.119 & 0.090 & 0.153 & 0.007  & 80 \\ \midrule
			\textbf{dml-norm} & None   & 10'000 & -0.001 & 0.106 & 0.132 & 0.132 & 0.000  & 95 \\
			\textbf{grf-cent} &        &        & 0.002  & 0.104 & 0.130 & 0.130 & 0.002  & 95 \\
			\textbf{mcf-cent} &        &        & 0.004  & 0.050 & 0.050 & 0.063 & -0.005 & 85 \\ \hdashline
			\textbf{dml-norm} & Med-   & 10'000 & 0.010  & 0.113 & 0.140 & 0.141 & -0.001 & 95 \\ 
			\textbf{grf-cent} & ium    &        & 0.016  & 0.106 & 0.131 & 0.132 & 0.000  & 95 \\
			\textbf{mcf-cent} &        &        & 0.012  & 0.055 & 0.051 & 0.069 & -0.002 & 84 \\ \hdashline
			\textbf{dml-norm} & Strong & 10'000 & 0.079  & 0.164 & 0.192 & 0.209 & -0.015 & 89 \\
			\textbf{grf-cent} &        &        & 0.043  & 0.115 & 0.130 & 0.145 & -0.002 & 92 \\
			\textbf{mcf-cent} &        &        & -0.016 & 0.091 & 0.050 & 0.104 & 0.005  & 63 \\ \bottomrule
		\end{tabular}
}
\vspace{-8pt}
\captionsetup{justification=justified,singlelinecheck=false}
\caption*{\footnotesize \textsf{Note: RMSE abbreviates the Root Mean Squared Error. \emph{CovP (95)} denotes the (average) probability that the true value is part of the estimated 95\% confidence interval. 1'000 / 250 replications are used for 2'500~/ 10'000 observations.}}
\end{table}

Concerning the point estimate, \emph{mcf} outperforms \emph{grf} and
\emph{dml} once there are 10 and more ($N$=2'500) or 20 and more
groups ($N$=10'000), independent of the strength of selectivity. It
is not surprising that \emph{mcf} performs better for strong
selectivity: the previous section showed that for strong selectivity the
\emph{mcf} dominates even for a GATE with only one group, i.e., the ATE.

We observe a differential dependence of the standard deviation on the
number of groups: its increase is much slower for the \emph{mcf} than
for \emph{dml} and \emph{grf}. The reason is the way the GATE estimators
are constructed by the different methods. As mentioned above, \emph{dml}
and \emph{grf} are averaging double-robust scores within the cells of
the discrete $Z$. Since the estimators are
$\sqrt N - $convergent, we
expect that when the sample is reduced to one quarter of the original
sample, the standard deviations of such estimators double. As the cells
defined by the group variable in the DGPs are of approximately equal
size, the number of groups and observations per group are inverse
proportionally related. The resulting doubling of the standard deviation
of the \emph{grf} and \emph{dml} estimators is exactly what is observed
in Tables \ref{tab:c28} to \ref{tab:c30} when comparing results for 5 vs. 20 groups, or 10
vs. 40 groups, respectively.

The \emph{mcf} aggregates IATEs within these cells. However, as the
IATEs also potentially use ``leaf-populating data'' outside of these
cells, the standard deviation of the \emph{mcf} increases much slower
than for \emph{grf} and \emph{dml}.\footnote{The aggregation of IATEs
	that potentially weigh information from observations belonging to
	other groups into a GATE resembles smoothing across (adjacent)
	categories in the context of categorical regressors. \citet{heiler2021shrinkage}
	show that optimal smoothing parameters in a
	non-parametric regression involving categorical regressors do not
	vanish asymptotically. Furthermore, they prove that the variance of
	the smoothed estimator is a weighted sum of the asymptotic variances
	from other categories. This finding provides a plausible rationale for
	the observed steadiness in the variance of the \emph{mcf} estimator
	across varying group sizes.} In fact, we observe that when the number
of groups increases further, the RMSE of the \emph{mcf} approaches the
one of the \emph{mcf} IATEs from below, while the RMSE of \emph{dml} and
\emph{grf} increases substantially and sometimes exceeds the RMSE of the
IATEs which are of much higher dimension.

Concerning inference, the findings are a bit more pronounced: \emph{dml}
and \emph{grf} have the correct coverage rates for no and medium
selectivity even for 40 groups, while the coverage rates for \emph{mcf}
are too low.\footnote{Note that when selection into treatment gets
	stronger, splitting for the \emph{mcf} (and \emph{grf}) stops earlier,
	as leaves must have both treated and controls after a possible split,
	thus bias increases.} The figures contained at the end of Tables \ref{tab:c28},
\ref{tab:c29}, and \ref{tab:c30} in \ref{appendix-c2} indicate that this problem comes mainly
from a bias of the GATEs with the smallest true values, while for the
other GATEs coverage is close to the nominal level. In fact, the good
performance in terms of RMSE and the problems with coverage are two
sides of the same coin. Due to their aggregation from IATEs that are
averaged over trees, \emph{mcf} estimators share the strengths and the
weaknesses of many smoothing methods: the variance is reduced at the
cost of an increasing bias. To improve on the inference in finite
samples, the simulations indicate that it needs some debiasing. This is
however beyond the scope of this paper.

\FloatBarrier

\subsubsection{Conditional average treatment effects at the finest aggregation level (IATEs)}\label{533}

Table \ref{tab5} contains the results for the IATEs, averaged over all $N$
IATEs. As inference is usually not the main goal when computing IATEs,
for the centred \emph{mcf} the more efficient version is included as
well.

\begin{table}[h!]
	\sffamily
	\scriptsize
	\caption{Simulation results for individualized average treatment effects (IATEs)}
	\label{tab5}
	\resizebox{\textwidth}{!}{
		\begin{tabular}{lcccccc}
			\toprule
			\multicolumn{3}{c}{} & \multicolumn{3}{c}{\textbf{Estimation of effects}} & \multicolumn{1}{c}{\textbf{Inference}} \\
			\midrule
			\textbf{Estimator} & Selectivity & Sample size & Bias & Mean abso- & RMSE &  CovP (95) in  \\
			& & & & lute error & & \% \\
			\midrule
			\multicolumn{1}{c}{(1)} & (2) & (3) & (4) & (5) & (7) & (9) \\
			\midrule
			\textbf{dml-ols}          & None             & 2'500  & -0.002 & 0.229 & 0.286 & 23 \\
			\textbf{dml-rf}           &                  &        & -0.002 & 0.270 & 0.342 & -  \\
			\textbf{grf-cent}         &                  &        & 0.004  & 0.155 & 0.187 & 77 \\
			\textbf{mcf-cent}         &                  &        & 0.002  & 0.162 & 0.199 & 77 \\
			\textbf{mcf-cent-eff} &  &        & 0.004  & 0.151 & 0.184 & -  \\ \hdashline
			\textbf{dml-ols}          & Med-             & 2'500  & 0.020  & 0.236 & 0.295 & 24 \\
			\textbf{dml-rf}           & ium              &        & 0.013  & 0.284 & 0.362 & -  \\
			\textbf{grf-cent}         &                 &        & 0.034  & 0.175 & 0.210 & 72 \\
			\textbf{mcf-cent}         &                 &        & 0.037  & 0.174 & 0.212 & 75 \\
			\textbf{mcf-cent-eff} &  &        & 0.040  & 0.163 & 0.197 & -  \\ \hdashline
			\textbf{dml-ols}          & Strong           & 2'500  & 0.140  & 0.285 & 0.357 & 28 \\
			\textbf{dml-rf}           &                  &        & 0.117  & 0.349 & 0.461 & -  \\
			\textbf{grf-cent}         &                 &        & 0.092  & 0.254 & 0.297 & 55 \\
			\textbf{mcf-cent}         &                  &        & 0.046  & 0.212 & 0.252 & 68 \\
			\textbf{mcf-cent-eff} &  &        & 0.048  & 0.205 & 0.240 & -  \\ \midrule
			\textbf{dml-ols}          & None             & 10'000 & -0.001 & 0.179 & 0.219 & 7  \\
			\textbf{dml-rf}           &                  &        & 0.000  & 0.201 & 0.256 & -  \\
			\textbf{grf-cent}         &                  &        & 0.002  & 0.082 & 0.103 & 91 \\
			\textbf{mcf-cent}         &                  &        & 0.004  & 0.096 & 0.121 & 83 \\
			\textbf{mcf-cent-eff} &  &        & 0.003  & 0.088 & 0.110 & -  \\ \hdashline
			\textbf{dml-ols}          & Med-             & 10'000 & 0.010  & 0.181 & 0.222 & 8  \\
			\textbf{dml-rf}           & ium              &        & 0.005  & 0.214 & 0.274 & -  \\
			\textbf{grf-cent}         &                  &        & 0.016  & 0.089 & 0.111 & 89 \\
			\textbf{mcf-cent}         &                 &        & 0.013  & 0.106 & 0.131 & 81 \\
			\textbf{mcf-cent-eff}             &          &        & 0.013  & 0.100 & 0.122 & -  \\ \hdashline
			\textbf{dml-ols}          & Strong           & 10'000 & 0.079  & 0.205 & 0.256 & 12 \\
			\textbf{dml-rf}           &                  &        & 0.058  & 0.275 & 0.418 & -  \\
			\textbf{grf-cent}         &                  &        & 0.053  & 0.123 & 0.162 & 82 \\
			\textbf{mcf-cent}         &                  &        & -0.014 & 0.156 & 0.185 & 66 \\
			\textbf{mcf-cent-eff}             &          &        & -0.014 & 0.151 & 0.151 & - \\ \bottomrule
		\end{tabular}
}
\vspace{-8pt}
\captionsetup{justification=justified,singlelinecheck=false}
\caption*{\footnotesize \textsf{Note: \emph{dml} is the normalized \emph{dml} (denoted as \emph{dml-norm} in Tables \ref{tab3} and \ref{tab4}). RMSE abbreviates the Root Mean Squared Error. \emph{CovP (95)} denotes the (average) probability that the true value is part of the estimated 95\% confidence interval. 1'000 / 250 replications are used for 2'500 / 10'000 observations.}}
\end{table}

For the point estimates the ordering is clear-cut. \emph{grf} and
\emph{mcf} are similar and outperform both \emph{dml} versions
(\emph{ols}, \emph{rf}). In many cases, the efficient centred \emph{mcf}
(\emph{mcf-cent-eff}) performs best in terms of RMSE. The only exception
to this rule seems to be, again, the case of a quadratic function for
the IATE, in which all methods seem to do (almost) equally bad with
large RMSEs and large mean absolute errors. With respect to the other
estimators, in a couple of cases OLS outperforms the more sophisticated
CML estimators (but substantially underperforms in many other
scenarios).

For inference, the results show that all estimators have coverage
probabilities that are substantially below their nominal levels.
However, given the level of granularity of the IATEs, this finding is of
course not surprising. At least for the \emph{mcf}, we conjecture that
this problem comes from the biases of the individual IATEs. It is less
likely that it comes from a too small estimated standard error, because
weights-based standard error estimation is performed in a very similar
way as for the ATE and the GATEs, in which it turned out to be almost
unbiased.

\subsubsection{Summary}\label{summary}

\emph{dml} performs particularly well when the target is low dimensional, such as the ATE
and GATEs with few groups, and selectivity is not too strong. If
selectivity is strong or the number of groups becomes too large, then
its performance quickly deteriorates.\footnote{A potential alternative
	\emph{dml}-based estimator in the strong selectivity setting could be
	the automatic \emph{dml} that would estimate inverse probability
	weights instead of propensity scores as nuisance parameters.}

\emph{grf} shows a similar good performance for the low dimensional
parameters, as well as a similarly bad performance for GATEs with many
groups.\footnote{Additional smoothing in a setting with many groups
	could be considered in empirical settings for \emph{dml}-based and
	\emph{grf} -based GATE estimators to further improve their RMSE.}
However, it performs well for IATEs. It is also less affected by strong
selectivity than \emph{dml}. These results however only hold for the
modified version of the \emph{grf} in which the outcome variable in the
training data is explicitly centred before it enters the \emph{grf}
algorithm. The original uncentred version of \emph{grf} as suggested in
\citet{athey2019grf} and implemented in their package underperforms in
many scenarios due its bias problem (as does the uncentred version of
the \emph{mcf}).\footnote{As described in \autoref{522}, labels centred
	and uncentred refer to whether the outcome variable enters the
	estimation procedure residualized by $K$-fold cross-fitting or
	not, respectively. In both cases, the \emph{grf} procedure still
	performs local centring using out-of-bag predictions as described in
\autoref{42}.}

Compared to \emph{dml} and \emph{grf}, the \emph{mcf} generally shows a
robust and competitive performance in many scenarios. The price to pay for
this robustness and competitiveness is the somewhat higher standard
deviation for the very low dimensional parameters like ATE and GATEs.
Depending on the sample and the resulting precision of the estimators,
this price may well be worthwhile paying. A further important advantage
of the \emph{mcf} is that its estimates are internally consistent over
aggregation levels as ATE and GATEs are computed as averages of IATEs.
This advantage becomes particularly apparent when the number of groups
for the GATEs increases. While \emph{dml} and \emph{grf} have
substantially higher RMSEs than for the more fain-grained IATEs, the
uncertainty (and bias) in the \emph{mcf} estimators smoothly increases
with the number of groups, until it reaches the level of the most
fine-grained heterogeneity parameter, the IATE.

\section{Conclusion}\label{6-conclusion}

Estimation of causal effects at different levels of granularity is of
great importance for informed decision-making and tailored
interventions. Lower levels of granularity capture the effect of a
policy or intervention on a large population, guiding decisions on
policies that cannot be targeted at individuals but must be deployed
universally. Higher levels of granularity capture effects at a group or
individual level that can serve for decisions how policies can be
tailored more individually or for better understanding of the effects of
large-scale policies at more granular levels. The complexity of such
interventions requires methods that can estimate fine-grained
heterogeneities of causal effects flexibly, such as some Causal Machine
Learning (CML) methods.

In this paper, we investigate such CML methods subject to the
restrictions that (i) they provide estimators of the causal effects at
all aggregations levels, (ii) are essentially non-parametric, (iii) can
predict IATE based on covariates only and (iv) that they allow for
classical repeated sampling inference. Ideally, they are also internally
consistent that aggregation of lower-level effects lead to the
higher-level effects. Double/debiased Machine Learning (\emph{dml}), the
Generalized Random Forest (\emph{grf}), and the Modified Causal Forest
(\emph{mcf}) fulfil these criteria and thus belong to the group of
estimation methods which we call `Comprehensive Causal Machine Learners'
(CCMLs).

Here, we describe these estimators and their proven theoretical
guarantees. For \emph{dml} and the \emph{grf}, they are already known,
but not so for the \emph{mcf}. Therefore, we explicitly provide them.
The large-scale simulation study reveals scenarios in which the methods
perform well. \emph{dml} with normalized weights performs well in terms
of RMSE and coverage probability when the target is low-dimensional,
i.e., ATE and GATEs with few groups, and when the selection into
treatment is not strong. \emph{grf} shows similar behaviour as
\emph{dml}, however its performance for IATEs in terms of RMSE is much
better. \emph{mcf} has similar performance as \emph{grf} in case of
IATEs and outperforms \emph{dml} and \emph{grf} in scenarios with many
GATEs or strong selection.

The results of the simulation study offer several practical
recommendations. For low-dimensional targets when selection into
treatment is moderate, \emph{dml} is preferred including statistical
inference. For IATEs, Causal Forest based methods perform well in terms
of point estimation. When inference is not a priority, the more
efficient version of \emph{mcf} is recommended to estimate IATEs. For
large groups, \emph{mcf} is recommended for point estimation of GATEs.
When sample is large enough and slight loss of efficiency is not
detrimental, \emph{mcf}~can be used to estimate all effects due to its
robustness to strong selection into treatment and large GATE groups.
When internal consistency of the effects is important, only \emph{mcf}
can guarantee it due to its aggregation strategy.

The practical use of these three CCMLs is supported by the availability
of well-maintained software packages: \emph{dml} is available as Python
and R packages, the g\emph{rf} is available as an R package, and the
\emph{mcf} is available as a Python package.\footnote{As examples can
	serve: \citet{bach2022doubleml}, \citet{bach2024dmlr}, and \citet{knaus2022double}
	for \emph{dml}; \citet{athey2019estimating} for \emph{grf}; and
	\citet{bodory2022high} for \emph{mcf}. The R packages
	can be downloaded from CRAN. The Python packages are available on
	PyPI.} On a practical note, the results in this paper indicate that
users of the \emph{grf} package are encouraged to perform local centring
of the outcome variable via $K$-fold cross-fitting to remove any
potential bias (the same holds for \emph{mcf}, where this step is
already implemented in the package).

The simulation study also points to topics for further research. For
example, the sensitivity of the \emph{dml} estimators to strong
selectivity opens a topic of how to make them more robust to extreme
propensity scores without impairing its efficiency and statistical
inference. Due to the smoothing character of Causal Forests, in finite
samples \emph{mcf} estimation of GATEs and IATEs balances the
bias-variance trade-off yielding low RMSEs but impairing the coverage
probability due to a small bias and low variance. Finding a way how to
further de-bias the estimates would improve coverage probability in
finite samples.

%% file: App.tex
\section{Details of the \emph{mcf} and \emph{grf}}\label{appendix-a}

In this appendix, we present the proven theoretical guarantees of the
\emph{mcf}, its inference procedure, as well as some more details of the
Causal Forest algorithms.

\subsection{Theoretical guarantees for \emph{mcf}} \label{appendix-a1}

\subsubsection{Proof for Proposition \ref{prop_1}} \label{appendix-a11}

\setcounter{prop}{0}
\begin{prop} Let $J$ denote the number of groups for
which GATEs are to be estimated and let $\mathit{\Pi}$ be a $\mathit{J \times 1}$
vector containing the non-zero population shares of observations in the
corresponding groups. Suppose that the total number of groups increases from $J$ to $J'$ and
the population share of group $j$ under $J$ groups, $\pi_j$, decreases to $\pi_{j'}$ under $J'$ groups.
Then, the asymptotic variances of the efficient \emph{dml} and \emph{grf} GATE
estimators increase for each group $j$ within $J$
proportionally to $\pi_j/\pi_{j'}$, while
the variance of \emph{mcf} estimator of the GATE does not depend on the
number of groups and remains stable. This implies $Var\left( {\widehat {GAT{E^{mcf}}}\left( {m,l;{z_j}} \right)} \right) \le Var\left( {\widehat {GAT{E^{dml/grf}}}\left( {m,l;{z_j}} \right)} \right)$
for large $J$.
\label{prop_1}
\end{prop}

\proof{
Based on the pointwise asymptotic results from \citet{semenova2021debiased},
the change in asymptotic variance for the
\emph{dml} and \emph{grf} GATE estimator for each group $j$ within
$J$ will be proportional to $\pi_j/\pi_{j'}$ with $J$ increasing
to $J'$. The behaviour of the \emph{mcf} estimator of the GATE
resembles in this situation a behaviour of a smoothing estimator.
Inherently, the weights-based variance of the \emph{mcf} estimator does
not depend on $J$.}

\subsubsection{Assumptions and IATE}\label{appendix-a12}

Next, we present the asymptotic properties of the \emph{mcf}
(abstracting from local centring).\footnote{More details of the
	assumptions needed as well as the formal proofs are contained in
	\ref{appendix-a14} below.}

The plain-vanilla \emph{mcf} procedure aggregates individual trees
$T$ into a forest in the following way: The
data are split into the training set (``split-construction data''),
$\Im_{tr}$, and
estimation set (``leaf-populating data''),
$\Im_{est}$, to form
two fixed non-overlapping halves of the full data set of sizes
$N_1=N_2=N/2$. The
subsampling rates for each tree are
$\beta_1$ and $\beta_2$, $1/2 < \beta_1 < \beta_2 \le 1$, for the training and estimation set
respectively, leading to a subsample size of $s_1 =c_1N_1^{\beta_1}$ for the
training set and of $s_2 =c_2N_2^{\beta_2}$ for the
estimation set, where the constant $c_1$ represents
how the subsample is further split for the tree building and
$c_2$ represents further potential split for the
estimation set.\footnote{If $N$ is an odd number, one of the two
halves will contain one observation more than the other one without
any consequences. Note also that the dependence of the subsample sizes
$s_1$ and $s_2$ on $N$ is suppressed in most of the following notation. Setting
$c_1$ to lower values helps to build decorrelated
trees, setting $c_1$ to higher values helps to
reduce bias. Practically, for high values of
$\beta_1$, $c_1$ should be set to
lower values such as 1/2. Further results indicate that larger
estimation samples help to reduce variance, therefore
$c_2$ and $\beta_2$ are recommended to be set to 1.} 
We name this procedure `two-sample
honesty' as estimation data will not become training data and vice
versa. This honesty concept underpins the weights-based inference
introduced in \citet{lechner2018modified}, which is covered in \ref{appendix-a2} below.

Each tree of the form $T(m,l,x;\xi,\Im_{tr},\Im_{est})$, where $\xi
\sim \Xi$ is a source of
auxiliary randomness in the tree building process (such as random choice
of splitting variables), can be used to estimate
$IATE(m,l;x)$. When not important for
proofs, $\xi$ and/or
$\Im_{tr}$ and $\Im_{est}$ will be
suppressed in the notation for better readability. \emph{mcf} is an
average over $B$ trees:
\begin{equation*}
	F(m,l,x;\Im_{tr},\Im_{est}) = \frac{1}{B}\sum\limits_{b = 1}^B T_b(m,l,x;{\xi_b},\Im_{tr,b},\Im _{est,b}) ,
\end{equation*}
where $\Im_{tr,b}$ and
$\Im_{est,b}$ are
drawn without replacement from $\Im_{tr}$ and $\Im_{est}$ at subsampling rates
$\beta_1$
and $\beta_2$, respectively.
$\xi_b$ is a random draw
from $\Xi$. The
$b$ subscript will be supressed when it will
not lead to any confusion. Like in \citet[WA18]{wager2018estimation}, the
trees $T$ need to be \emph{honest}, as defined in \defref{def1}:

\begin{mydef}
	A tree grown on a training sample $\Im_{tr,b}$ is \emph{honest} 
	if the tree does not use
	the responses $Y$ from the estimation sample
	$\Im_{est,b}$ to place its splits.
	\label{def1}
\end{mydef}

Honesty is crucial for inference and to bound the bias of the
\emph{mcf}. The splitting and subsampling procedure described above
guarantees that the trees are honest. The main difference to the
definition of honesty of \citetalias{wager2018estimation} and \citet{athey2019grf} is the reverse
order of steps. \citetalias{wager2018estimation} first subsample and then split the data for their
double-sample trees. Here, the data set is split first and then the
training and estimation sets are subsampled from the given split. This
allows to better control bias and variance rates as the size of the
training set will codetermine the size of the final leaves translating
into bias and the size of the estimation set will influence the variance
rate.

To guarantee consistency, the final leaves must asymptotically shrink in
all dimensions similarly as in \citet{meinshausen2006quantile} and \citetalias{wager2018estimation}, invoking the
following definition of a random-split tree:

\begin{mydef}
	A tree is considered a \emph{random-split} tree if
	at every splitting step, marginalizing over
	$\xi$, there is a guaranteed
	minimum probability $\pi/p$
	that the next split occurs along the $u$-th covariate for some
	$0< \pi \leq 1$, for all $u=1,..., \allowbreak p$. \label{def2}
\end{mydef}

There are several options how to obtain a random-split tree. For the
strategy implemented in \emph{mcf}, see \ref{appendix-b31}. 

The following Definition \ref{def3} controls the shape of the leaves. Definition
\ref{def4} imposes \emph{symmetry}. Both are required to derive the asymptotic
results.

\begin{mydef}
	A tree predictor grown by recursive partitioning
	is $(\alpha,\nu)$\emph{-regular} for some
	$\alpha > 0$ if
	(1) each split leaves at least a fraction $\alpha$ of the available
	training observations of each treatment on each side of the split, (2)
	the final leaf containing $x$ has at least
	$\nu$ observations from each
	of the $M$ treatment groups for some $\nu
	\in \mathbb N$, and (3) the
	final leaf containing $x$ has at least one treatment with less than
	$2\nu - 1$ observations.
	\label{def3}
\end{mydef}

Regarding the role of the splitting rule on
$(\alpha,\nu)$-regularity, the algorithm first
determines splits that do not violate the regularity condition. For
these, the splitting criterion is calculated and the split that achieves
the minimum value of the objective function is chosen as the best one.
By following this procedure, any influence of the splitting criterion
(including the penalty) on the regularity of the final leaves can be
ruled out.

\begin{mydef}
	A predictor is \emph{symmetric} if the output of
	the predictor does not depend on the order in which the observations are
	indexed in the training and estimation samples.
	\label{def4}
\end{mydef}

Consider $B$ trees satisfying Definitions \ref{def1}-\ref{def4}, that training and estimation data are obtained from a two-sample
honesty procedure, and let $\hat \theta_{m,l}^{mcf}(x)$ denote an
\emph{mcf} estimator of $IATE(m,l;x)$ obtained as
\begin{align*}
\widehat{IATE^{mcf}}(m,l;x) & =
\hat \theta_{m,l}^{mcf}(x) =
\frac{1}{B}\sum\limits_{b
= 1}^{B} \sum\limits_{i=1}^{N_2}
w_{i,b}^{mcf}(d_i,x_i;x,m,l)y_i \\ & = 
\sum\limits_{i = 1}^{N_2}w_i^{mcf}(d_i,x_i;x,m,l)y_i,
\end{align*}
where $w_{i,b}^{mcf}(d_i,x_i;x,m,l) = \left(\frac{\underline{1}(d_i = m)}{N_{S_b(x)}^m} - \frac{\underline{1}(d_i = l)}{N_{S_b(x)}^l}\right)\underline{1}(x_i \in S_b(x))$ represent the weights for the IATE estimate in tree $b$. This
estimator is differencing the average outcomes of the treatment groups
evaluated on the estimation subsample in the tree leaf containing point
$x$, $S_b(x)$. Averaging across
trees allows for a weighted representation of the forest estimator of
the IATE with forest weights
$w_i^{mcf}(d_i,x_i;x,m,l) = \frac{1}{B}\sum\limits_{b= 1}^{B} w_{i,b}^{mcf}(d_i,x_i;x,m,l)$.

Furthermore, in order to derive the bias bound and asymptotic
distribution of the estimator, the common support assumption must hold
in its stricter version as in \citetalias{wager2018estimation}, i.e.~$0 <
\varepsilon < p_d(x) < 1 - \varepsilon$ for all $d$ and $x$
and some small $\varepsilon$.
Given these concepts, we state the main theorems guaranteeing
consistency and asymptotic Gaussianity of the \emph{mcf} estimator of
the IATE. Further assumptions necessary for achieving this asymptotic
distribution in the case of i.i.d. sampling are (i) Lipschitz continuity
of first and second order moments of the outcome variable conditional on
the covariates, (ii) using subsampling to obtain the training data for
tree building (subsamples should increase with $N$, slower than
$N$, but not too slow), and (iii) conditions on the covariates
(independent, continuous with bounded support, $p$ is fixed, i.e.~low-dimensional). 
All proofs are collected in the \ref{appendix-a14}. Each
section in \ref{appendix-a14} contains all the necessary proofs and
intermediate results for the corresponding main theorem. One of the main
differences to the proofs in \citetalias{wager2018estimation} is caused by the different splitting
and subsampling approach to achieve honesty. The Causal Forest estimator
in \citetalias{wager2018estimation} approximates a U-statistic. Thus, their proofs are based on the
corresponding Gaussian theory. Since the \emph{mcf} allows for
$\beta_2 = 1$, the estimator cannot be generally
interpreted as a U-statistic and so the results in \citetalias{wager2018estimation} do not apply.
Instead, the weighted representation of the forest estimate is leveraged
to obtain asymptotic properties that also cover the case of
$\beta_2 = 1$.

The first result is the bound on the bias of the forest estimator. The
proof is like the proof in \citetalias{wager2018estimation}. In the first step, we show that the
leaves get small in volume as $s_1$
gets large. In the second step, we show that the estimation observations
in the final leaf can be seen as a subset of nearest neighbours around
the point $x$. Thus, their expected distance and Lipschitz
continuity help to bound the bias. \lemref{lem1} below in \ref{appendix-a141}
gives rates at which the Lebesgue measure of the final leaves in a
regular tree shrinks under the assumption of the covariates being
independent from each other and uniformly distributed.

\begin{thm}
	In addition to the conditions of \lemref{lem1} (regular
	trees and independent uniformly distributed covariates),
	suppose that trees $T$ are random-split and
	honest and all $E\left[ {\left.	{{Y^d}} \right| X = x} \right]$ are Lipschitz continuous. Then,
	the absolute bias of the \emph{mcf} IATE estimator at a given value of
	$x$ is bounded by $$\left| E\left[\hat \theta_{m,l}^{mcf}(x)\right] - \theta_{m,l}^{0}(x)\right| = O\left(s_1^{-\log(1 - \alpha)/p\log(\alpha)}\right).$$   \vspace{-\baselineskip} \label{thm1-noproof}
\end{thm}

Note that the bias rate of the forest is the same as the bias rate of a
single tree, as a forest prediction is the average of the tree
predictions. The bias rate is mainly driven by the shrinkage of the
Lebesgue measure of the leaf that is influenced by the choice of
parameter $\alpha$. The upper bound in \thmref{thm1-noproof} resembles the bias rate
of nearest-neighbours regression estimators under Lipschitz continuity
in a $p$-dimensional space for which the rate of the expected
number of nearest-neighbours is influenced by the parameter $\alpha$.
This stems from the fact that the final leaves can be bounded by balls
whose volume shrink at the same rate as the volume of the final leaves,
as shown in the proof in the \ref{appendix-a141}. Note that this rate is
rather conservative as it stems from bounding the shallowest final leaf
with a leaf that would always end up with a $(1-\alpha)$ share of
observations at the same level of depth. The rate is faster than in \citetalias{wager2018estimation}
because we bound by the expected distance between the nearest neighbours
and the point $x$ instead of the longest expected diameter of the
leaf.

The second result is the asymptotic distribution of the IATE estimator
at point $x$.

\begin{thm}
	Assume that there is a sample of size $N$
	containing i.i.d.~data $(X_i,Y_i,\allowbreak D_i) \in \left[0,1\right]^p \times	\mathbb{R} \times \{0,1,...,M-1\}$ for a given value of
	$x$. Moreover, the covariates are independently and uniformly
	distributed $X \sim U\left(\left[0,1\right]^p\right)$.
	Let $T$ be an honest, regular, and symmetric random-split tree.
	Further assume that $E\left[\left. Y^d \right| X = x \right]$ and
	$E\left[\left. ( Y^d )^2 \right| X = x \right]$ are Lipschitz
	continuous and
	$Var \left( \left. Y^d \right| X = x \right) > 0$. Then for
	$1/2 < \beta_1 < \beta_2 < \frac{p +
	2}{p} \frac{\log(1 - \alpha)}{\log(\alpha)} \beta_1$,
	$$
	\frac{\hat \theta_{m,l}^{mcf}(x) - \theta_{m,l}^{0}(x)}{\sqrt{Var\left(\hat \theta_{m,l}^{mcf}(x)\right)}} \overset{d}{\longrightarrow} N(0,1).
	$$
	\vspace{-\baselineskip} 
\label{thm2-noproof}
\end{thm}

The restrictions on the sampling rates require that $(2+p)
\log(1-\alpha) / (p \log(\alpha)) >1$. This suggests
that $\alpha$ needs to be set closer and closer to 0.5 for larger
$p$. This is a consequence of the conservative bias bound and the
curse of dimensionality as values of $\alpha$ closer to 0.5 make sure that the
shallowest final leaves get tighter upper bounds ensuring that the bias
vanishes fast enough. Additionally, the relationship between the
subsampling rates further ensures that the final leaf does not end up
with too many honest observations and the squared bias--variance ratio
converges to 0. The results further indicate that variance is converging
to zero the fastest when $\beta_2=1$. Similarly, higher
values of $\beta_1$ make the bias converge to zero
faster. Due to the conservativeness of the bias bound, it is not
straightforward to suggest a value for parameter $\alpha$. Following the
conservative bound, $\alpha$ should be set closer and closer to 1/2 for
higher dimensions (increasing $p$). Practically, lower values of
$\alpha$ increase the flexibility of the splitting procedure allowing
daughter leaves to contain at least an $\alpha$ share of the
observations in the parent node, i.e.~still allowing for splits that
would be enforced by higher $\alpha$. The convergence of the
bias-variance ratio to 0 together with a side result of the variance
converging to 0 guarantees the consistency of the \emph{mcf} estimator
of the IATE as captured in the following corollary.

\setcounter{corr}{1}
\begin{corr}
	Let all assumptions from \thmref{thm2-noproof} hold. Then, $\hat \theta_{m,l}^{mcf}(x) \overset{p}{\longrightarrow} \theta_{m,l}^{0}(x).$
\end{corr}

\subsubsection{GATEs and ATE}\label{appendix-a13}

Estimators for GATEs and ATEs are obtained by averaging the IATEs in the
respective subsamples defined by $z$ (assuming discrete $Z$).
Although estimating ATEs and GATEs directly instead of aggregating IATEs
could lead to more efficient estimators (see \autoref{41} and \autoref{42}), the
computational burden would also be higher, in particular if the number
of GATEs of interest is large, as is common in many empirical studies.
Furthermore, there is no guarantee that the results are internally
consistent, i.e., the respective averages of the IATEs are indeed close
to their ATE and GATE counterparts. Therefore, letting
$\hat \theta_{m,l}^{mcf}(x)$ be an estimator of
$IATE(m,l,x)$, the \emph{mcf} estimator computes GATEs and ATEs as
appropriate averages of $\hat \theta_{m,l}^{mcf}(x)$'s:
\begin{align*}
	\widehat{GATE}^{mcf}(m,l;z) & = \hat \theta_{m,l}^{mcf}(z) = \frac{1}{N_2^z}\sum\limits_{i= 1}^{N_2} \underline{1}(z_i=z) \hat \theta_{m,l}^{mcf}(x_i) \\ & = \sum\limits_{i = 1}^{N_2}w_i^{mcf}(z,m,l)y_i; \\
	w_i^{mcf}(z,m,l) & = \frac{1}{N_2^z}\sum\limits_{j= 1}^{N_2} \underline{1}(z_j=z) w_i^{mcf}(d_i,x_i;x_j,m,l); \quad \quad N_2^z=\sum_{i=1}^{N_2} \underline{1}(z_i=z). \\
	\widehat{ATE}^{mcf}(m,l) & = \hat \theta_{m,l}^{mcf} = \frac{1}{N_2}\sum\limits_{i= 1}^{N_2} \hat \theta_{m,l}^{mcf}(x_i) \\ & = \sum\limits_{i = 1}^{N_2}w_i^{mcf}(m,l)y_i; \\
	w_i^{mcf}(m,l) & = \frac{1}{N_2}\sum\limits_{j= 1}^{N_2} w_i^{mcf}(d_i,x_i;x_j,m,l).
\end{align*}

These expressions show that ATEs and GATEs have the same weights-based
representation as the IATEs. Hence, asymptotic Gaussianity can be
established in a similar way.

\begin{thm}
	Let all assumptions of \thmref{thm2-noproof} hold. Then,
	$$
	\frac{\hat \theta_{m,l}^{mcf} - \theta_{m,l}^{0}}{\sqrt{Var\left(\hat \theta_{m,l}^{mcf}\right)}} \overset{d}{\longrightarrow} N(0,1).
	$$
	\vspace{-\baselineskip} 
	\label{thm3-noproof}
\end{thm}

Utilizing the same Central Limit Theorem for triangular arrays as in
\thmref{thm3-noproof} and assuming that $Z$ is among the splitting variables
yields the following corollary:

\begin{corr}
	Let all assumptions from \thmref{thm2-noproof} hold
	and assume that a discrete $Z$ is among the splitting variables. Then for $Z = z$,
	$$
	\frac{\hat \theta_{m,l}^{mcf}(z) - \theta_{m,l}^{0}(z)}{\sqrt{Var\left(\hat \theta_{m,l}^{mcf}(z)\right)}} \overset{d}{\longrightarrow} N(0,1).
	$$
\end{corr}

\subsubsection{Proofs}\label{appendix-a14}

The following notation as in \citetalias{wager2018estimation} is used for the asymptotic scaling:
$f(s) \gtrsim g(s)$ means that $\liminf \limits_{s \to
\infty} f(s)/g(s) \ge 1$,
and $f(s) \lesssim g(s)$ means that $\liminf \limits_{s \to \infty} f(s)/g(s) \le 1$.
Further, $f(s) = \Omega(g(s))$ means that $\liminf\limits_{s \to \infty} \left|f(s)\right|/g(s)>0$, i.e., that
$\left|f(s)\right|$ is bounded below by $g(s)$ asymptotically.

\paragraph{Proofs for Theorem \ref{thm1-proof}}\label{appendix-a141}

In this section, we collect all necessary results for the bias bound. In
the first step, we show that the volume of the final tree leaf
containing point $x$ shrinks with larger subsample size
$s_1$. We focus on the volume instead
of the diameter in comparison to \citetalias{wager2018estimation} as the volume is important to
determine the expected value of estimation samples in the leaf as the
subsampling rates in the training and estimation set differ.

\begin{lem}
	(based on Lemma 1 in \citetalias{wager2018estimation}) Let
	$S(x)$ be a final leaf containing the point
	$x$ in a regular tree according to the definitions above and let
	$\lambda(S(x))$ be its
	Lebesgue measure. Suppose that
	${X} \sim
	U\left([0,1]^p\right)$ independently. Then for
	$\alpha \le
	0.5$, the expected value of the Lebesgue measure of the
	final leaf has the following bounds
	\begin{gather*}
	E\left[\lambda(S(x))\right] = O\left({{s_1}^{- \log (1 -\alpha )/\log (\alpha )}}\right), \\
	E\left[\lambda (S(x))\right] = \Omega \left({s_1^{-1}}\right).
	\end{gather*}
\vspace{-\baselineskip} 
\label{lem1}
\end{lem}
\proof{
Let $c(x)$ denote number of
splits leading to the leaf $S(x)$ and
$s_1^d$ be the number of observations
treated with treatment $d$ in the training subsample. By the
results in \citet{wager2015adaptive}, in particular using their Lemma 12,
Lemma 13 and Corollary 14, with high probability and simultaneously for
all but last $O(\log (\log
{s_1}))$ parent nodes above
$S(x)$, the number of training observations
in the node divided by ${s_1}$ is within
a factor $1 + o(1)$ of the Lebesgue measure
of the node. Therefore, for large enough
${s_1}$ with probability greater than
$1 - 1/{s_1}$ it holds that
$$\lambda (S(x)) \le {(1 -\alpha + o(1))^{c(x)}}.$$

To further evaluate the upper bound for
$\lambda (S(x))$, the smallest
number of splits that could lead to a leaf
$S(x)$ needs to be determined. Let
${s_{min}} = \mathop
{\min }\limits_d
{s_{1,d}}$ denote the smallest treatment in the
training subsample. By regularity, the following holds
${s_{min }}{\alpha^{c(x)}} \le 2\nu - 1$. Since 
$s_{min} \gtrsim s_1 \varepsilon$, then $c(x) \ge \log((2\nu - 1)/({s_1}\varepsilon))/\log\left(\alpha \right)$ for large ${s_1}$ and $\lambda(S(x))$ is bounded by
$$\lambda (S(x)) = O\left({{s_1}^{-\log(1 - \alpha)/\log (\alpha )}}\right).$$
This also translates into the upper bound for $E\left[ {\lambda (S(x))}\right]$.

Let ${N_{S(x)}}$ and
$N_{S(x)}^d$ denote the number of
observations in leaf $S(x)$ and the number
of observations treated with $d$ in leaf
$S(x)$, respectively. Regarding the lower
bound, by regularity, $N_{S(x)}^d \ge
\nu$ holds for all treatments. It can be shown that the
probability of ${N_{S(x)}} \ge
\nu M$ when $\lambda (S(x))
\le \nu M/{s_1}$ decays at least at a rate
of $1/{s_1}$ for uniformly distributed
covariates. Therefore, $\nu M/{s_1} \le
\lambda (S(x))$ with probability
$1 - O(1/{s_1})$ yielding
$E\left[ {\lambda (S(x))}
\right] = \Omega \left(
{s_1^{-1}} \right)$. The upper
bound is above the lower bound for $\alpha \le 1/2$.
}

\setcounter{thm}{0}
\begin{thm}
	Under the conditions of \lemref{lem1}, suppose
	that trees $T$ are random-split and
	honest and $E\left[ {\left.
	{{Y^d}} \right| X = x}
	\right]$ are Lipschitz continuous. Then,
	the bias of the \emph{mcf} estimator of the IATE for a given value of
	$x$ is bounded by
	$$\left|{} {E\left[
	{\hat \theta _{m,l}^{mcf}(x)}
	\right] - \theta _{m,l}^0(x)}
	\right|{} = O\left( {{s_1}^{
	- \log (1 - \alpha )/p\log
	(\alpha )}} \right).$$
\label{thm1-proof}
\vspace{-\baselineskip} 
\end{thm}

\proof{
As a forest prediction is an average of the tree
predictions, the rate of the bias of a tree prediction is also the rate
of the bias of the forest prediction. Let
${T_b}(x,d)$ denote a tree prediction of
a potential outcome of treatment $d$ at point $x$ of a form
$${T_b}(x,d) =
\sum\limits_{j = 1}^{{s_2}}
{w_{bj,b}^{mcf}({d_{bj}},{x_{bj}};x,d){y_{bj}}},$$
where $b_{j}$ denotes the $j$-th observation in
the $b$-th estimation subsample. The respective weights are given
by:
$$w_{bj,b}^{mcf}({d_{bj}},{x_{bj}};x,d)
= \begin{cases} \left({N_{{S_b}(x)}^d}\right)^{-1} & \text{if} \quad x_{bj} \in {S_b}(x)  \quad \text{and} \quad {{d_{bj}} = d} \\ 0 & \text{else} \end{cases}.$$

The tree estimator of the treatment effect is $T(m,l,x) =
T(x,m) - T(x,l)$. As the treatment effect estimator is a
difference of two potential outcome estimators, the absolute bias can be
bounded by the rate of the bias of the potential outcome estimators as
$$\left| E\left[T(m,l,x)\right] - IATE(m,l;x)\right|  \le \sum\limits_{d \in \{m,l\} } \left|E\left[T(x,d)\right] - E\left[\left. Y^d \right| X = x \right] \right|.$$
Therefore, in the following, the bias of the tree estimator for the
potential outcome will be analysed.

The next observation is that if $E\left[
{{Y^d}\left|{} {X = x}
\right.} \right]$ is
Lipschitz, then $E\left[
{Y\left|{} {X = x,D = d} \right.}
\right]$ is also Lipschitz as the two
expectations coincide under the CIA, CS and observation rule. By using
Jensen's inequality and Lipschitz continuity, the absolute bias can be
bounded by

\bigskip

\resizebox{0.97\textwidth}{!}{$
\begin{aligned}
	 \left|{E\left[{{T_b}(x,d)} \right] -
E\left[ {{Y^d}\left|{} {X =
x} \right.} \right]}
\right| & = \left|{}
{E\left[ {E\left[
{\left.
{\frac{1}{{N_{{S_b}(x)}^d}}\sum\limits_{i
\in {S_b}(x),{D_i} = d} {{Y_i}} }
\right| N_{{S_b}(x)}^d}
\right]} \right] -
E\left[ {{Y^d}\left|{} {X =
x} \right.} \right]}
\right| \\ & =
\left|{} {E\left[
{\frac{1}{{N_{{S_b}(x)}^d}}\sum\limits_{i
\in {S_b}(x),{D_i} = d} {E\left[
{{Y_i}\left|{} {N_{{S_b}(x)}^d}
\right.} \right]} }
\right] - E\left[
{{Y^d}\left|{} {X = x}
\right.} \right]}
\right| \\ & 
\le E\left[
{\frac{1}{{N_{{S_b}(x)}^d}}\sum\limits_{i
\in {S_b}(x),{D_i} = d}
{\left|{} {E\left[
{{Y_i}\left|{} {N_{{S_b}(x)}^d}
\right.} \right] -
E\left[ {{Y^d}\left|{} {X =
x} \right.} \right]}
\right|} }
\right]\\ & =
E\left[
{\frac{1}{{N_{{S_b}(x)}^d}}\sum\limits_{i
\in {S_b}(x),{D_i} = d}
{\left|{} {E\left[
{{Y_i}\left|{} {N_{{S_b}(x)}^d}
\right.} \right] -
E\left[ {{Y_i}\left|{}
{{X_i} = x,{D_i} = d} \right.}
\right]} \right|} }
\right]\\ & \le
E\left[
{\frac{1}{{N_{{S_b}(x)}^d}}\sum\limits_{i
\in {S_b}(x),{D_i} = d}
{{C_d}\left\|{{X_i} - x}\right\|} }
\right].\end{aligned}$}

\bigskip
Using Lipschitz continuity moves the analysis of the bound from the
$Y| X$ space to the covariate space. This allows bounding
the bias using the random-split assumption. In the rest of the proof, we
can assume that $\pi = 1$ since the rate on a distance between
conditional expectation of $Y$ on the leaf and conditional
expectation of $Y$ at $x$ is the slowest for $\pi = 1$. The
rationale for the bias speeding up for $0 < \pi < 1$ is that less relevant dimensions are split less frequently and
partitioning focuses on dimensions more relevant for approximating
$E[Y| X=x]$. As the $\alpha$ regularity would yield a very loose bound on the
expected distance, we take a different approach here based on the
nearest neighbours (NN) of the point $x$ that also contain all the
observations in the final leaf. Each final leaf can be bounded by a ball
with the centre at $x$ and radius equal to the longest segment of
the leaf which we denote as
$diam({S_b}(x))$. Then the number of
treated observations in the ball, denoted as
$N_{B({S_b}(x))}^d$, follows a
binomial distribution for ${s_2}$
observations and the success probability being the Lebesgue measure of
the ball that can be seen as a function of the
$diam({S_b}(x))$ since the covariates are
independent and uniformly distributed.\footnote{Without loss of
	generality, we assume constant treatment probability for all $x$.
	The proof can be generalized to treatment probabilities varying with
	$x$, leading to a different constant in the following bounds.} At
the same time the number of observations in the final leaf
$N_{{S_b}(x)}^d$ follows also a
binomial distribution for ${s_2}$
observations and the success probability being the Lebesgue measure of
the final leaf that can be seen as a function of the
$diam({S_b}(x))$ and the angles to the
vertices from the ``origin'' vertex using a high-dimensional polar
system. As both random variables depend on the same power of
$diam({S_b}(x))$ that shrinks to 0 due to
random-split property such that we can conclude that
$E[N_{B({S_b}(x))}^d] \le
O\left( {E[N_{{S_b}(x)}^d]}
\right)$ with a constant larger than 1. As
the ball contains the final leaf, $\sum\limits_{i
\in {S_b}(x),{D_i} = d}
{\left\|{{X_i} - x}\right\|} \le \sum\limits_{i \in B({S_b}(x)),{D_i} = d} {\left\|{{X_i} - x}\right\|}$.

\bigskip
When we randomly split all data points in the estimation subsample with
treatment $d$ into
$N_{B({S_b}(x))}^d + 1$ segments,
the first $N_{B({S_b}(x))}^d$
segments will have a length
${s_{2,d}}/N_{B({S_b}(x))}^d$.
Denote $\tilde X_j^x$ as
the first nearest neighbour in the $j$\textsuperscript{th} segment.
Then,
$$\sum\limits_{i
\in B({S_b}(x)),{D_i} = d}
{\left\|{{X_i} - x}\right\|} \le \sum\limits_{j = 1 \atop D_j = d}^{N_{B({S_b}(x))}^d}
{\left\|{} {\tilde X_j^x - x} \right\|}.$$

Therefore, the absolute bias can be further bounded by
\bigskip

\resizebox{0.97\textwidth}{!}{$
\begin{aligned}\left|
{E\left[ {{T_b}(x,d)} \right] -
E\left[ {{Y^d}\left| {X =
x} \right.} \right]}
\right| & \le
{C_d}E\left[
{\frac{1}{{N_{{S_b}(x)}^d}}\sum\limits_{j= 1 \atop D_j = d}^{N_{B({S_b}(x))}^d}
{\left\|{\tilde
X_j^x - x} \right\|} }
\right] \\ & =
{C_d}E\left[ {E\left[
{\left.
{\frac{1}{{N_{{S_b}(x)}^d}}\sum\limits_{j= 1\atop D_j =d}^{N_{B({S_b}(x))}^d}
{\left\|{\tilde
X_j^x - x} \right\|} }
\right| N_{{S_b}(x)}^d,N_{B({S_b}(x))}^d}
\right]}
\right] \\ & =
{C_d}E\left[
{\frac{{N_{B({S_b}(x))}^d}}{{N_{{S_b}(x)}^d}}E\left[
{\left. {\left\|
{\tilde X_1^x - x}
\right\|}
\right| N_{B({S_b}(x))}^d}
\right]}
\right]\\ & \le
{C_{d,B}}E\left[ {E\left[
{\left. {\left\|{}
{\tilde X_1^x - x}
\right\|}
\right| N_{B({S_b}(x))}^d}
\right]} \right] +
O\left( {1/{s_2}}
\right)\\ & =
{C_{d,B}}E\left[ {E\left[
{\left. {\left\|{}
{{X_{\left(
{1,\left\lfloor
{{s_{2,d}}/N_{B({S_b}(x))}^d}
\right\rfloor } \right)}} -
x} \right\|}
\right| N_{B({S_b}(x))}^d}
\right]} \right] +
O\left( {1/{s_2}}
\right),\end{aligned}$}

\bigskip
\noindent where ${X_{\left( {1,N}\right)}}$ denotes the first nearest
neighbour among $N$ observations and
${C_{d,B}}$ collects the Lipschitz
constant and the constant from the ratio of the expected number of
observations in the ball and the final leaf. The result is obtained by
applying a second order Taylor approximation of
$N_{B({S_b}(x))}^d/N_{{S_b}(x)}^d$
around
$(E[N_{B({S_b}(x))}^d],E[N_{{S_b}(x)}^d])$,
the bound on the ratio of expected number of observations in the ball
and the final leaf, the fact that
$\left\|{}
{\tilde X_1^x - x}
\right\|$ is bounded
and Cauchy-Schwarz inequality. For a fixed
$N_{B({S_b}(x))}^d$ in a ball
containing a leaf from a regular, random-split tree, we can use results
in \citet{Gyoerfi2002} for nearest neighbour (NN)
estimators that also use the expected distance of the first neighbours
in their Theorem 6.2 and Lemma 6.4 yielding the bound
$${C_{d,B}}E\left[
{E\left[ {\left.
{\left\|{}
{{X_{\left(
{1,\left\lfloor
{{s_{2,d}}/N_{B({S_b}(x))}^d}
\right\rfloor } \right)}} -
x} \right\|}
\right| N_{B({S_b}(x))}^d}
\right]} \right] \le
\frac{{{{\tilde
c}_d}}}{{s_{2,d}^{1/p}}}E\left[
{N{{_{{S_b}(x)}^d}^{\frac{1}{p}}}}
\right],$$
where ${\tilde c_d}$
collects the constant ${C_{d,B}}$ and
all constants that emerge in the NN proof. The upper bound for the last
expectation can be then found by applying Jensen's inequality,
\begin{align*}E\left[
{E\left[ {\left.
{{{\left(
{\frac{{N_{{S_b}(x)}^d}}{{{s_{2,d}}}}}
\right)}^{\frac{1}{p}}}}
\right|\lambda ({S_b}(x))}
\right]} \right] &\le
E\left[ {{{\left(
{\frac{{E\left[ {\left.
{N_{{S_b}(x)}^d}
\right|\lambda ({S_b}(x))}
\right]}}{{{s_{2,d}}}}}
\right)}^{\frac{1}{p}}}}
\right]\\ & \le
{{\tilde c}_\varepsilon
}E\left[ {{{\left(
{\lambda ({S_b}(x))}
\right)}^{\frac{1}{p}}}}
\right] \le {{\tilde
c}_\varepsilon }{\left(
{E\left[ {\lambda ({S_b}(x))}
\right]}
\right)^{\frac{1}{p}}},\end{align*}
where ${\tilde c_\varepsilon
}$ collects constants stemming from the common support
assumption. Using the results from \lemref{lem1}, the result at the tree level
is
$$\left|{} {E\left[
{{T_b}(x,d)} \right] - E\left[
{{Y^d}\left|{} {X = x}
\right.} \right]}
\right|{} = O\left( {{s_1}^{
- \log (1 - \alpha )/p\log
(\alpha )}} \right).$$
The final constant is a function of the Lipschitz constant, the constant
from the ratio of the observations in the ball and the final leaf, the
common support parameter $\varepsilon
$, and $\nu $,
the regularity parameter controlling the number of the observations in
the final leaf. As the forest estimator is the average of the tree
estimators, the result above holds also for the forest estimators.
}

\paragraph{Proofs for Theorem \ref{thm2-proof}}\label{appendix-a142}

The proofs of asymptotic Gaussianity build on a central limit theorem
for weakly dependent random variables of a form
${A_{i,{N_2}}} = {W_i}{Y_i} - E\left[ {{{W}_i}{Y_i}} \right]$, introduced
in \citet{neumann2013central}. In this section, we prove that
${A_{i,{N_2}}}$ satisfy the
conditions of the CLT yielding the first necessary result. As not all
${Y_i}$ are unbiased estimators of
${\mu _d}(x)$, the weighted
average is not an unbiased estimator. Therefore, in the second step it
is necessary to show that the ratio of bias and variance converges to
zero as the sample gets larger and the final leaf shrinks
asymptotically.

For the analysis of the variance of the forest estimator of the IATE, we
observe the following:

\setcounter{corr}{0}
\begin{corr}
	For any forest estimator $F$ that averages
	$B$ tree estimators $T$, the rate of the forest variance
	$Var(F)$ is bounded from above and below by the rate of the
	individual tree variance $Var(T)$.
	\label{cor1}
\end{corr}

\proof{
The variance of a forest estimator in a simplified notation is

$$Var(F) =
\frac{1}{{{B^2}}}\left[
{\sum\limits_{b = 1}^B
{Var({T_b}) + } \sum\limits_{b =
1}^B
{\sum\limits_{b\textquotesingle{}
\ne b} {Cov({T_b},{T_{b\textquotesingle}})} }
} \right].$$
As a forest is an average of trees, the worst upper bound of the
variance of the forest estimator is the variance of an individual tree
estimator. \lemref{lem3} shows the upper bound for the tree estimator that
converges to zero for ${\beta_2} > {\beta_1}$.

Since any covariance has to go to zero at least as fast as the variance,
the lower bound of the rate of the variance is also determined by the
variance of an individual tree.}

In the following, we therefore focus on deriving the properties of the
tree weights.

\begin{lem}
	Let ${W_{i,b}}: = W_{bi,b}^{mcf}(D,X;x,d)$ where the subscript
	$bi$ represents $i$-th observation in the estimation sample of
	size $N_2$ that was subsampled to the subsample
	$b$ of size $s_2$. Suppose that the assumptions
	from \lemref{lem1} hold and the tree is honest and symmetric. Moreover
	${\beta_2} > {\beta_1}/2 > 1/4$. Then, the moments of the tree weights have the following rates and bounds:
	\begin{enumerate}
		\def\labelenumi{\alph{enumi})}
		\item
		$E\left[ {{W_{i,b}}}
		\right] \sim
		\frac{1}{{{N^{{\beta
		_2}}}}}$ ,
		\item $E\left[ {W_{i,b}^2}
		\right] = \Omega \left(
		{{N^{\frac{{\log (1 -
		\alpha )}}{{\log (\alpha
		)}}{\beta _1} - 2{\beta _2}}}}
		\right)$ and $E\left[ {W_{i,b}^2}
		\right] = O\left(
		{{N^{{\beta _1} - 2{\beta
		_2}}}} \right)$,
		\item
		$Var({W_{i,b}})$ has the same bounds
		as $E\left[ {W_{i,b}^2}
		\right]$,
		\item
		$\left|{}
		{Cov\left( {{W_{i,b}},{W_{j,b}}}
		\right)} \right|{} =
		\Omega \left(
		{{N^{\frac{{\log (1 -
		\alpha )}}{{\log (\alpha
		)}}{\beta _1} - 3{\beta _2}}}}
		\right)$ and
		$\left|{}
		{Cov\left( {{W_{i,b}},{W_{j,b}}}
		\right)} \right|{} =
		O\left( {{N^{{\beta _1} -
		3{\beta _2}}}}
		\right)$.
	\end{enumerate}
	\label{lem2}
\end{lem}

\proof{
\begin{enumerate}
	\def\labelenumi{\alph{enumi})}
	\item
	Due to symmetry, the expected value of the first moment of the tree
	weights can be expressed as:
\begin{align*}E\left[
{{W_{i,b}}} \right] & = E\left[
{E\left[ {{W_{i,b}}\left|{}
{N_{{S_b}(x)}^d} \right.}
\right]}
\right]\\ & =
E\left[ {\frac{{{s_2} -
N_{{S_b}(x)}^d}}{{{s_2}}} \cdot 0 +
\frac{{N_{{S_b}(x)}^d}}{{{s_2}}}\frac{1}{{N_{{S_b}(x)}^d}}}
\right]\\ & =
\frac{1}{{{s_2}}} \sim
\frac{1}{{{N^{{\beta
_2}}}}}.\end{align*}

	\item
	Let ${p_d}(S(x))$ be the propensity
	score of getting treatment $d$ when in leaf $S(x)$. Due to
	the common support assumption
	${p_d}(S(x))$ must lie in an interval
	$\left( {\varepsilon ,1 -
	\varepsilon } \right)$ for
	large $N$. Thus, under symmetry, the expected value of the second
	moment of the tree weights can be expressed as:
	\begin{align*}E\left[
	{W_{i,b}^2} \right] & = E\left[
	{E\left[ {W_{i,b}^2\left|{}
	{N_{{S_b}(x)}^d} \right.}
	\right]}
	\right]\\ & =
	E\left[ {\frac{{{s_2} -
	N_{{S_b}(x)}^d}}{{{s_2}}} \cdot 0 +
	\frac{{N_{{S_b}(x)}^d}}{{{s_2}}}\frac{1}{{N{{_{{S_b}(x)}^d}^2}}}}
	\right]\\ & =
	\frac{1}{{{s_2}}}E\left[
	{\frac{1}{{N_{{S_b}(x)}^d}}}
	\right] =
	\frac{1}{{{s_2}}}E\left[
	{\frac{{1 - {{(1 - \lambda
	(S(x)){p_d}(S(x)))}^{{s_2}}}}}{{{s_2}\lambda
	(S(x)){p_d}(S(x))}}}
	\right],\end{align*}
	where the last equality uses the law of iterated expectations and the
	fact that $N_{{S_b}(x)}^d$ is a
	positive binomial random variable.
	Since $E\left[
	{1/N_{{S_b}(x)}^d} \right] >{}
	0$, the lower bound is determined by $E\left[
	{1/{s_2}\lambda (S(x)){p_d}(S(x))}
	\right]$. The
	lower bound can then be derived as follows using Jensen's inequality and
	results in \lemref{lem1}:
	$$E\left[ {W_{i,b}^2}
	\right] = \Omega \left(
	{\frac{1}{{{s_2}{s_2}E\left[
	{\lambda (S(x))} \right]}}}
	\right) = \Omega \left(
	{{N^{ - 2{\beta _2} +
	\frac{{\log (1 - \alpha
	)}}{{\log (\alpha
	)}}{\beta _1}}}}
	\right).$$
	
	The upper bound can be similarly derived for
	${\beta _2} >
	{\beta _1}/2 > 1/4$ as

	\bigskip
	
	\resizebox{0.93\textwidth}{!}{$
	\begin{aligned}E\left[
	{\frac{1}{{N_{{S_b}(x)}^d}}}
	\right] & = E\left[
	{\frac{{1 - {{(1 - \lambda
	(S(x)){p_d}(S(x)))}^{{s_2}}}}}{{{s_2}\lambda
	(S(x)){p_d}(S(x))}}}
	\right]\\ & =
	E\left[ {\left.
	{\frac{{1 - {{(1 - \lambda
	(S(x)){p_d}(S(x)))}^{{s_2}}}}}{{{s_2}\lambda
	(S(x)){p_d}(S(x))}}}
	\right|\lambda (S(x)) >{}
	\nu M/{s_1}} \right]\Pr
	\left( {\lambda (S(x)) <
	\nu M/{s_1}} \right)\\ & +
	E\left[ {\left.
	{\frac{{1 - {{(1 - \lambda
	(S(x)){p_d}(S(x)))}^{{s_2}}}}}{{{s_2}\lambda
	(S(x)){p_d}(S(x))}}}
	\right|\lambda (S(x))
	\ge \nu M/{s_1}} \right]\Pr
	\left( {\lambda (S(x)) \ge
	\nu M/{s_1}} \right)\\ &
	\le 1 \cdot O(1/{s_1}) +
	O({s_1}/{s_2}) \cdot 1 = O\left(
	{1/{N^{{\beta _2} - {\beta
	_1}}}}
	\right).\end{aligned}$}

	\bigskip
	
	The two results yield
	\begin{gather*}
	E\left[ {W_{i,b}^2}
	\right] = \Omega \left(
	{{N^{ - 2{\beta _2} +
	\frac{{\log (1 - \alpha
	)}}{{\log (\alpha
	)}}{\beta _1}}}}
	\right), \\
	E\left[ {W_{i,b}^2}
	\right] = O\left( {{N^{ -
	2{\beta _2} + {\beta _1}}}}
	\right).
	\end{gather*}
	Both rates have constants that depend on
	$\varepsilon $, the common
	support parameter and regularity parameter $\nu$. The constant for
	the upper bound also depends on the number of treatments as the number
	of treatments influences the smallest expected Lebesgue measure of the
	leaf.

	\item
	Since $E\left[ {W_{i,b}^2}
	\right] \to 0$ for ${\beta _2} > {\beta _1}/2 > 1/4$, the
	bounds for $Var({W_{i,b}})$ are the	same as for $E\left[ {W_{i,b}^2}\right]$.
	\item
	The covariance can be expressed as
	$$Cov({W_{i,b}},{W_{j,b}}) =
	Cov\left( {{W_{i,b}},1 -
	\sum\limits_{k \ne j}
	{{W_{k,b}}} } \right) = - Var({W_{i,b}}) -
	\sum\limits_{k \ne i,j}
	{Cov({W_{i,b}},{W_{k,b}})} $$
	yielding
	$$Cov({W_{i,b}},{W_{j,b}}) = -
	\frac{{Var({W_{i,b}})}}{{{s_2} -
	1}}.$$
	The bounds for the covariance are
	\begin{gather*}
	\left| {Cov\left(
	{{W_{i,b}},{W_{j,b}}} \right)}
	\right| = \Omega
	\left( {{N^{ - 3{\beta _2} +
	\frac{{\log (1 - \alpha
	)}}{{\log (\alpha
	)}}{\beta _1}}}}
	\right), \\
	\left| {Cov\left(
	{{W_{i,b}},{W_{j,b}}} \right)}
	\right| = O\left( {{N^{ -
	3{\beta _2} + {\beta _1}}}}
	\right).
	\end{gather*}
	Note that
	$$\sum\limits_{k
	\ne j} {Cov({W_{i,b}},{W_{k,b}})} =
	\frac{{Var({W_{i,b}})}}{{{s_2} -
	1}}.$$
\end{enumerate}
}

\begin{lem}
	Suppose that the tree conditions from \lemref{lem2} hold
	i.e., we build regular, honest, symmetric trees. Additionally,
	$E\left[ {\left.
	{{Y^d}} \right| X = x}
	\right]$ is Lipschitz continuous.
	Moreover, assume that $E\left[
	{\left. {{{( {{Y^d}})}^2}} \right| X = x}
	\right]$ is also Lipschitz continuous
	and $Var\left(
	{{Y^d}\left|{} {X = x}
	\right.} \right) >{}
	0$. Further, the sampling rates satisfy
	${\beta _2} > {\beta _1} > 1/2$. Then
	the tree variance has the following rates
	\begin{gather*}Var\left(
	{\sum\limits_{i = 1}^{{s_2}}
	{{W_{i,b}}{Y_i}} } \right) =
	O\left( {{N^{{\beta _1} -
	{\beta _2}}}}
	\right),\\ Var\left(
	{\sum\limits_{i = 1}^{{s_2}}
	{{W_{i,b}}{Y_i}} } \right) =
	\Omega \left(
	{{N^{\frac{{\log (1 -
	\alpha )}}{{\log (\alpha
	)}}{\beta _1} - {\beta _2}}}}
	\right).\end{gather*}
	\label{lem3}
	\vspace{-\baselineskip} 
\end{lem}

\proof{
The variance of a tree estimator of
${\mu _d}(x)$ can be
decomposed as
$$Var\left(
{\sum\limits_{i = 1}^{{s_2}}
{{W_{i,b}}{Y_i}} } \right) =
\sum\limits_{i = 1}^{{s_2}}
{Var\left( {{W_{i,b}}{Y_i}}
\right)} + \sum\limits_{i =
1}^{{s_2}} {\sum\limits_{j
\ne i} {Cov\left(
{{W_{i,b}}{Y_i},{W_{j,b}}{Y_j}}
\right)} } .$$
The upper bound for $Var\left(
{{W_{i,b}}{Y_i}} \right)$ is
\begin{align*}Var\left(
{{W_{i,b}}{Y_i}} \right) & =
E\left[ {W_{i,b}^2Y_i^2}
\right] - {E^2}\left[
{{W_{i,b}}{Y_i}}
\right]\\ & \le
E\left[ {W_{i,b}^2Y_i^2}
\right] = E\left[
{E\left[ {\left. {W_{i,b}^2}
\right|{X_i}}
\right]E\left[ {\left.
{Y_i^2} \right|{X_i}}
\right]}
\right]\\ & \le
E\left[ {E\left[ {\left.
{W_{i,b}^2} \right|{X_i}}
\right]E\left[ {\mathop
{\sup }\limits_{d,x}
E\left[ {\left.
{{{\left( {{Y_i}} \right)}^2}}
\right|{X_i} = x,{D_i} = d}
\right]} \right]}
\right]\\ & =
E\left[ {E\left[ {\left.
{W_{i,b}^2} \right|{X_i}}
\right]E\left[ {\mathop
{\sup }\limits_{d,x}
E\left[ {\left.
{{{\left( {Y_i^d}
\right)}^2}} \right|{X_i} =
x} \right]} \right]}
\right]\\ & \le
E\left[ {W_{i,b}^2}
\right]{C_2} = O\left( {{N^{ -
2{\beta _2} + {\beta _1}}}}
\right),\end{align*}
using the identifying assumptions, Lipschitz continuity of
$E\left[ {\left.
{{{\left( {Y_i^d}
\right)}^2}} \right|{X_i}}
\right]$ on a bounded covariate space and
noting that ${W_{i,b}}$ and
${Y_i}$ are independent conditional on
${X_i}$. This yields that
$$\sum\limits_{i =
1}^{{s_2}} {Var\left( {{W_{i,b}}{Y_i}}
\right)} \le
{s_2}E\left[ {W_{i,b}^2}
\right]{C_2} = O\left(
{{N^{{\beta _1} - {\beta
_2}}}} \right).$$
Note that this upper bound converges to zero for
${\beta _2} > {\beta _1} > 1/2$. In order
to derive the upper bound for the covariance part, we rewrite the
covariance as
\begin{align*}\left|
{Cov({W_{i,b}}{Y_i},{W_{j,b}}{Y_j})}
\right| & = \left|
{Cov\left( {{W_{i,b}}{Y_i},\left(
{1 - \sum\limits_{k \ne j}
{{W_{k,b}}} } \right){Y_j}}
\right)}
\right|\\ & =
\left| { -
\sum\limits_{k \ne j}
{Cov\left(
{{W_{i,b}}{Y_i},{W_{k,b}}{Y_j}}
\right)} }
\right|\\ & =
\left|
{\sum\limits_{k \ne j}
{Cov\left(
{{W_{i,b}}{Y_i},{W_{k,b}}{Y_j}}
\right)} }
\right|.\end{align*}
Next, we derive the upper bound of $\sum\limits_{k
\ne j} {Cov\left(
{{W_{i,b}}{Y_i},{W_{k,b}}{Y_j}}
\right)}$
{\small
\begin{align*} \sum\limits_{k
\ne j} {Cov\left(
{{W_{i,b}}{Y_i},{W_{k,b}}{Y_j}}
\right)} & = \sum\limits_{k
\ne j} {E\left[
{{W_{i,b}}{Y_i}{W_{k,b}}{Y_j}} \right]
- E\left[ {{W_{i,b}}{Y_i}}
\right]E\left[
{{W_{k,b}}{Y_j}} \right]}
 \\ & =
\sum\limits_{k \ne j}
{E\left[ {{W_{i,b}}{Y_i}{W_{k,b}}}
\right]E\left[ {{Y_j}}
\right] - E\left[
{{W_{i,b}}{Y_i}}
\right]E\left[ {{W_{k,b}}}
\right]E\left[ {{Y_j}}
\right]}\\& =
\sum\limits_{k \ne j}
{\left( {E\left[
{{W_{i,b}}{Y_i}{W_{k,b}}} \right] -
E\left[ {{W_{i,b}}{Y_i}}
\right]E\left[ {{W_{k,b}}}
\right]} \right)E\left[
{{Y_j}} \right]} \\ &=
\sum\limits_{k \ne j}
{\left( {E\left[
{E\left[\left.
{W_{i,b}}{W_{k,b}}\right| {{X_i}},{X_k}
\right]E\left[
{{Y_i}\left| {{X_i}}
\right.} \right]}
\right]} \right.}
\\ &  - \left.
{E\left[ {E\left[
{{W_{i,b}}\left| {{X_i}}
\right.} \right]E\left[
{{Y_i}\left| {{X_i}}
\right.} \right]}
\right]E\left[ {{W_{k,b}}}
\right]} \right)E\left[
{{Y_j}} \right] \\ &  =
\sum\limits_{k \ne j}
{\left( {E\left[ {\left(
{E\left[\left.
{W_{i,b}}{W_{k,b}}\right| {{X_i}},{X_k} \right] -
E\left[ {{W_{i,b}}\left|
{{X_i}} \right.}
\right]E\left[ {{W_{k,b}}}
\right]} \right)E\left[
{{Y_i}\left| {{X_i}}
\right.} \right]}
\right]} \right)E\left[
{{Y_j}} \right]}  \\ & 
\le \sum\limits_{k
\ne j} {\left( {E\left[
{{W_{i,b}}{W_{k,b}}} \right] -
E\left[ {{W_{i,b}}}
\right]E\left[ {{W_{k,b}}}
\right]} \right)C_1^2}
\\ & =
\sum\limits_{k \ne j}
{Cov({W_{i,b}},{W_{k,b}})C_1^2 =
\frac{{Var({W_{i,b}})}}{{{s_2} -
1}}C_1^2},\end{align*}}

\noindent where the inequality uses the same supremum argument as the proof for
the variance, the fact that the expected potential outcomes are
Lipschitz continuous on a bounded covariate space, i.e.,
$E\left[
{Y_i^d\left| {{X_i}}
\right.} \right] \in
\left[ { - {C_1},{C_1}}
\right]$ and
$E\left[ {Y_i^d}
\right] \in \left[ { -
{C_1},{C_1}} \right]$ for some
positive constant ${C_1}$, and the fact
that the final sum of covariances is positive and therefore the product
certainly bounds the original sum from above. This yields an upper bound
for
\begin{align*}\sum\limits_{i
= 1}^{{s_2}} {\sum\limits_{j
\ne i} {\left|
{Cov\left(
{{W_{i,b}}{Y_i},{W_{j,b}}{Y_j}}
\right)} \right|} }
& \le {s_2}({s_2} -
1)\frac{{Var({W_{i,b}})}}{{{s_2} -
1}}C_1^2\\ & = O\left(
{{N^{{\beta _1} - {\beta
_2}}}}
\right).\end{align*}
The overall upper bound for the tree variance is
$$Var\left(
{\sum\limits_{i = 1}^{{s_2}}
{{W_{i,b}}{Y_i}} } \right) =
O\left( {{N^{{\beta _1} -
{\beta _2}}}} \right).$$
The constant depends on $\varepsilon
$, the common support parameter, regularity parameter
$\nu $, number of treatments
$M$ and a constant related to Lipschitz
continuity of the outcome variable.

Due to non-negativity of the variance, it is enough to find the lower
bound for $E\left[ {W_{i,b}^2Y_i^2} \right]$ to
analyse the lower bound of $Var\left(
{{{W}_{i,b}}{Y_i}}
\right)$.
\begin{align*}E\left[
{W_{i,b}^2Y_i^2} \right] & =
E\left[ {E\left[
{W_{i,b}^2\left| {{X_i}}
\right.} \right]E\left[
{Y_i^2\left| {{X_i}}
\right.} \right]}
\right]\\ & \ge
E\left[ {E\left[
{W_{i,b}^2\left| {{X_i}}
\right.} \right]E\left[
{\mathop {\inf
}\limits_{d,x} E\left[
{Y_i^2\left| {{X_i}}
\right. = x,{D_i} = d} \right]}
\right]}
\right]\\ & =
E\left[ {E\left[
{W_{i,b}^2\left| {{X_i}}
\right.} \right]}
\right]E\left[ {\mathop
{\inf }\limits_{d,x}
{\mkern 1mu} E\left[
{{{\left( {Y_i^d}
\right)}^2}\left| {{X_i} =
x} \right.} \right]}
\right]\\ & =
\Omega \left(
{{N^{\frac{{\log (1 -
\alpha )}}{{\log (\alpha
)}}{\beta _1} - 2{\beta _2}}}}
\right).\end{align*}
Therefore, $\sum\limits_{i
= 1}^{{s_2}} {Var\left(
{{W_{i,b}}{Y_i}} \right)} =
\Omega \left(
{{N^{\frac{{\log (1 -
\alpha )}}{{\log (\alpha
)}}{\beta _1} - {\beta _2}}}}
\right)$.

\bigskip
The lower bound for the covariance can be derived using the triangular
inequality,
\begin{align*}\left|
{Cov({W_{i,b}}{Y_i},{W_{j,b}}{Y_j})}
\right| & = \left|
{E\left[
{{W_{i,b}}{Y_i}{W_{j,b}}{Y_j}} \right]
- E\left[ {{W_{i,b}}{Y_i}}
\right]E\left[
{{W_{j,b}}{Y_j}} \right]}
\right|\\ &
\ge \bigg|
\left| {E\left[
{{W_{i,b}}{Y_i}{W_{j,b}}{Y_j}}
\right]} \right| -
\left| {E\left[
{{W_{i,b}}{Y_i}}
\right]E\left[
{{W_{j,b}}{Y_j}} \right]}
\right|
\bigg|\end{align*}
and finding lower bounds for the individual elements:
\bigskip

\resizebox{0.97\textwidth}{!}{$
	\begin{aligned}
\left|
{E\left[
{{W_{i,b}}{Y_i}{W_{j,b}}{Y_j}}
\right]} \right| & =
\left| {E\left[
{E\left[
{{W_{i,b}}{Y_i}{W_{j,b}}{Y_j}\left|
{\lambda (S(x))} \right.}
\right]} \right]}
\right|\\ & =
\left| {E\left[
{E\left[ {1/{{\left(
{N_{S(x)}^d}
\right)}^2}{Y_i}{Y_j}\left|
{\lambda (S(x)),{X_i},{X_j} \in
S(x)} \right.} \right]\Pr
\left( {{X_i},{X_j} \in
S(x)\left| {\lambda (S(x))}
\right.} \right)}
\right]}
\right|\\ &
\ge \frac{1}{{{{\left(
{{s_{2,d}}}
\right)}^2}}}\mathop
{\inf }\limits_{d,x}
E\left[ {\left.
{\left| {{Y_i}{Y_j}}
\right|}
\right|{D_i}=d,{D_j} = d,{X_i}=x,{X_j} =
x} \right]E\left[
{{{\left( {\lambda (S(x))}
\right)}^2}}
\right]\\ & =
\Omega \left( {{N^{ -
2{\beta _2} - 2{\beta _1}}}}
\right) \\ 
\left|
{E\left[ {{W_{i,b}}{Y_i}}
\right]} \right| & =
\left| {E\left[
{E\left[
{{W_{i,b}}{Y_i}\left|
{\lambda (S(x))} \right.}
\right]} \right]}
\right|\\ &=
\left| {E\left[
{E\left[
{{W_{i,b}}{Y_i}\left| {{X_i}
\in S(x),\lambda (S(x))}
\right.} \right]\Pr
\left( {{X_i} \in
S(x)\left| {\lambda (S(x))}
\right.} \right)}
\right]}
\right|\\&
\ge
\frac{1}{{{s_{2,d}}}}\mathop
{\inf }\limits_{d,x}
E\left[ {\left.
{\left| {{Y_i}}
\right|} \right|{D_i} =
d,{X_i} = x} \right]E\left[
{\lambda (S(x))}
\right]\\& =
\Omega \left( {{N^{ -
{\beta _2} - {\beta _1}}}}
\right)\end{aligned}$}

\bigskip
\noindent Since $\left|
{E\left[
{{W_{i,b}}{Y_i}{W_{j,b}}{Y_j}}
\right]} \right| =
\Omega \left( {{N^{ -
2{\beta _2} - 2{\beta _1}}}}
\right)$ and $\left| {E\left[
{{W_{i,b}}{Y_i}} \right]}
\right|\left|
{E\left[ {{W_{j,b}}{Y_j}}
\right]} \right| =
\Omega \left( {{N^{ -
2{\beta _2} - 2{\beta _1}}}}
\right)$, the final lower bound is
$$\left|
{Cov({W_{i,b}}{Y_i},{W_{j,b}}{Y_j})}
\right| = \Omega
\left( {{N^{ - 2{\beta _2} -
2{\beta _1}}}}
\right).$$
Putting these results together yields a lower bound for the covariance:
$$\sum\limits_{i =
1}^{{s_2}} {\sum\limits_{j
\ne i} {\left|
{Cov\left(
{{W_{i,b}}{Y_i},{W_{j,b}}{Y_j}}
\right)} \right|} } =
\Omega \left( {{N^{ -
2{\beta _1}}}}
\right).$$
As the lower bound for the covariance exceeds the lower bound for the
variance, the lower bound for the tree variance is determined by the one
converging slower to zero i.e.
$$Var\left(
{\sum\limits_{i = 1}^{{s_2}}
{{W_{i,b}}{Y_i}} } \right) =
\Omega \left(
{{N^{\frac{{\log (1 -
\alpha )}}{{\log (\alpha
)}}{\beta _1} - {\beta _2}}}}
\right).$$
The constant depends on $\varepsilon
$, the common support parameter, regularity parameter
$\nu $ and a constant related
to Lipschitz continuity of the outcome variable.
}

\begin{lem}
Let the assumptions from \lemref{lem3} hold and
${A_{i,{N_2}}}: = {W_i}{Y_i} -
E\left[ {{W_i}{Y_i}}
\right]$, where
${W_i}$ are the forest weights of the
\emph{mcf}. Then ${({A_{i,{N_2}}})_{i =
1,...,{N_2}}}$ is a triangular array satisfying:
\begin{enumerate}
	\def\labelenumi{\alph{enumi})}
	\item
	$E[{A_{i,{N_2}}}] = 0$,
	\item
	$\sum\limits_{i =
	1}^{{N_2}} {E[A_{i,{N_2}}^2]} <
	\infty $ for all
	$N$ and $i$,
	\item $\sigma^2_N := Var(A_{1,N_2} + ... + A_{N_2,N_2}) \underset{N\to \infty}{\xrightarrow{\hspace{40pt}}} 0$,
	\item $\sum\limits_{i= 1}^{N_2} E\left[A_{i, N_2}^2\underline{1}\left(|A_{i,N_2}|>\tilde{\varepsilon}\right)\right] \underset{N\to \infty}{\xrightarrow{\hspace{40pt}}} 0$
	for all $\tilde \varepsilon	> 0$,
	\item
	There is a summable sequence ${\left(
	{{\tilde \pi _r}} \right)_{r
	\in \mathbb{N}}}$ such
	that for all $\tilde u \in
	\mathbb{N}$ and all indices
	$1 \le {q_1} < {q_2}
	< ... < {q_{\tilde u}}
	< {q_{\tilde u}} + r = {t_1}
	\le {t_2} \le {N_2}$,
	the following upper bounds for covariances hold true for all
	measurable functions
	$g:{\mathbb{R}^{\tilde
	u}} \to \mathbb{R}$ with
	${\left\| g
	\right\|_\infty } =
	{\sup _{x \in
	{\mathbb{R}^{\tilde
	u}}}}\left| {g(x)}
	\right| \le 1$:
$$\left|
{Cov(g({A_{{q_1},{N_2}}},...,{A_{{q_{\tilde
u}},{N_2}}}){A_{{q_{\tilde
u}},{N_2}}},{A_{{t_1},{N_2}}})}
\right| \le \left(
{E[A_{{q_{\tilde u}},{N_2}}^2] +
E[A_{{t_1},{N_2}}^2] + N_2^{ - 1}}
\right){\tilde \pi _r}$$
and
$$\left|
{Cov(g({A_{{q_1},{N_2}}},...,{A_{{q_{\tilde
u}},{N_2}}}),{A_{{t_1},{N_2}}},{A_{{t_2},{N_2}}})}
\right| \le \left(
{E[A_{{t_1},{N_2}}^2] +
E[A_{{t_2},{N_2}}^2] + N_2^{ - 1}}
\right){\tilde \pi _r}.$$
\end{enumerate}
\label{lem4}
\end{lem}

\proof{
\begin{enumerate}
	\def\labelenumi{\alph{enumi})}
	\item
	$E\left[ {{W_i}{Y_i} -
	E\left[ {{W_i}{Y_i}} \right]}
	\right] = 0$.
	\item $E\left[ {A_{i,{N_2}}^2}
	\right] = Var\left( {{W_i}{Y_i}}
	\right) = {E\left[
	{W_i^2Y_i^2} \right] -
	{E^2}\left[ {{W_i}{Y_i}}
	\right]}$

	To derive upper bounds for
	$E[{W_i}{Y_i}]$ and
	$E[W_i^2Y_i^2]$, we use that
	${W_i}$ and
	${Y_i}$ are independent conditional on
	${X_i}$ and $E\left[ {\left.
	{{Y^d}} \right| X = x}
	\right]$ and
	$E\left[ {\left.
	{{{( {{Y^d}})}^2}} \right| X = x}
	\right]$ are Lipschitz continuous and
	therefore can be bounded from above by ${C_1}
	< \infty $ and
	${C_2} < \infty $
	respectively on a bounded covariate space. The first and second moment
	of the forest weights are
$$E\left[ {{W_i}}
\right] = \frac{{{N_2} -
{s_2}}}{{{N_2}}} \cdot 0 +
\frac{{{s_2}}}{{{N_2}}}E\left[
{\frac{1}{B}\sum\limits_{b
= 1}^B {{W_{i,b}}} } \right] =
\frac{{{s_2}}}{{{N_2}}}\frac{1}{B}B\frac{1}{{{s_2}}}
= \frac{1}{{{N_2}}} \sim {N^{
- 1}}$$
	and
	\begin{align*}E\left[
	{W_i^2} \right] & = \frac{{{N_2}
	- {s_2}}}{{{N_2}}} \cdot 0 +
	\frac{{{s_2}}}{{{N_2}}}E\left[
	{{{\left(
	{\frac{1}{B}\sum\limits_{b
	= 1}^B {{W_{i,b}}} } \right)}^2}}
	\right]\\ &=
	\frac{{{s_2}}}{{{N_2}}}\frac{1}{{{B^2}}}\left[
	{\sum\limits_{b = 1}^B
	{E\left[ {W_{i,b}^2} \right] +
	} \sum\limits_{b = 1}^B
	{\sum\limits_{b\textquotesingle
	\ne b} {E\left[
	{{W_{i,b}}{W_{i,b\textquotesingle}}}
	\right]} } }
	\right]\\ &=
	\frac{{{s_2}}}{{{N_2}}}\frac{1}{{{B^2}}}\left[
	{BE\left[ {W_{i,b}^2} \right] +
	B(B - 1)\left(
	{Cov({W_{i,b}},{W_{i,b\textquotesingle}}) +
	{{\left( {E\left[ {{W_{i,b}}}
	\right]} \right)}^2}}
	\right)}
	\right]\\ &\le
	\frac{{{s_2}}}{{{N_2}}}\frac{1}{{{B^2}}}\left[
	{BE\left[ {W_{i,b}^2} \right] +
	B(B - 1)\left( {Var({W_{i,b}}) +
	{{\left( {E\left[ {{W_{i,b}}}
	\right]} \right)}^2}}
	\right)}
	\right]\\& =
	\frac{{{s_2}}}{{{N_2}}}\frac{1}{{{B^2}}}\left[
	{{B^2}E\left[ {W_{i,b}^2}
	\right]} \right] =
	O\left( {{N^{ - 1 - {\beta _2} +
	{\beta _1}}}}
	\right).\end{align*}
	
	These results together with identifying assumptions and Lipschitz
	continuity can be used to further bound the following expectations:
	\begin{gather*}
		E\left[ {{W_i}{Y_i}}
		\right] = E\left[
		{E\left[ {{W_i}\left|
		{{X_i}} \right.}
		\right]E\left[
		{{Y_i}\left| {{X_i}}
		\right.} \right]}
		\right] = O\left( {{N^{ - 1}}}
		\right), \\
		E[W_i^2Y_i^2] = E\left[
		{E\left[ {W_i^2\left|
		{{X_i}} \right.}
		\right]E\left[
		{Y_i^2\left| {{X_i}}
		\right.} \right]}
		\right] = O\left( {{N^{ - 1 -
		{\beta _2} + {\beta _1}}}}
		\right)
	\end{gather*}
It follows that
$$\sum\limits_{i =
1}^{{N_2}} {E[A_{i,{N_2}}^2]} =
O\left( {{N^{ - {\beta _2} +
{\beta _1}}}} \right) <
\infty.$$
%
	\item \begin{align*}Var({A_{1,{N_2}}}
	+ ... + {A_{{N_2},{N_2}}})& = Var\left(
	{\sum\limits_{i = 1}^{{N_2}}
	{{W_i}{Y_i}} - \sum\limits_{i =
	1}^{{N_2}} {E\left[ {{W_i}{Y_i}}
	\right]} } \right) =
	Var\left( {\sum\limits_{i
	= 1}^{{N_2}} {{W_i}{Y_i}} }
	\right)\\& =
	\sum\limits_{i = 1}^{{N_2}}
	{Var\left( {{W_i}{Y_i}} \right)}
	+ \sum\limits_{i = 1}^{{N_2}}
	{\sum\limits_{j \ne i}
	{Cov\left( {{W_i}{Y_i},{W_j}{Y_j}}
	\right)} } \\&
	\le \sum\limits_{i =
	1}^{{N_2}} {Var\left( {{W_i}{Y_i}}
	\right)} + \sum\limits_{i
	= 1}^{{N_2}} {\sum\limits_{j
	\ne i} {\left|
	{Cov\left( {{W_i}{Y_i},{W_j}{Y_j}}
	\right)} \right|} }
	\end{align*}

	Using the results from b), the first element is bounded at rate
	$O\left( {{N^{ - {\beta
	_2} + {\beta _1}}}}
	\right)$ and by a similar logic as in the
	tree case the sum of all covariances has to decay to zero as fast as the
	sum of variances. These results yield that $Var(A_{1,N_2} + ... + A_{N_2,N_2}) \underset{N\to \infty}{\xrightarrow{\hspace{40pt}}} 0$.

	\item
	Due to monotonicity and the result in b):
	$$\sum\limits_{i= 1}^{N_2} E\left[A_{i, N_2}^2\underline{1}\left(|A_{i,N_2}|>\tilde{\varepsilon}\right)\right] \le \sum\limits_{i= 1}^{N_2} E\left[A_{i, N_2}^2\right] \to 0 \text{ for all } \tilde \varepsilon > 0.$$
	\item
	Due to exchangeability which implies strict stationarity and the
	result in c), it is possible to interpret
	${\left( {{A_{i,{N_2}}}}
	\right)_{i = 1,...,{N_2}}}$as a
	$\rho $-mixing process.
	Since every $\phi $-mixing
	process is also $\rho
	$-mixing, then along the Lemma 20.1 in \citet{billingsley1968convergence} for $\phi$-mixing
	processes, we can also bound the covariances of a
	$\rho $-mixing process as
	follows:\footnote{\citet{Bradley1986} showed that the coefficients of the
		two processes fulfil the following inequality:
		${\rho _r} \le 2\sqrt
		{{\phi _r}} $.}
	\begin{multline*}\left|
	{Cov\left( {g\left(
	{{A_{{q_1},{N_2}}},...,{A_{{q_{\tilde
	u}},{N_2}}}}
	\right){A_{{q_{\tilde
	u}},{N_2}}},{A_{{t_1},{N_2}}}}
	\right)}
	\right|\\
	\le 2\sqrt {{\phi
	_{{t_1} - {q_{\tilde u}}}}}
	\sqrt {E\left[
	{{g^2}\left(
	{{A_{{q_1},{N_2}}},...,{A_{{q_{\tilde
	u}},{N_2}}}}
	\right)A_{{q_{\tilde
	u,{N_2}}}}^2} \right]} \sqrt
	{E\left[ {A_{{t_1},{N_2}}^2}
	\right]} \end{multline*}
	and
	\begin{multline*}\left|
	{Cov(g({A_{{q_1},{N_2}}},...,{A_{{q_{\tilde
	u}},{N_2}}}),{A_{{t_1},{N_2}}}{A_{{t_2},{N_2}}}}
	\right| \le 2{\phi
	_{{t_1} - {q_{\tilde
	u}}}}E\left[ {\left|
	{{A_{{t_1},{N_2}}}{A_{{t_2},{N_2}}}}
	\right|} \right] \\ \le
	2{\phi _{{t_1} - {q_{\tilde
	u}}}}E\left[ {A_{{t_1},{N_2}}^2}
	\right]\end{multline*}
	where the last inequality follows from the stationarity of the process.
	The two results can be further bounded
	\begin{align*}\left|
	{Cov(g({A_{{q_1},{N_2}}},...,{A_{{q_{\tilde
	u}},{N_2}}}){A_{{q_{\tilde
	u}},{N_2}}},{A_{{t_1},{N_2}}}}
	\right| & \le 2\sqrt
	{{\phi _{{t_1} - {q_{\tilde
	u}}}}} \sqrt {E\left[
	{A_{{q_{\tilde u}},{N_2}}^2}
	\right]} \sqrt {E\left[
	{A_{{t_1},{N_2}}^2} \right]}
	\\ & \le \sqrt
	{{\phi _{{t_1} - {q_{\tilde
	u}}}}} \left( {E\left[
	{A_{{q_{\tilde u}},{N_2}}^2}
	\right] + E\left[
	{A_{{t_1},{N_2}}^2} \right]}
	\right)\\& \le
	\sqrt {{\phi _{{t_1} -
	{q_{\tilde u}}}}} \left(
	{E\left[ {A_{{q_{\tilde
	u}},{N_2}}^2} \right] +
	E\left[ {A_{{t_1},{N_2}}^2}
	\right] + N_2^{ - 1}}
	\right)\end{align*}
	by property of the function $g()$ and
	inequality of arithmetic and geometric mean and
	$$\left|
	{Cov(g({A_{{q_1},{N_2}}},...,{A_{{q_{\tilde
	u}},{N_2}}}),{A_{{t_1},{N_2}}}{A_{{t_2},{N_2}}}}
	\right| \le \sqrt
	{{\phi _{{t_1} - {q_{\tilde
	u}}}}} \left( {E\left[
	{A_{{t_1},{N_2}}^2} \right] +
	E\left[ {A_{{t_2},{N_2}}^2}
	\right] + N_2^{ - 1}}
	\right)$$
	by stationarity and the fact that the mixing coefficients are smaller or
	equal to 1. Then the weak dependence conditions are fulfilled for
	${\tilde \pi _r} = \sqrt
	{{\phi _r}} $.
\end{enumerate}
}

\begin{lem}
Define
${A_{i,{N_2}}}$ as in \lemref{lem4}, then $\hat \theta
_{m,l}^{mcf}(x) - E\left[ {\hat
\theta _{m,l}^{mcf}(x)} \right] =
\sum\limits_{i = 1}^{{N_2}}
{{A_{i,{N_2}}}} $ and

$$\frac{{\hat
\theta _{m,l}^{mcf}(x) - E\left[
{\hat \theta _{m,l}^{mcf}(x)}
\right]}}{{\sqrt
{Var\left(\hat \theta _{m,l}^{mcf}(x)\right)}
}} \to N(0,1).$$
\label{lem5}
\end{lem}

\proof{
The normality proof for triangular arrays satisfying the
conditions in \lemref{lem4} is proven in \citet{neumann2013central} and can be applied on
estimation of potential outcomes ${\mu
_m}(x)$ and ${\mu
_l}(x)$. As those are estimated on the estimation data
set, the difference of the two quantities will also follow a normal
distribution.}

\begin{thm}
	Assume that there is a sample of size $N$
	containing i.i.d.~data $(X_i,Y_i,\allowbreak D_i) \in \left[0,1\right]^p \times	\mathbb{R} \times \{0,1,...,M-1\}$ for a given value of
	$x$. Moreover, the covariates are independently and uniformly
	distributed $X \sim U\left(\left[0,1\right]^p\right)$.
	Let $T$ be an honest, regular, and symmetric random-split tree.
	Further assume that $E\left[\left. Y^d \right| X = x \right]$ and
	$E\left[\left. ( Y^d )^2 \right| X = x \right]$ are Lipschitz
	continuous and
	$Var \left( \left. Y^d \right| X = x \right) > 0$. Then for
	$1/2 < \beta_1 < \beta_2 < \frac{p +
		2}{p} \frac{\log(1 - \alpha)}{\log(\alpha)} \beta_1$,
	$$
	\frac{\hat \theta_{m,l}^{mcf}(x) - \theta_{m,l}^{0}(x)}{\sqrt{Var\left(\hat \theta_{m,l}^{mcf}(x)\right)}} \overset{d}{\longrightarrow} N(0,1).
	$$
	\vspace{-\baselineskip} 
\label{thm2-proof}
\end{thm}

\proof{
Given the result in \lemref{lem5}, it remains to show that
$$\frac{{E\left[
{\hat \theta _{m,l}^{mcf}(x)}
\right] - \theta
_{m,l}^0(x)}}{{\sqrt {Var\left(\hat
\theta _{m,l}^{mcf}(x)\right)} }} \to
0.$$
The final result will follow then from Slutsky's lemma. By \thmref{thm1-proof}, we
have
$$\left| {E\left[
{\hat \theta _{m,l}^{mcf}(x)}
\right] - \theta _{m,l}^0(x)}
\right| = O\left( {{s_1}^{
- \log (1 - \alpha )/p\log
(\alpha )}} \right).$$
From \corref{cor1} and \lemref{lem3}, we get
$$Var\left(\hat \theta
_{m,l}^{mcf}(x)\right) = \Omega \left(
{{N^{\frac{{\log (1 -
\alpha )}}{{\log (\alpha
)}}{\beta _1} - {\beta _2}}}}
\right).$$
It follows that
$$\frac{{\left(
{E\left[ {\hat \theta
_{m,l}^{mcf}(x)} \right] - \theta
_{m,l}^0(x)} \right)}}{{\sqrt
{Var\left(\hat \theta _{m,l}^{mcf}(x)\right)}
}} = O\left( {{N^{ - \left(
{\frac{1}{p} + \frac{1}{2}}
\right)\frac{{\log (1 -
\alpha )}}{{\log (\alpha
)}}{\beta _1} +
\frac{{{\beta _2}}}{2}}}}
\right).$$
The ratio converges to zero when
$$\frac{p}{{2 +
p}}{\beta _2} <
\frac{{\log (1 - \alpha
)}}{{\log (\alpha
)}}{\beta _1}.$$}

\paragraph{Proofs for Theorem \ref{thm3-proof}}\label{appendix-a143}

\begin{thm}
Let all assumptions from \thmref{thm2-proof} hold and define
$\hat \theta_{m,l}^{mcf}$ as an average of all corresponding
$\hat \theta_{m,l}^{mcf}(x)$. Then,
$$\frac{{\hat
\theta _{m,l}^{mcf} - \theta
_{m,l}^0}}{{\sqrt {Var\left(\hat
\theta _{m,l}^{mcf}\right)} }} \to
N(0,1).$$
\vspace{-\baselineskip}
\label{thm3-proof}
\end{thm}

\proof{
Using the CLT for triangular arrays of weakly dependent
random variables requires to check that all requirements in \lemref{lem4} hold
for ${A_{i,{N_2}}}: = W_i^{m,l}{Y_i} -
E\left[ {W_i^{m,l}{Y_i}}
\right]$, where
$W_i^{m,l}: = W_i^{mcf}(m,l)$
represents the weights of the ATE. The proof uses the observation that
the rates for the $ATE$ cannot be worse than for the $IATE$
and, therefore, the conditions in \lemref{lem4} will be satisfied for the
weights of the ATE. The convergence rate may be affected due to the
pointwise convergence at different points $x$.\footnote{Nevertheless, the
	simulation results show that the RMSE for the \emph{mcf} estimator of
	the ATE decreases at a $\sqrt N -
	$rate for many DGPs, suggesting that the estimator
	exhibits $\sqrt N -
	$convergence.}

\begin{enumerate}
	\def\labelenumi{\alph{enumi})}
	\item
	$E\left[ {{A_{i,{N_2}}}}
	\right] = 0$ holds trivially.
	\item
	$\sum\limits_{i =
	1}^{{N_2}} {E\left[
	{A_{i,{N_2}}^2} \right]} =
	\sum\limits_{i = 1}^{{N_2}}
	{Var\left( {W_i^{m,l}{Y_i}}
	\right)} $. The upper bound on the
	individual variances is the upper bound of
	$E\left[ {{{\left(
	{W_i^{m,l}} \right)}^2}Y_i^2}
	\right]$:
\begin{align*}E\left[
{{{\left( {W_i^{m,l}}
\right)}^2}Y_i^2} \right] & =
E\left[ {{{\left(
{\frac{1}{{{N_2}}}\sum\limits_{j
= 1}^{{N_2}}
{W_i^{mcf}({D_i},{X_i};{X_j},m,l)} }
\right)}^2}Y_i^2} \right]
\\ & \le
\frac{1}{{{{\left( {{N_2}}
\right)}^2}}}E\left[
{{{\left( {{N_2}}
\right)}^2}E\left[
{\left. {{{\left(
{W_i^{mcf}({D_i},{X_i};{X_j},m,l)}
\right)}^2}} \right|{X_i}}
\right]E\left[
{Y_i^2\left| {{X_i}}
\right.} \right]}
\right]\\ & =
O\left( {{N^{ - 1 - {\beta _2} +
{\beta _1}}}}
\right).\end{align*}
The second requirement is also satisfied as
	$\sum\limits_{i =
	1}^{{N_2}} {Var\left(
	{W_i^{m,l}{Y_i}} \right)} =
	O\left( {{N^{ - {\beta _2} +
	{\beta _1}}}} \right) <
	\infty.$

	\item
	\begin{align*}Var({A_{1,{N_2}}}
	+ ... + {A_{{N_2},{N_2}}}) & = Var\left(
	{\sum\limits_{i = 1}^{{N_2}}
	{W_{i}^{m,l}{Y_i}} -
	\sum\limits_{i = 1}^{{N_2}}
	{E\left[ {W_{i}^{m,l}{Y_i}}
	\right]} } \right) \\ & =
	Var\left( {\sum\limits_{i
	= 1}^{{N_2}} {W_{i}^{m,l}{Y_i}} }
	\right)\\ & \le
	\sum\limits_{i = 1}^{N_2}
	{Var\left( {W_{i}^{m,l}{Y_i}}
	\right)} + \sum\limits_{i
	= 1}^{N_2} {\sum\limits_{j
	\ne i} {\left|
	{Cov\left(
	{W_{i}^{m,l}{Y_i},W_{j}^{m,l}{Y_j}}
	\right)} \right|} }
	\end{align*}
	Using the results from b), the first element is bounded at rate
	$O\left( {{N^{ - {\beta
	_2} + {\beta _1}}}}
	\right)$ and by a similar logic as in the
	tree case, the sum of all covariances must decay to zero as fast as the
	sum of the variances. These results yield that $Var(A_{1,N_2} + ... + A_{N_2,N_2}) \underset{N\to \infty}{\xrightarrow{\hspace{40pt}}} 0$.

	\item
	Due to monotonicity and result in b):
	$$\sum\limits_{i= 1}^{N_2} E\left[A_{i, N_2}^2\underline{1}\left(|A_{i,N_2}|>\tilde{\varepsilon}\right)\right] \le \sum\limits_{i= 1}^{N_2} E\left[A_{i, N_2}^2\right] \to 0 \text{ for all } \tilde \varepsilon > 0.$$

	\item
	The proof follows the same logic as in \lemref{lem4}.
\end{enumerate}
With this, all requirements for the CLT for triangular arrays of weakly
dependent random variables hold, so that we get
$$\frac{{\hat
\theta_{m,l}^{mcf} - \theta_{m,l}^0}}{{\sqrt {Var\left(\hat
\theta_{m,l}^{mcf}\right)} }} \to
N(0,1).$$
Note that $E\left[
{\sum\limits_{i = 1}^{{N_2}}
{W_i^d{Y_i}} } \right] =
E\left[ {{Y^d}}
\right]$ holds due to exchangeability
and the fact that the weights sum to 1. Therefore, we applied the CLT
directly to the quantity of interest.
}

\setcounter{corr}{2}
\begin{corr}
Let all assumptions from \thmref{thm2-proof} hold
and assume that a discrete $Z$ is among the splitting variables. Then for
$Z = z$,
$$
\frac{\hat \theta_{m,l}^{mcf}(z) - \theta_{m,l}^{0}(z)}{\sqrt{Var\left(\hat \theta_{m,l}^{mcf}(z)\right)}} \overset{d}{\longrightarrow} N(0,1).
$$
\end{corr}

\proof{
In a similar fashion as in \thmref{thm2-proof}, by setting
${A_{i,{N_2}}}: = W_i^{z,m,l}{Y_i} -
E\left[ {W_i^{z,m,l}{Y_i}}
\right]$, where
$W_i^{z,m,l}: =
W_i^{mcf}(z,m,l)$, and checking that all
requirements in \lemref{lem4} hold for
${A_{i,{N_2}}}$, the CLT for
triangular arrays of weakly dependent random variables can be applied to
show
$$
\frac{\hat \theta_{m,l}^{mcf}(z) - E\left[ \hat \theta_{m,l}^{mcf}(z) \right]}{\sqrt{Var\left(\hat \theta_{m,l}^{mcf}(z)\right)}} \overset{d}{\longrightarrow} N(0,1).
$$

Similarly as for $IATE$, when some
${Y_i}$ with a $z_{i}$
that does not belong to group $z$ has a non-zero weight, the final
estimator is not an unbiased estimator of
${\mu _d}(z)$. Assuming
that $Z$ is among the splitting variables guarantees that there
exists some sample size $\tilde
N$ such that for all $N >
\tilde N$ all trees split on $Z$. This
implies that only ${Y_i}$ which belong to
group $z$ have non-zero weights and
$E\left[ {\hat
\theta
_{m,l}^{mcf}(z)} \right] = \theta
_{m,l}^0(z).$}

\subsection{Weights-based inference}\label{appendix-a2}

There are several suggestions in the literature on how to conduct
inference and how to compute standard errors of Random Forest based
predictions \citep[e.g.,][and the references therein]{wager2014confidence, wager2018estimation}.
Although these methods can be used to
conduct inference on the IATEs, it is yet unexplored how these methods
could be readily generalized to take account of the aggregation steps
needed for the GATE and ATE parameters.

Therefore, the \emph{mcf} uses an alternative inference method useful
for estimators that have a representation as weighted averages of the
observed outcome. This perspective is attractive for Random Forest based
estimators as they consist of trees that first stratify the data (when
building a tree), and subsequently average over these strata (when
building the forest). Thus, the \emph{mcf} exploits the weights-based
representation explicitly for inference \citep[see also][for a related approach]{abadie2006large}.

Considering a weights-based estimator with random weights
$W_{i}$ (that are normalized such that all weights add
up to $N$) for $\hat
\theta $ (i.e., IATEs, GATEs, or ATEs):
$$\hat \theta =
\frac{1}{N}\sum\limits_{i
= 1}^N {{W_i}{Y_i}} ;\quad \quad \quad
Var(\hat \theta ) =
Var\left(
{\frac{1}{N}\sum\limits_{i
= 1}^N {{W_i}{Y_i}} }
\right).$$
The variance of the estimator can be rewritten as:
$$Var\left(
{\frac{1}{N}\sum\limits_{i
= 1}^N {{W_i}{Y_i}} } \right) =
{E_W}\left(
{\frac{1}{{{N^2}}}\sum\limits_{i
= 1}^N {{W_i}^2\sigma
_{Y| W}^2({W_i})} } \right) +
Va{r_W}\left(
{\frac{1}{N}\sum\limits_{i
= 1}^N {{W_i}\mu_{Y|W}({W_i})} }
\right),$$
where ${\mu
_{Y\left| W \right.}}({W_i})
= E({Y_i}|{W_i})$ and
$\sigma_{Y| W}^2({W_i}) =
Var({Y_i}|{W_i}).$ The derivation exploits
the combination of two-sample honesty with an i.i.d sample. Remember
that under two-sample honesty, observations can be either used for
training or for estimation, without switching the roles.\footnote{In
	contrast, the `one-sample honesty' in \citet{wager2018estimation} is based
	on continuously switching the role of observations used for tree
	building and effect estimation in their Causal Forest. Under this
	splitting procedure, each weight may still depend on many
	observations, and the conditioning set cannot be reduced as in the
	case of two-sample honesty.} Further recall that the forest weights of
the \emph{mcf} are computed as a function of training data determining
the final leaves and value $x_{i}$ picks the
corresponding weight, leading to the following representation:
${W_i} = w({x_i},{\Im
_{tr}})$. Therefore, under i.i.d. sampling,
$Y_{i}$ and $W_{j}$ are independent.
Thus, the conditioning set
${W_1},...,{W_N}$ can be reduced to
${W_i}$ for each conditional mean and
conditional variance.

This leads to the following expression of the variance of the proposed
estimators:\footnote{Note that the weighting estimator uses only the
	honest data, therefore the weights here sum up to
	$N_2$.}
$$Var\left(
{\frac{1}{{{N_2}}}\sum\limits_{i
= 1}^{{N_2}} {{W_i}{Y_i}} } \right) =
{E_W}\left(
{\frac{1}{{N_2^2}}\sum\limits_{i
= 1}^{{N_2}} {{W_i}^2\sigma
_{Y| W}^2({W_i})} } \right) +
Va{r_W}\left(
{\frac{1}{{{N_2}}}\sum\limits_{i
= 1}^{{N_2}} {{W_i}\mu
_{Y| W}({W_i})} }
\right).$$
The above expression suggests using the following estimator:
$$\widehat {Var(\hat
\theta )} =
\frac{1}{{N_2^2}}\sum\limits_{i
= 1}^{{N_2}} {w_i^2\hat \sigma
_{Y| W}^2({w_i})} +
\frac{1}{{{N_2}({N_2} -
1)}}\sum\limits_{i = 1}^{{N_2}}
{{{\left( {{w_i}{{\hat
\mu }_{Y|\hat W}}({w_i}) -
\frac{1}{{{N_2}}}\sum\limits_{i
= 1}^{{N_2}} {{w_i}{{\hat \mu
}_{Y| W}}({w_i})} } \right)}^2}}.
$$

The conditional expectations and variances may be computed by standard
non-parametric or ML methods, as this is a one-dimensional problem for
which many well-established estimators exist. \citet{bodory2020finite}
investigate $k$-nearest
neighbour estimators to obtain estimates for these quantities. They
found good results in a binary treatment setting for the ATET. The same
method is used here.\footnote{They also found a considerable robustness
	on how exactly to compute the conditional means and variances. Note
	that since their results relate to aggregate treatment effect
	parameters, their generalisability to the level of IATE's is unclear.}
As both, ${\mu
_{Y| W}}$ and
$\sigma
_{Y| W}^2$, are bounded and the number of
nearest neighbours in their estimation is chosen such that it grows
slower than $N_2$, consistency of the conditional
expectations is guaranteed, see, e.g., \citet{devroye1994strong}. Additionally, for larger $N$
the non-zero weights concentrate at a close neighbourhood of the given
point $x$ as the leaves are shrinking towards a point. It is,
however, beyond the scope of this paper to analyse rigorously the exact
statistical conditions needed for this estimator to lead to valid
inference.

\subsection{Comparison of \emph{mcf} and \emph{grf}}\label{appendix-a3}
Comparing forest weights of \emph{grf},
$w_i^{grf}(x)$, and the \emph{mcf},
$w_i^{mcf}({d_i},{x_i};x,m,l)$,
both represent how important observation $i$ is for estimation of
any quantity at point $x$. Meanwhile, the \emph{mcf} weights weigh
the observed outcome directly, while \emph{grf} applies the forest
weights in a locally weighted moment function. Thus, a weighted
representation of the final \emph{grf} estimator of the IATE at point
$x$ weighs the observed outcome $y_{i}$ by
weights that combine $w_i^{grf}(x)$
with treatment assignments in a non-linear fashion (see the result in
\autoref{42}).

A further difference between the \emph{grf} and the \emph{mcf} is
related to the number of forest weights used in the IATE estimator, and
how many trees contribute to the calculation of each weight. Both are
determined by the corresponding honesty procedures. \autoref{42}
summarizes that the \emph{grf} forest weights represent by how much each
observation is relevant for the $IATE(m,l;x)$. This means that
there will be $N$ estimated forest weights because each half-sample
is drawn from the full sample. For large enough $B$, each
observation has a chance to end up in the estimation sample on which the
weights are estimated. However, the half-sampling and further honesty
split lead every forest weight
$w_i^{grf}(x)$ to be based on
approximately $B/4$ trees. On the other hand, \emph{mcf} computes
only $N/2$ of forest weights due to the primary half-split of data
into training and estimation data sets. However, since the same
$N/2$ observations are used in each tree to estimate the weights
for $\beta_{2}=1$, each forest weight is averaged across
all $B$ trees. For given $B$ and $N$, the \emph{grf} has
an advantage in smoothing over more observations and the \emph{mcf} in
the precision of the weights, potentially influencing finite sample
properties of both methods.

\begin{figure}[H]
	\caption{Diagram of \emph{grf}}
\includegraphics[width=\textwidth]{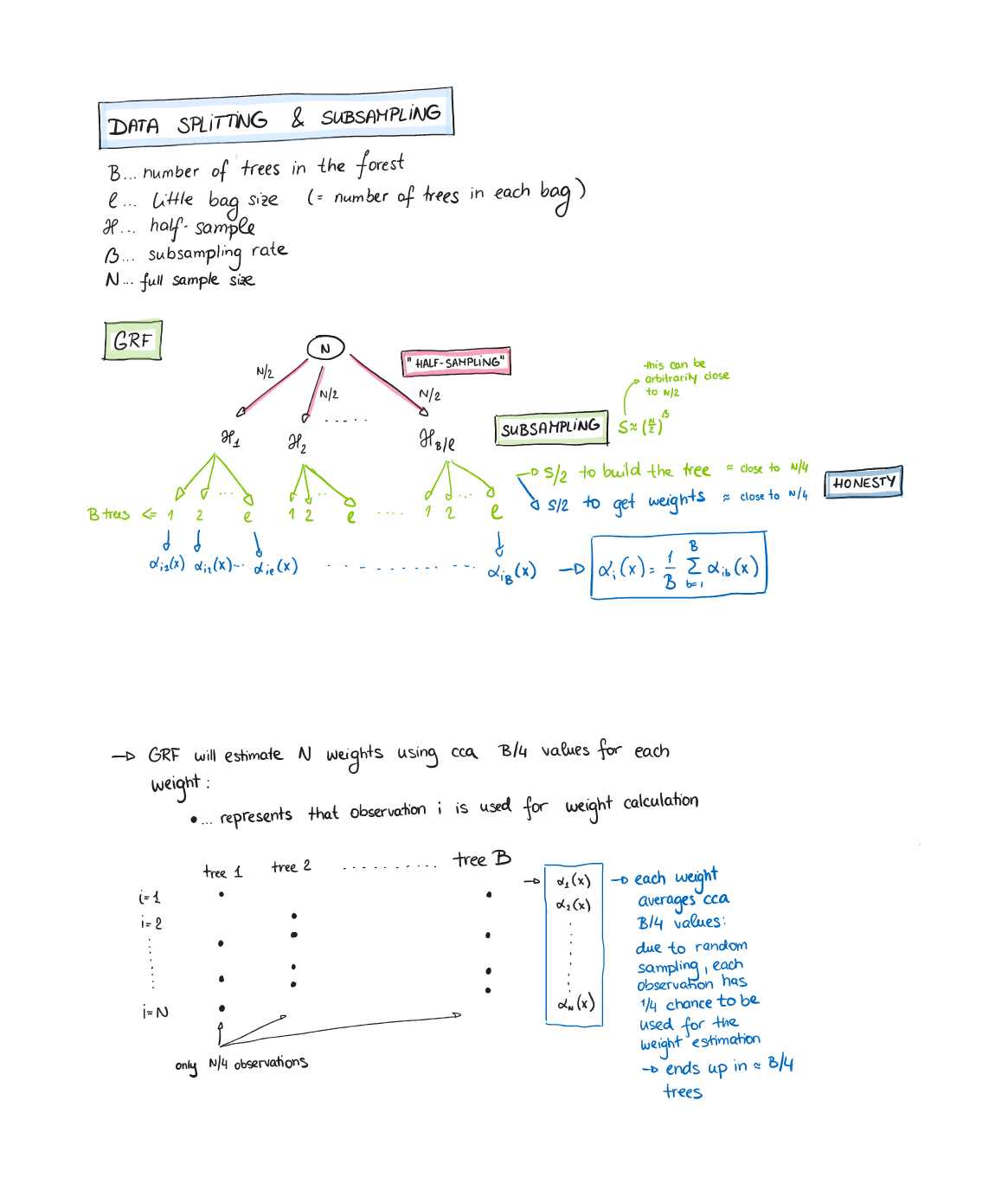}	
\label{fig:a1}
\end{figure}

\begin{figure}[H]
		\caption{Diagram of \emph{mcf}}
	\includegraphics[width=\textwidth]{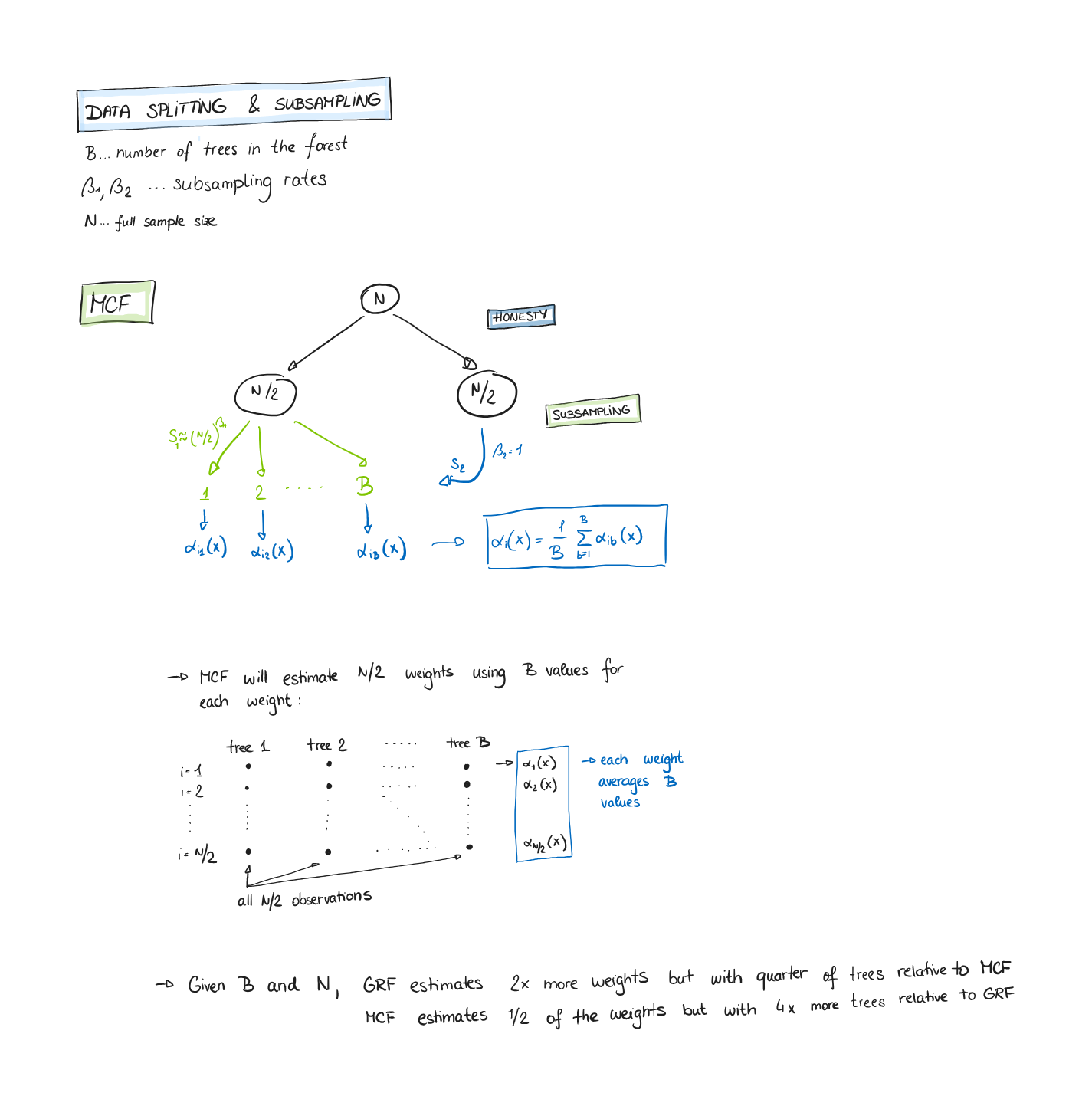}	
	\label{fig:a2}
\end{figure}

Figures \ref{fig:a1} and \ref{fig:a2} capture graphically the main differences between the
\emph{mcf} and \emph{grf} procedures regarding the one-sample and
two-sample honesty and estimation of the weights.

\subsection{Results for \emph{grf} estimation of ATE}\label{appendix-a4}

The ATE estimator for treatment pair $m$ and 0 implemented in
\emph{grf} is based on the following score function:
$$\psi
_{m,0}^{grf}(O;{\theta
_{m,0}},\eta _{ATE(m,0)}^{grf}(X)) =
\Gamma _{m,0}^{grf}(O;\eta
_{ATE(m,0)}^{grf}(X)) - {\theta
_{m,0}},$$
\begin{align*}
	\Gamma _{m,0}^{grf}(O;\eta
	_{ATE(m,0)}^{grf}(X)) & = \theta_{m,0}(X) + \frac{\underline{1}(D=m)(Y - \mu(X) + \sum\limits_{d>0}p_d(X)\theta_{d,0}(X)-\theta_{m,0}(X))}{p_m(X)} \\
	& - \frac{\underline{1}(D=0)(Y-\mu(X)+\sum\limits_{d>0}p_d(X)\theta_{d,0}(X))}{1-\sum\limits_{d>0}p_d(X)},
\end{align*}
where $\mu(x)=E[Y|X=x]$ and
$\eta^{grf}_{ATE(m,0)}(X)=(\theta_{1,0}(X),\ldots, \theta_{M-1,0}(X), \mu(X), p_{1}(X),\ldots, \allowbreak p_{M-1}(X)).$ Since
\begin{align*}{\mu
_d}(x) & = E\left[ {Y\left| {X
= x} \right.,D = d}
\right],\\ \mu
(x) & = \sum\limits_{d = 0}^{M - 1}
{{p_d}(x){\mu _d}(x)}
,\\{\theta _{d,0}}(x)& =
{\mu _d}(x) - {\mu
_0}(x),\\{\mu _0}(x)& =
\mu (x) - \sum\limits_{d
> 0} {{p_d}(x){\theta _{d,0}}(x)}
,\\{\mu _m}(x) &=
\mu (x) - \sum\limits_{d
> 0} {{p_d}(x){\theta _{d,0}}(x) +
{\theta _{m,0}}(x)}
,\end{align*}
it directly follows that $\psi
_{m,0}^{grf}(O;{\theta
_{m,0}},\eta _{ATE(m,0)}^{grf}(X)) =
\psi _{m,0}^{dml}(O;{\theta
_{m,0}},\eta _{m,0}^{dml}(X))$
which is the DR efficient score of \citet{robins1995semiparametric} for the
ATE. The identification result $E\left[
{\psi _{m,0}^{grf}(O;\theta
_{m,0}^0,\eta _{ATE(m,0)}^{grf,0}(X))}
\right] = 0$ follows directly as well.
Since the \emph{dml} score is asymptotically efficient, the \emph{grf}
score can be expected to be asymptotically efficient when all nuisance
parameters are consistently estimated.

The following proofs investigate consistent estimation of the ATE and
its robustness to misspecification of nuisance parameters. Assume that
$\hat \eta
_{ATE(m,0)}^{grf}$ converges in probability to
some values ${\bar \eta
_{ATE(m,0)}}$. If ${\bar
\eta _{ATE(m,0)}} = \eta
_{ATE(m,0)}^0$, all nuisance parameters are
consistently estimated. Since the ATE estimator is a sample average,
\begin{align*}
	\widehat{ATE^{grf}}(m,0) &= \frac{1}{N} \sum_{i=1}^{N} \Gamma^{grf}_{m,0} (o_i; \widehat{\eta}^{grf}_{ATE(m,0)} (x_i)) \\
	&= \frac{1}{N} \sum_{i=1}^{N} \widehat{\theta}_{m,0} (x_i) + \frac{\underline{1}(d_i = m)}{\widehat{p}_m (x_i)} \left( y_i - \widehat{\mu} (x_i) + \sum_{d>0} \widehat{p}_d (x_i) \widehat{\theta}_{d,0} (x_i) - \widehat{\theta}_{m,0} (x_i) \right) \\
	&\quad - \frac{\underline{1}(d_i = 0)}{1 - \sum\limits_{d>0} \widehat{p}_d (X)} \left( y_i - \widehat{\mu} (x_i) + \sum_{d>0} \widehat{p}_d (x_i) \widehat{\theta}_{d,0} (x_i) \right),
\end{align*}
it converges in probability to
\begin{multline*}
	E \left[ \bar{\theta}_{m,0}(X) + \frac{\underline{1}(D = m)}{\bar{p}_m(X)}
	\left( Y - \bar{\mu}(X) + \sum_{d>0} \bar{p}_d(X) \bar{\theta}_{d,0}(X) - \bar{\theta}_{m,0}(X) \right) \right. \\
	\left. - \frac{\underline{1}(D = 0)}{1 - \sum\limits_{d>0} \bar{p}_d(X)}
	\left( Y - \bar{\mu}(X) + \sum_{d>0} \bar{p}_d(X) \bar{\theta}_{d,0}(X) \right) \right].
\end{multline*}
If the propensity scores are consistently estimated, i.e.,
${\bar p_d}(X) =
p_d^0(X)$ for all $d>0$, then by using
the Law of Iterated Expectations and the identification assumptions, the
probability limit of the ATE estimator becomes

\bigskip

\resizebox{0.98\textwidth}{!}{$
	\begin{aligned}
 & E\left[
{{{\bar \theta }_{m,0}}(X) +
\mu _m^0(X) - \bar \mu (X)
+ \sum\limits_{d > 0}
{p_d^0(X){{\bar \theta
}_{d,0}}(X)} - {{\bar \theta
}_{m,0}}(X) - \mu _0^0(X) + \bar
\mu (X) - \sum\limits_{d
> 0} {p_d^0(X){{\bar
\theta }_{d,0}}(X)} }
\right]\\ & =
E\left[ {\mu _m^0(X) -
\mu _0^0(X)} \right] =
ATE_{m,0}^0.\end{aligned}$}

\bigskip

\noindent This implies that the ATE estimator remains consistent when propensity
scores are estimated consistently, even if the response function
$\mu(x)$ and the treatment effects
$\theta_{1,0}(X),  \allowbreak \ldots, \theta_{M-1,0}(X)$ are not
consistently estimated. The presence of the propensity score in the
estimation formula for the potential outcomes yields an inconsistent
estimator of the ATE when the propensity scores are not consistently
estimated, even if $\mu(x)$ and $\theta_{1,0}(X),
	\ldots, \theta_{M-1,0}(X)$ are consistently estimated.
Therefore, consistent estimation of the ATE based on the \emph{grf}
score is not robust to misspecification in propensity scores.

Proving Neyman-orthogonality of $\psi
_{m,0}^{grf}(O;{\theta
_{m,0}},\eta
_{ATE(m,0)}^{grf}(X))$, it needs to be shown
that at $r = 0$:
$$\frac{{\partial
E\left[ {\psi
_{m,0}^{grf}\left(O;\theta
_{m,0}^0,\eta _{ATE(m,0)}^{grf,0}(X) +
r\left(\eta _{ATE(m,0)}^{grf}(X) - \eta
_{ATE(m,0)}^{grf,0}(X)\right)\right)}
\right]}}{{\partial r}} =
0.$$
For better readability, denote $\delta : =
\eta _{ATE(m,0)}^{grf}(X) - \eta
_{ATE(m,0)}^{grf,0}(X)$ as the vector containing
the deviations of all nuisance parameters from their true values. The
individual elements, denoted as ${\delta
_{\theta (d,0)}}$,
${\delta _\mu
}$, and ${\delta
_{p(d)}}$, capture the deviations of the treatment
effects, response function $\mu(x)$, and propensity scores,
respectively, for all treatments $d > 0$. The
\emph{grf} score evaluated at $(O;\theta
_{m,0}^0,\eta _{ATE(m,0)}^{grf,0}(X) +
r\delta )$ takes the form:
\small
\begin{align*}
	&E \Bigg[ \psi_{m,0}^{grf} (O; \theta_{m,0}^{0}, \eta_{ATE(m,0)}^{grf,0}(X) + r\delta) \Bigg] =
	E \Bigg[ \theta_{m,0}^{0}(X) + r\delta_{\theta(m,0)} + \\
	&\frac{\underline{1}(D = m) \left( Y - \mu^{0}(X) - r\delta_{\mu} + \sum\limits_{d>0} (p_d^{0}(X) + r\delta_{p(d)}) (\theta_{d,0}^{0}(X) + r\delta_{(d,0)}) - \theta_{m,0}^{0}(X) - r\delta_{\theta(m,0)} \right)}
	{p_m^{0}(X) + r\delta_{p(m)}} \\
	&\quad - \frac{\underline{1}(D = 0) \left( Y - \mu^{0}(X) - r\delta_{\mu} + \sum\limits_{d>0} (p_d^{0}(X) + r\delta_{p(d)}) (\theta_{d,0}^{0}(X) + r\delta_{\theta(d,0)}) \right)}
	{1 - \sum\limits_{d>0} \left(p_d^{0}(X) + r\delta_{p(d)}\right)} - \theta_{m,0}^{0} \Bigg].
\end{align*}
\normalsize
Noting that
\begin{align*}{\delta
_{p(0)}} &= - \sum\limits_{d
> 0} {{\delta _{p(d)}}}
,\\{\delta
_{\mu (0)}}& = {\delta
_\mu } - \sum\limits_{d
> 0} {p_d^0(X){\delta
_{\theta (d,0)}}} -
\sum\limits_{d > 0}
{{\delta _{p(d)}}\theta
_{d,0}^0(X)} - r\sum\limits_{d
> 0} {{\delta
_{p(d)}}{\delta _{\theta (d,0)}}}
,\\{\delta
_{\mu (m)}}&= {\delta
_\mu } - \sum\limits_{d
> 0} {p_d^0(X){\delta
_{\theta (d,0)}}} -
\sum\limits_{d > 0}
{{\delta _{p(d)}}\theta
_{d,0}^0(X)} - r\sum\limits_{d
> 0} {{\delta
_{p(d)}}{\delta _{\theta (d,0)}}}
+ {\delta _{\theta
(m,0)}},\end{align*}
differentiating under the expectation sign at $r = 0$ leads
to:\footnote{The identifying assumptions and regularity conditions, such as Lipschitz continuity, guarantee that sufficient
	conditions for interchanging the expectation and derivation signs are
	fulfilled.}
\begin{align*}
	&\frac{\partial E \left[ \psi_{m,0}^{grf} (O; \theta_{m,0}^{0}, \eta_{ATE(m,0)}^{grf,0}(X) + r\delta) \right]}{\partial r}
	= E \Bigg[ \delta_{\theta(m,0)} \\
	& + \frac{\underline{1}(D = m) \delta_{\mu(m)} p_m^{0}(X) - \underline{1}(D = m) (Y - \mu_m^{0}(X)) \delta_{p(m)}}{\left(p_m^{0}(X)\right)^2} \\ &- \frac{\underline{1}(D = 0) \delta_{\mu(0)} p_0^{0}(X) - \underline{1}(D = 0) (Y - \mu_0^{0}(X)) \delta_{p(0)}}{\left(p_0^{0}(X)\right)^2} \Bigg].
\end{align*}
Using the law of iterated expectations and the identifying assumptions
yields at $r = 0$:
$$\frac{{\partial
E\left[ {\psi
_{m,0}^{grf}(O;\theta
_{m,0}^0,\eta _{ATE(m,0)}^{grf,0}(X) +
r\delta )}
\right]}}{{\partial r}} =
E\left[ {{\delta
_{\theta (m,0)}} + {\delta
_{\mu (m)}} - {\delta
_{\mu (0)}}} \right] =
0,$$
proving Neyman-orthogonality.

\section{Additional details of the Monte Carlo study}\label{appendix-b}

\ref{appendix-b1} explains the general simulation protocol used, while \ref{appendix-b2}
explains the data generating processes in detail. \ref{appendix-b3} gives
the details of the implementation of the various versions of the
different estimators used.

\subsection{Simulation protocol}\label{appendix-b1}

Table \ref{tab:b1} shows the protocol employed in the Monte Carlo study.

\begin{table}[h]
	\centering
	\caption{Protocol of the Monte Carlo Study}
	\sffamily
	\begin{tabularx}{\textwidth}{@{}p{1cm}X@{}}
		\toprule
        1 & \hangindent3em \hangafter1 Specify the data generating process with respect to (i) sample size, (ii) strength of selectivity into treatments, (iii) type of covariates, (iv) influence of covariates on non-treatment potential outcomes (including the degree of sparsity), (v) size of treatment effect and its heterogeneity, (vi) treatment share, and (vii) number of treatments. \\ 
		2 & Draw training data of size $N$. \\ 
		3 & Draw prediction data of (same) size $N$. \\ 
		4 & \hangindent3em \hangafter1 Compute the true values of ATE and GATE on a large sample (1,000,000 observations) and get the true IATE from the DGP. \\ 
		5 & Train the different estimators on the training data. \\ 
		6 & Predict ATE, GATEs, and IATEs on the prediction data. \\ 
		7 & Repeat steps 2 to 6 $R$ times ($R = 1'000 \times (2'500 / N)$). \\ 
		8 & Compute the performance measures. \\ 
		9 & Repeat steps 1 to 8 for different specifications. \\ 
		\bottomrule
	\end{tabularx}
	\vspace{-8pt}
	\caption*{\footnotesize \textsf{Note: The number of replications $(R)$ declines such that the simulation noise remains approximately constant if the estimator is $\sqrt{N}-$convergent.}}
	\label{tab:b1}
\end{table}

\subsection{Data generating	processes}\label{appendix-b2}

The description of the data generating process covers the covariates,
the selection (treatment) process, the outcome process, and the effects
and their heterogeneity.

\subsubsection{Covariates}\label{appendix-b21}
The covariates are independent from each other and either normally or
uniformly distributed,\footnote{The reason for the uniform distribution
	is that the results for the \emph{grf} and the \emph{mcf} assume
	uniformly distributed covariates.} both with mean zero and variance
one.
\begin{align*}
& x_i^U \sim uniform(-\sqrt{12}/2,\sqrt{12}/2), & \dim (x_i^U) = {p^U} \\ 
& x_i^N \sim N(0,1), & \dim (x_i^N) = {p^N} \\ 
& p = {p^U} + {p^N},\quad \forall i = 1,...,N. & \end{align*}

We also consider a scenario where the first 5 variables of $X^N$
and $X^U$ are
generated as dummy variables, with ${X^D} =
2\cdot \underline 1 (X \textgreater 0) -
1$, also with mean zero.

In the simulations, we consider values of the following triples
($p^{N}$, $p^{U}$,
$p^{D}$): (10, 10, 0), (5, 5, 0), (25, 25, 0), (20,
0, 0), (0, 20, 0), and (5, 5, 10). Thus, we capture cases of 10 to 50
covariates that may be of different types. (10, 10, 0) is the base
specification.

\subsubsection{Selection (treatment) process}\label{appendix-b22}

We consider cases of two and four treatments where all treatments have
equal shares. As an extension, in the binary treatment case there are
also treatment shares of 25\%. The treatment indicators are obtained
from the quantiles of the following index function:
\small
\begin{align*}
	\breve{d}_i &= \frac{\breve{\lambda} \, x_i' \beta}{\sqrt{\sum\limits_{j=1}^{p} \beta_j^2 / 1.25}} + u_i, 
	& \beta &= (1, 1 - 1/k, \dots, 1/k, 0, \dots, 0), 
	& \dim(\beta) &= p, 
	& \breve{\lambda} &\in \{0, 0.42, 1.25\} \\
	& & u_i & \sim N(0,1), & & & \forall i & = 1, \dots, N.
\end{align*}
\normalsize

The impact of the covariates declines with their order. The parameter $\breve{\lambda}$
determines the strength of the selection process, from a randomized
experiment ($\breve{\lambda}$=0) to the case of strong selection ($\breve{\lambda}$=1.25 for uniform and
mixed covariates; in case of normally distributed covariates grid for $\breve{\lambda}$
is 0, 0.45, and 1.5). In the intermediate setting, a machine learning
estimation (using 1'000'000 observations) of the selection equation will
obtain an out-of-sample $R^{2}$ of about 10\%, while
in the extreme case this value rises to about 42\%. As it turns out that
the strength of selectivity is an important parameter when considering
the performance of estimators, the base scenario considers all three
selectivity levels. The baseline scenario in \ref{appendix-c2} is based on
the medium selectivity when estimating effects at all granularity
levels.

\subsubsection{IATEs}\label{appendix-b23}

The IATEs are deterministic functions of the covariates. All
specifications of the IATEs are based on the following linear index,
which is like the one used for the selection process:
\small
\begin{align*}
	\breve{\Delta}_i &= \frac{\zeta \, x_i' \beta}{\sqrt{\sum\limits_{j=1}^{p} \beta_j^2 / 1.25}}, 
	& \beta &= (1, 1 - 1/k, \dots, 1/k, 0, \dots, 0), 
	& \dim(\beta) &= p, 
	& \zeta &\in \{0,\zeta^{med}\} \\
	& & \forall i & = 1, \dots, N.
\end{align*}
\normalsize

The value of ${\zeta
^{med}}$ captures the strength of heterogeneity. Its
exact value depends on the type of non-linearity and the
$R^{2}$ of the potential non-treatment outcome
process to ensure that the implications of the different effect
strengths on the outcome remain stable.

The following variations of heterogeneity are considered for the IATE:
\begin{table*}[h]
	\centering
	\renewcommand{\arraystretch}{1.5} 
	\begin{tabularx}{0.9\textwidth}{@{}X X@{}}  
		IATE Formula & As a Function of \(X\beta\) \\ 
		\( IATE_i = \breve{\Delta}_i + 1 \) & Linear \\ 
		\( IATE_i = F^{\text{logistic}} (\breve{\Delta}_i) + 0.5 \) & Nonlinear \\ 
		\( IATE_i = \frac{\breve{\Delta}_i^2 - 1.25}{\sqrt{3}} + 1 \) & Quadratic \\ 
		\( IATE_i = f^{WA}(\tilde{x}_{1,i}) f^{WA}(\tilde{x}_{2,i}) - 2.8 \) & Step function \citep[like][]{wager2018estimation} \\ 
	\end{tabularx}
	\caption*{$F^{\text{logistic}}: \text{c.d.f. of logistic distribution}, \quad
			f^{WA}(x) = 1 + \frac{1}{1 + e^{-20(x - 1/3)}}, \quad \forall i = 1, \dots, N.$}
\end{table*}

They represent a linear function and three non-linear functions. Note
that the last function is very similar to the DGP used in \citetalias{wager2018estimation}. For the
latter, the heterogeneity depends only on two variables (which are the
ones that are most important for the selection and the outcome
processes). ${\tilde
x_{1,i}}$ and ${\tilde
x_{2,i}}$ are transformed versions of
$x_{1,i}$ and $x_{2,i}$, where the
exact transformation used depends on the distribution of each variable.
The baseline scenario in \ref{appendix-c2} is based on the step function
only.

The resulting ATE in these cases is always one, independent of the
heterogeneity specification. In addition to these cases, we also
consider a case of the IATEs all being equal to zero.

\subsubsection{Outcome processes}\label{appendix-b24}

The outcome process consists first of specifying a process for the
non-treatment potential outcome. The potential outcome with treatment is
obtained by adding the IATE plus noise to the non-treatment outcome.

The following non-linear outcome process is specified for the potential
non-treatment outcome:
\small
\begin{align*}
	\breve{y}_i &= \frac{x_i' \beta}{\sqrt{\sum\limits_{j=1}^{p} \beta_j^2 / 1.25}}, 
	& \beta &= (1, 1 - 1/k, \dots, 1/k, 0, \dots, 0), 
	& \dim(\beta) &= p, \\
	y_i^0 & = \delta \sin(\breve{y_i}) + \varepsilon_i^0, & \delta & \in \{0, \delta^{med}, \delta^{strong}\}, \quad \varepsilon_i^0 \sim N(0,1) \\
	y_i^1 & = \delta \sin(\breve{y_i}) + IATE_i + \varepsilon_i^1, & \phantom{\delta} & \phantom{\in \{0, \delta^{med}, \delta^{strong}\},i} \quad \varepsilon_i^1 \sim N(0,1)	& \forall i & = 1, \dots, N. \\
\end{align*}
\normalsize
${\delta ^{med}}$ and
${\delta ^{strong}}$
are chosen such that the $R^{2}$ in the outcome
process of the potential non-treatment outcome is about 10\% (base
specification) and 45\%.

\begin{figure}[ht]
	\caption{Shape of potential outcomes for different shapes of IATEs ($\delta^{med}$)}
	\includegraphics[width=\textwidth]{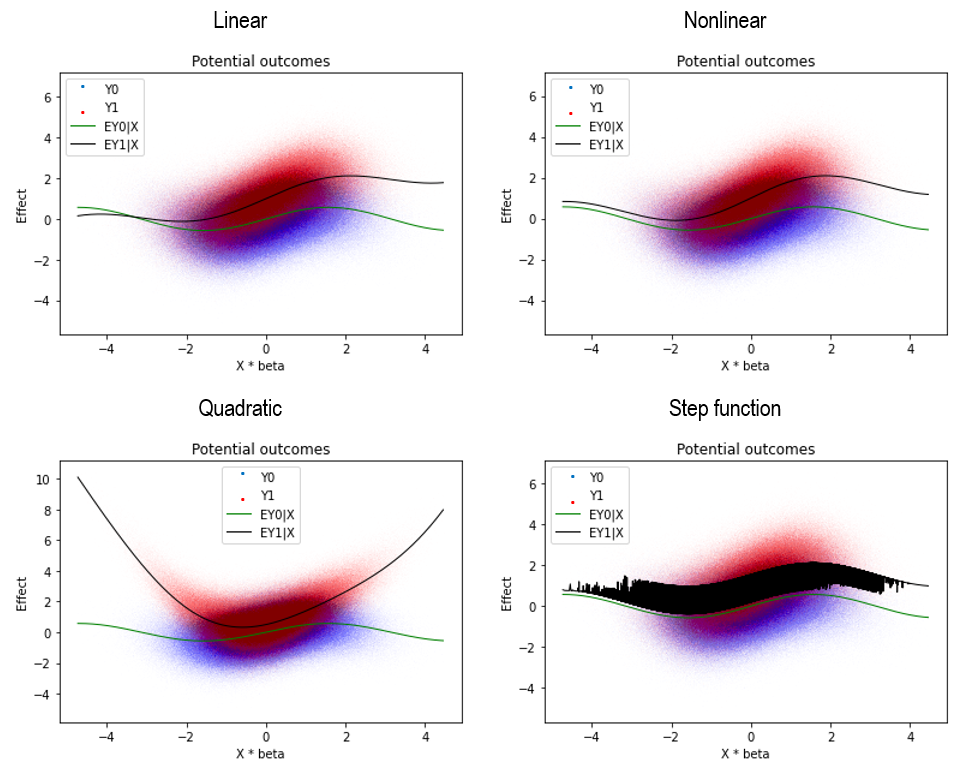}
	\label{fig:b1}
	\caption*{\footnotesize \textsf{Note: Figures are based on 1'000'000 observations.}}
\end{figure}

\begin{figure}[ht]
	\caption{Shape of potential outcomes for different shapes of IATEs ($\delta^{strong}$)}
	\includegraphics[width=\textwidth]{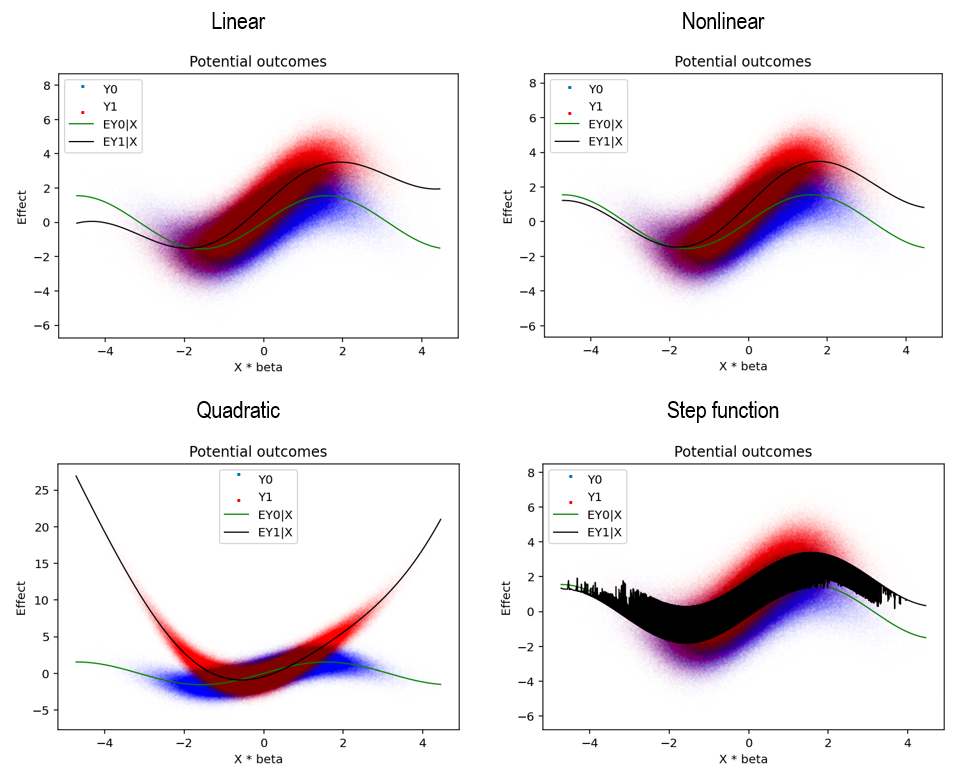}
	\label{fig:b2}
	\caption*{\footnotesize \textsf{Note: Figures are based on 1'000'000 observations.}}
\end{figure}

Figure \ref{fig:b1} shows the relation of the potential outcomes and their
expectations with the index ${x_i'}\beta
$ for the case of
$p^{N}=p^{U}=10$, $p=20$,
$k^{N}=k^{U}=5$, $k=10$,
and ${\delta ^{med}}$
with respect to the different heterogeneities. Figure \ref{fig:b2} shows the same
relationship for the stronger effect size
${\delta ^{strong}}$.

Figures \ref{fig:b1} show that the differences between the linear and the
non-linear case are rather small, at least compared to the quadratic
IATEs and those based on the step function. In particular the latter,
for which the IATEs depend only on the first two most important
covariates, show behaviour that is not being well approximated by any
parsimonious parametric function of the linear index. Thus, we
conjecture that this type of effect heterogeneity will be most difficult
to estimate well.

Increasing the predictiveness of the covariates for the non-treatment
potential outcome (Figure \ref{fig:b2}), denoted by the blue dots, and its
conditional expected values, denoted by the green line, leads to
substantially more pronounced non-linearity with respect to the linear
index. This is true for all specifications of the IATEs, but
particularly so for the quadratic one.

\subsubsection{Overview of simulations and outcome tables}\label{appendix-b25}

Table \ref{tab:b2} gives an overview where to find the simulations results for
the different data generating processes in \ref{appendix-c}.

\begin{table}[h!]
	\centering
	\caption{Overview of specifications and their locations}
	\sffamily
	\begin{tabularx}{\textwidth}{@{}l c c c c c c c c c c@{}}
		\toprule
		\textbf{Table} & \textbf{Hetero-} & \textbf{Selec-} & $\boldsymbol{k}$ & $\boldsymbol{p}$ & $\boldsymbol{R^2}$ & $\boldsymbol{X}$ & $\boldsymbol{N}$ & \textbf{No of} & \textbf{Treat-} & \textbf{GATE} \\
		& \textbf{geneity} & \textbf{tivity} & & & $\boldsymbol{(y^0)}$ & & & \textbf{treat-} & \textbf{ment} & \textbf{groups} \\
		& & & & & & & & \textbf{ments} & \textbf{share} & \\
		\midrule
		C.1  & \textbf{none}   & none    & 10  & 20  & 10\%  & U, N  & stand  & 2  & 50\%  & 5  \\
		C.2  & \textbf{none}   & middle  & 10  & 20  & 10\%  & U, N  & stand  & 2  & 50\%  & 5  \\
		C.3  & \textbf{none}   & strong  & 10  & 20  & 10\%  & U, N  & stand  & 2  & 50\%  & 5  \\
		C.4  & \textbf{linear} & none    & 10  & 20  & 10\%  & U, N  & stand  & 2  & 50\%  & 5  \\
		C.5  & \textbf{linear} & middle  & 10  & 20  & 10\%  & U, N  & stand  & 2  & 50\%  & 5  \\
		C.6  & \textbf{linear} & strong  & 10  & 20  & 10\%  & U, N  & stand  & 2  & 50\%  & 5  \\
		C.7  & \textbf{nonlin} & none    & 10  & 20  & 10\%  & U, N  & stand  & 2  & 50\%  & 5  \\
		C.8  & \textbf{nonlin} & middle  & 10  & 20  & 10\%  & U, N  & stand  & 2  & 50\%  & 5  \\
		C.9  & \textbf{nonlin} & strong  & 10  & 20  & 10\%  & U, N  & stand  & 2  & 50\%  & 5  \\
		C.10 & \textbf{quadrat} & none   & 10  & 20  & 10\%  & U, N  & stand  & 2  & 50\%  & 5  \\
		C.11 & \textbf{quadrat} & middle & 10  & 20  & 10\%  & U, N  & stand  & 2  & 50\%  & 5  \\
		C.12 & \textbf{quadrat} & strong & 10  & 20  & 10\%  & U, N  & stand  & 2  & 50\%  & 5  \\
		C.13 & step   & none    & 10  & 20  & 10\%  & U, N  & stand  & 2  & 50\%  & 5  \\
		C.14 & step   & middle  & 10  & 20  & 10\%  & U, N  & stand  & 2  & 50\%  & 5  \\
		C.15 & step   & strong  & 10  & 20  & 10\%  & U, N  & stand  & 2  & 50\%  & 5  \\
		C.16 & step   & middle  & \textbf{5}   & \textbf{10}  & 10\%  & U, N  & stand  & 2  & 50\%  & 5  \\
		C.17 & step   & middle  & \textbf{25}  & \textbf{50}  & 10\%  & U, N  & stand  & 2  & 50\%  & 5  \\
		C.18 & step   & middle  & \textbf{4}   & 20  & 10\%  & U, N  & stand  & 2  & 50\%  & 5  \\
		C.19 & step   & middle  & \textbf{16}  & 20  & 10\%  & U, N  & stand  & 2  & 50\%  & 5  \\
		C.20 & step   & middle  & 10  & 20  & \textbf{0\%}   & U, N  & stand  & 2  & 50\%  & 5  \\
		C.21 & step   & middle  & 10  & 20  & \textbf{45\%}  & U, N  & stand  & 2  & 50\%  & 5  \\
		C.22 & step   & middle  & 10  & 20  & 10\%  & \textbf{U}     & stand  & 2  & 50\%  & 5  \\
		C.23 & step   & middle  & 10  & 20  & 10\%  & \textbf{N}     & stand  & 2  & 50\%  & 5  \\
		C.24 & step   & middle  & 10  & 20  & 10\%  &\textbf{U,N,D}  & stand  & 2  & 50\%  & 5  \\
		C.25 & step   & middle  & 10  & 20  & 10\%  & U, N  & stand  & \textbf{4}  & equal  & 5  \\
		C.26 & step   & middle  & 10  & 20  & 10\%  & U, N  & \textbf{40'000}  & 2  & 50\%  & 5  \\
		C.27 & step   & middle  & 10  & 20  & 10\%  & U, N  & stand  & 2  & \textbf{25\%}  & 5  \\
		C.28 & step   & none    & 10  & 20  & 10\%  & U, N  & stand  & 2  & 50\%  & \textbf{5-40} \\
		C.29 & step   & middle  & 10  & 20  & 10\%  & U, N  & stand  & 2  & 50\%  & \textbf{5-40}  \\
		C.30 & step   & strong  & 10  & 20  & 10\%  & U, N  & stand  & 2  & 50\%  & \textbf{5-40}  \\
		\bottomrule
	\end{tabularx}
	\vspace{-8pt}
	\caption*{\footnotesize \textsf{\textbf{Note:} Deviations from the base benchmark specifications are in bold font. $N$: \emph{stand} implies that $N = 2'500$ and $N = 10'000$ are in the same table. $X$: \emph{U, N, D} denotes uniformly and normally distributed covariates, as well as dummy variables.}}
	\label{tab:b2}
\end{table}

\subsection{Implementation of the estimators}\label{appendix-b3}
To avoid additional computational costs in an already computationally
very expensive simulation study, none of the estimators has been
explicitly tuned. Either default values are used for the tuning
parameters, or tuning parameters have been set to a fixed value
beforehand. In an empirical application, the quality of estimation could
be improved by hyperparameter tuning. For more details on hyperparameter
tuning for \emph{dml}, see \citet{bach2024hyperparameter}. Forest estimators often minimize out-of-bag error to
find optimal hyperparameters. For details on the tuning procedures,
please refer to the documentation of \emph{grf} and \emph{mcf}.

\subsubsection{MCF}\label{appendix-b31}
The results of the \emph{mcf} are computed with the Python versions of
the package which are available on PyPI. Most of the simulations have
been performed with version 0.3.3., but some of the simulations used
also the newer and faster versions up to 0.4.3. Although some default
values changed a bit between the versions, results were very similar. In
the following, sample A denotes one half of the data set used to build
the forest (``split-construction'' data) and sample B denotes the other
half of the sample used to estimate the effects (``leaf-populating''
data).

Contrary to the default values, the minimum leaf size in each tree
equals 5. The number of trees contained in any forest equals 1000. Trees
are formed on random subsamples drawn without replacement (subsampling)
with a sample size of 50\% of the size of sample A. Regarding estimation
of MCE and finding close neighbours, closeness is based on a Random
Forest based prognostic score \citep[see][]{hansen2008prognostic},
$\left[\hat \mu_0(x),....,\hat \mu_{M - 1}(x)\right]$ weighted by the Mahalanobis
distance, as forming the neighbours by simplified Mahalanobis matching
suffers in large-dimensional problems.

To construct random-split trees, the number of variables tested for a
split is determined through the following process: Initially, a random
number is drawn from a Poisson distribution with parameter
($0.35*p-1$). To this result, 1 is added, ensuring that at least 1
variable is considered for a split. If the final result exceeds
$p$, all variables are tested for a split. Otherwise, the specified
number of variables is randomly selected from the pool of all $p$
variables.

Two variants of this estimator are used, with and without local
centring. Local centring is implemented in the \emph{mcf} package by
running a regression random forest (using the Python package
scikit-learn) to predict $E[Y| X]$ ($X$ does not
include the treatment). These (out-of-training-sample) predictions are
obtained in a 5-fold cross-fitting scheme. The predictions are
subtracted from the observed $Y$ to obtain a centred outcome,
$Y^{cent}$. $Y^{cent}$ (instead
of $Y$) is subsequently used as outcome in the \emph{mcf}
algorithm. The simulations show that this centred version outperforms
the uncentred version when there is non-random selection. Recentring is
implemented in the following way:

\begin{enumerate}
	\def\labelenumi{\arabic{enumi})}
	\item
	Estimate the trees that define the Random Forest for
	$E[Y| X = x]$ in sample A.
	\item
	Recentring of outcomes in sample A: Split sample A randomly into
	$K$ equally sized parts, $A\text{--}1$ to $A\text{--}K$. Use the
	outcomes in the union of the $K-1$ folds $A\text{--}1$ to
	$A\text{--}(K-1)$ to obtain the Random Forest predictions given the
	forest estimated in step 1). Use these predictions to predict
	$E[Y| X = x]$ in fold $A\text{--}K$.
	Do this for all-possible combinations of folds (cross-fitting as used
	in $k$-fold cross-fitting). Subtract the predictions from the
	actual values of $Y$.
	\item Redo step 2 in sample B using the estimated forests of sample A.
\end{enumerate}

Concerning the specifics of the local centring algorithm, there are a
couple of points worth mentioning. First, to avoid overfitting, the
outcomes of observation `$i$' are not used to predict itself.
Therefore, the chosen implementation is based on cross-fitting.

Second, weights-based inference requires avoiding a dependence of the
weights in sample B on outcomes of sample A. However, since recentring
uses outcome variables independent of the treatment state, this could
induce a correlation between the recentred outcomes in different
treatment states. This finite sample correlation will be ignored here
\citep[as in][]{athey2019grf}.

Third, the number of folds is a tuning parameter that influences the
noise added to the recentred outcome by subtracting an estimated
quantity. The simulation results indicate that the computationally most
attractive choice of $K=2$ may be too small in medium sized samples
and that a somewhat larger number of folds may be needed to avoid much
additional noise to the estimators.

The nonparametric regressions that enter the estimation of the standard
errors are based on $k$-NN estimation with number of neighbours
equal to $\mathit{2}\ sqrt(N)$.

If inference is not of interest, a more efficient \emph{mcf} estimator
(denoted as IATE eff in \ref{appendix-c}) can be obtained by switching the
roles of the samples used for building the forest and populating the
leaves with outcome values, and subsequently averaging the two
estimates.

\subsubsection{GRF}\label{appendix-b32}

The reported results are based on the R package \emph{grf} (version
2.3.1). The default forest parameters are 2000 trees (twice more than in
\emph{mcf)}, minimum node size is 5 and number of variables tested for a
split is a random draw from a Poisson distribution with parameter
\textit{min(p, round\_up(sqrt(p+20))).}

The default implementation of the \emph{grf} uses out-of-bag predictions
of nuisance parameters to remove confounding bias in order to find the
best splits on which the IATEs are estimated. As this implementation is
not in line with the theory in \citet{athey2019grf}, a centred version of
the algorithm, using $K$-fold cross-fitting to calculate the
residualized outcome that is later fed as an input to the \emph{grf}, is
added to the simulation.

For later estimation of GATEs and ATEs, so-called DR scores need to be
constructed. GATE and ATE can be estimated directly via functions
\emph{average\_treatment\_effect()} and
\emph{best\_linear\_predictor()}. Both functions need the estimated
forest as an input to estimate the effects on the data used for building
the forest and estimating the forest weights. Here, the simulation
deviates from the protocol, as currently \emph{grf} does not provide an
option to estimate the DR scores on a new (prediction) data set. The
package provides function \emph{get\_scores()} which computes the
estimated component of a DR score on the data set that was used for the
forest if needed. The current implementation of GATE estimation omits
the \emph{best\_linear\_predictor()} and directly regresses the scores from
\emph{get\_scores()} on group dummies exploiting the homoscedastic error
terms for estimation of standard errors.

\subsubsection{DML}\label{appendix-b33}

The key element of all \emph{dml} estimators is the DR component of the
\emph{dml} score. For treatments $m$ and $l$, this DR
component is defined as:
\small
\begin{gather*}
\Gamma^{dml}_{m,l}(X,Y,D;\eta^{dml}_{m,l}(X)) = \mu_m(X) - \mu_l(X) 
+ \frac{\underline{1}(D=m)(Y - \mu_m(X))}{p_m(X)}
- \frac{\underline{1}(D=l)(Y - \mu_l(X))}{p_l(X)}, \\
\mu_d(x) = E\left[Y \left| {D =d,X = x} \right.\right], \quad p_d(x) = P(D =d\left| {X = x}\right.).
\end{gather*}
\normalsize

The nuisance parameters, $\eta
_{m,l}^{dml}(X) = ({\mu
_m}(X),{\mu
_l}(X),{p_m}(X),{p_l}(X))$, are estimated with
regression forests. As usual, the necessary cross-fitting is implemented
via 5-fold cross-fitting. To obtain effects, the estimated DR components
of the score for the desired treatment contrasts are formed. The ATE is
obtained by averaging these differences.

GATEs and IATEs are obtained by regression-type approaches in which the
estimated DR component of the score serves as dependent variable. The
GATEs are computed as OLS-coefficients of a (saturated) regression model
with the indicators for the GATEs groups acting as independent
variables. IATEs are estimated by using $X$ as independent
variables either in a regression random forest or in an OLS estimation.
When OLS regressions are used, inference is based on
heteroscedasticity-robust covariance matrix of the corresponding
coefficients. No inference is obtained for the random forest based
IATEs.

As the weights may lead to small sample issues, in particular when
selection probabilities get close to zero, two versions of the
\emph{dml} estimator are considered. The first one is taking the weights
as they are, while the second version normalises the sum of the weights
to one ($N$)\footnote{Similarly, as in the normalized DR learner in
	\citet{knaus2022double}.} and truncates `too large' weights. Too large here means
that a single weight is larger than 5\% of the sum of all weights (if
so, it is truncated at 5\%). Since the normalised versions appears to
outperform the non-normalized one in many simulations, the latter is
presented in the main body of the text.

\subsubsection{OLS}\label{appendix-b34}

As a benchmark estimator, a two-sample (for binary treatments) OLS
estimator is implemented in a standard way. However, since there are
substantial non-linearities in the DGP, it is not surprising that there
are many scenarios in which this estimator performs badly. Therefore,
OLS results are not reported in the main part of the paper.

\section{Detailed results of the Monte Carlo study}\label{appendix-c}

\subsection{Base specifications}\label{appendix-c1}
In this section, we vary the selectivity and the type of the IATE
jointly and keep the other parameters fixed at their base values (i.e.,
$k=10$, $p=20$,
$R^{2}(y^{0}) = 10\%$,
$X^{N}$ and $X^{U}$, binary
treatment, treatment share 50\%, $N=2'500$ and $N =10'000$).

\begin{table}[h!]
	\sffamily
	\small
	\caption{No IATE, no selectivity}
	\resizebox{\textwidth}{!}{
\begin{tabular}{lccccccccccc}
		\toprule
		\multicolumn{3}{c}{} & \multicolumn{6}{c}{\textbf{Estimation of effects}} & \multicolumn{3}{c}{\textbf{Estimation of std.}} \\
		\multicolumn{3}{c}{} & \multicolumn{6}{c}{} & \multicolumn{3}{c}{\textbf{errors}} \\
		\midrule
		& \textit{\# of} & Estim- & Bias & Mean & Std. & RMSE& Skew- & Ex. & Bias & CovP & CovP \\
		& \textit{groups} & ator & & abs. & dev. & & ness & Kurto- & (SE) & (95) & (80) \\
		& & & & error& & & & sis & & in \% & in \% \\
		\midrule
		& (1) & (2) & (3) & (4) & (5) & (6) & (7) & (8) & (9) & (10) & (11) \\
		\midrule
		& \multicolumn{11}{c}{N = 2'500} \\
		\midrule
		\textbf{ATE}      & \textit{1} & \textit{mcf}  & -0.003 & 0.048 & 0.060 & 0.060 & 0.01  & -0.05 & 0.000  & 96 & 79 \\
\bottomrule
	\end{tabular}
	}
	\vspace{-8pt}
	\captionsetup{justification=justified,singlelinecheck=false}
	\caption*{\footnotesize \textsf{\textbf{Note:} Table to be continued.}}
\end{table}

\captionsetup[ContinuedFloat]{labelformat=continued}
																					
\begin{table}[h!]
	\ContinuedFloat
	\sffamily
	\small
	\caption{No IATE, no selectivity}
	\resizebox{\textwidth}{!}{
		\begin{tabular}{lccccccccccc}
			\toprule
			\multicolumn{3}{c}{} & \multicolumn{6}{c}{\textbf{Estimation of effects}} & \multicolumn{3}{c}{\textbf{Estimation of std.}} \\
			\multicolumn{3}{c}{} & \multicolumn{6}{c}{} & \multicolumn{3}{c}{\textbf{errors}} \\
			\midrule
			& \textit{\# of} & Estim- & Bias & Mean & Std. & RMSE& Skew- & Ex. & Bias & CovP & CovP \\
			& \textit{groups} & ator & & abs. & dev. & & ness & Kurto- & (SE) & (95) & (80) \\
			& & & & error& & & & sis & & in \% & in \% \\
			\midrule
			& (1) & (2) & (3) & (4) & (5) & (6) & (7) & (8) & (9) & (10) & (11) \\
			\midrule
			& \multicolumn{11}{c}{N = 10'000} \\
			\midrule
			\textbf{ATE}      & \textit{1} & \textit{mcf}  & 0.001  & 0.022 & 0.028 & 0.028 & 0.12  & 0.03  & 0.003  & 96   & 82 \\
\bottomrule
		\end{tabular}
	}
	\vspace{-8pt}
	\captionsetup{justification=justified,singlelinecheck=false}
	\caption*{\footnotesize \textsf{\textbf{Note:} For GATE and IATE the results are averaged over all effects. \emph{CovP(95, 80)} denotes the (average) probability that the true value is part of the 95\%/80\% confidence interval. 1'000 / 500 replications used for 2'500 / 10'000 obs.}}
\end{table}

\begin{table}[h!]
	\sffamily
	\small
	\caption{No IATE, medium selectivity}
	\resizebox{\textwidth}{!}{
		\begin{tabular}{lccccccccccc}
			\toprule
			\multicolumn{3}{c}{} & \multicolumn{6}{c}{\textbf{Estimation of effects}} & \multicolumn{3}{c}{\textbf{Estimation of std.}} \\
			\multicolumn{3}{c}{} & \multicolumn{6}{c}{} & \multicolumn{3}{c}{\textbf{errors}} \\
			\midrule
			& \textit{\# of} & Estim- & Bias & Mean & Std. & RMSE& Skew- & Ex. & Bias & CovP & CovP \\
			& \textit{groups} & ator & & abs. & dev. & & ness & Kurto- & (SE) & (95) & (80) \\
			& & & & error& & & & sis & & in \% & in \% \\
			\midrule
			& (1) & (2) & (3) & (4) & (5) & (6) & (7) & (8) & (9) & (10) & (11) \\
			\midrule
			& \multicolumn{11}{c}{N = 2'500} \\
			\midrule
			\textbf{ATE}      & \textit{1} & \textit{mcf}  & 0.160 & 0.160 & 0.061 & 0.171 & 0.12  & 0.10  & 0.000  & 25 & 9  \\
\bottomrule
		\end{tabular}
	}
	\vspace{-8pt}
	\captionsetup{justification=justified,singlelinecheck=false}
	\caption*{\footnotesize \textsf{\textbf{Note:} Table to be continued.}}
\end{table}

\begin{table}[h!]
	\ContinuedFloat
	\sffamily
	\small
	\caption{No IATE, medium selectivity}
	\resizebox{\textwidth}{!}{
		\begin{tabular}{lccccccccccc}
			\toprule
			\multicolumn{3}{c}{} & \multicolumn{6}{c}{\textbf{Estimation of effects}} & \multicolumn{3}{c}{\textbf{Estimation of std.}} \\
			\multicolumn{3}{c}{} & \multicolumn{6}{c}{} & \multicolumn{3}{c}{\textbf{errors}} \\
			\midrule
			& \textit{\# of} & Estim- & Bias & Mean & Std. & RMSE& Skew- & Ex. & Bias & CovP & CovP \\
			& \textit{groups} & ator & & abs. & dev. & & ness & Kurto- & (SE) & (95) & (80) \\
			& & & & error& & & & sis & & in \% & in \% \\
			\midrule
			& (1) & (2) & (3) & (4) & (5) & (6) & (7) & (8) & (9) & (10) & (11) \\
			\midrule
			& \multicolumn{11}{c}{N = 10'000} \\
			\midrule
			\textbf{ATE}      & \textit{1} & \textit{mcf}  & 0.138 & 0.138 & 0.028 & 0.141 & 0.15  & -0.02 & 0.002  & 0    & 0  \\
\bottomrule
		\end{tabular}
	}
	\vspace{-8pt}
	\captionsetup{justification=justified,singlelinecheck=false}
	\caption*{\footnotesize \textsf{\textbf{Note:} For GATE and IATE the results are averaged over all effects. \emph{CovP(95, 80)} denotes the (average) probability that the true value is part of the 95\%/80\% confidence interval. 1'000 / 500 replications used for 2'500 / 10'000 obs.}}
\end{table}

\begin{table}[h!]
	\sffamily
	\small
	\caption{No IATE, strong selectivity}
	\resizebox{\textwidth}{!}{
		\begin{tabular}{lccccccccccc}
			\toprule
			\multicolumn{3}{c}{} & \multicolumn{6}{c}{\textbf{Estimation of effects}} & \multicolumn{3}{c}{\textbf{Estimation of std.}} \\
			\multicolumn{3}{c}{} & \multicolumn{6}{c}{} & \multicolumn{3}{c}{\textbf{errors}} \\
			\midrule
			& \textit{\# of} & Estim- & Bias & Mean & Std. & RMSE& Skew- & Ex. & Bias & CovP & CovP \\
			& \textit{groups} & ator & & abs. & dev. & & ness & Kurto- & (SE) & (95) & (80) \\
			& & & & error& & & & sis & & in \% & in \% \\
			\midrule
			& (1) & (2) & (3) & (4) & (5) & (6) & (7) & (8) & (9) & (10) & (11) \\
			\midrule
			& \multicolumn{11}{c}{N = 2'500} \\
			\midrule
			\textbf{ATE}      & \textit{1} & \textit{mcf}  & 0.375 & 0.375 & 0.063 & 0.380 & 0.03  & -0.03 & 0.000  & 0  & 0  \\
\bottomrule
		\end{tabular}
	}
	\vspace{-8pt}
	\captionsetup{justification=justified,singlelinecheck=false}
	\caption*{\footnotesize \textsf{\textbf{Note:} Table to be continued.}}
\end{table}

\begin{table}[h!]
	\ContinuedFloat
	\sffamily
	\small
	\caption{No IATE, strong selectivity}
	\resizebox{\textwidth}{!}{
		\begin{tabular}{lccccccccccc}
			\toprule
			\multicolumn{3}{c}{} & \multicolumn{6}{c}{\textbf{Estimation of effects}} & \multicolumn{3}{c}{\textbf{Estimation of std.}} \\
			\multicolumn{3}{c}{} & \multicolumn{6}{c}{} & \multicolumn{3}{c}{\textbf{errors}} \\
			\midrule
			& \textit{\# of} & Estim- & Bias & Mean & Std. & RMSE& Skew- & Ex. & Bias & CovP & CovP \\
			& \textit{groups} & ator & & abs. & dev. & & ness & Kurto- & (SE) & (95) & (80) \\
			& & & & error& & & & sis & & in \% & in \% \\
			\midrule
			& (1) & (2) & (3) & (4) & (5) & (6) & (7) & (8) & (9) & (10) & (11) \\
			\midrule
			& \multicolumn{11}{c}{N = 10'000} \\
			\midrule
			\textbf{ATE}      & \textit{1} & \textit{mcf}  & 0.334  & 0.334 & 0.030 & 0.336 & -0.03  & 0.60  & 0.001  & 0  & 0  \\
\bottomrule
		\end{tabular}
	}
	\vspace{-8pt}
	\captionsetup{justification=justified,singlelinecheck=false}
	\caption*{\footnotesize \textsf{\textbf{Note:} For GATE and IATE the results are averaged over all effects. \emph{CovP(95, 80)} denotes the (average) probability that the true value is part of the 95\%/80\% confidence interval. 1'000 / 500 replications used for 2'500 / 10'000 obs.}}
\end{table}

\begin{table}[h!]
	\sffamily
	\small
	\caption{Linear IATE, no selectivity}
	\resizebox{\textwidth}{!}{
		\begin{tabular}{lccccccccccc}
			\toprule
			\multicolumn{3}{c}{} & \multicolumn{6}{c}{\textbf{Estimation of effects}} & \multicolumn{3}{c}{\textbf{Estimation of std.}} \\
			\multicolumn{3}{c}{} & \multicolumn{6}{c}{} & \multicolumn{3}{c}{\textbf{errors}} \\
			\midrule
			& \textit{\# of} & Estim- & Bias & Mean & Std. & RMSE& Skew- & Ex. & Bias & CovP & CovP \\
			& \textit{groups} & ator & & abs. & dev. & & ness & Kurto- & (SE) & (95) & (80) \\
			& & & & error& & & & sis & & in \% & in \% \\
			\midrule
			& (1) & (2) & (3) & (4) & (5) & (6) & (7) & (8) & (9) & (10) & (11) \\
			\midrule
			& \multicolumn{11}{c}{N = 2'500} \\
			\midrule
			\textbf{ATE}      & \textit{1} & \textit{mcf}  & -0.004 & 0.049 & 0.062 & 0.062 & -0.02 & -0.06 & -0.021 & 96 & 80 \\
\bottomrule
		\end{tabular}
	}
	\vspace{-8pt}
	\captionsetup{justification=justified,singlelinecheck=false}
	\caption*{\footnotesize \textsf{\textbf{Note:} Table to be continued.}}
\end{table}

\begin{table}[h!]
	\ContinuedFloat
	\sffamily
	\small
	\caption{Linear IATE, no selectivity}
	\resizebox{\textwidth}{!}{
		\begin{tabular}{lccccccccccc}
			\toprule
			\multicolumn{3}{c}{} & \multicolumn{6}{c}{\textbf{Estimation of effects}} & \multicolumn{3}{c}{\textbf{Estimation of std.}} \\
			\multicolumn{3}{c}{} & \multicolumn{6}{c}{} & \multicolumn{3}{c}{\textbf{errors}} \\
			\midrule
			& \textit{\# of} & Estim- & Bias & Mean & Std. & RMSE& Skew- & Ex. & Bias & CovP & CovP \\
			& \textit{groups} & ator & & abs. & dev. & & ness & Kurto- & (SE) & (95) & (80) \\
			& & & & error& & & & sis & & in \% & in \% \\
			\midrule
			& (1) & (2) & (3) & (4) & (5) & (6) & (7) & (8) & (9) & (10) & (11) \\
			\midrule
			& \multicolumn{11}{c}{N = 10'000} \\
			\midrule
			\textbf{ATE}      & \textit{1} & \textit{mcf}  & 0.001  & 0.023 & 0.029 & 0.029 & 0.108 & -0.20 & 0.004  & 98 & 82 \\
\bottomrule
		\end{tabular}
	}
	\vspace{-8pt}
	\captionsetup{justification=justified,singlelinecheck=false}
	\caption*{\footnotesize \textsf{\textbf{Note:} For GATE and IATE the results are averaged over all effects. \emph{CovP(95, 80)} denotes the (average) probability that the true value is part of the 95\%/80\% confidence interval. 1'000 / 500 replications used for 2'500 / 10'000 obs.}}
\end{table}

\begin{table}[h!]
	\sffamily
	\small
	\caption{Linear IATE, medium selectivity}
	\resizebox{\textwidth}{!}{
		\begin{tabular}{lccccccccccc}
			\toprule
			\multicolumn{3}{c}{} & \multicolumn{6}{c}{\textbf{Estimation of effects}} & \multicolumn{3}{c}{\textbf{Estimation of std.}} \\
			\multicolumn{3}{c}{} & \multicolumn{6}{c}{} & \multicolumn{3}{c}{\textbf{errors}} \\
			\midrule
			& \textit{\# of} & Estim- & Bias & Mean & Std. & RMSE& Skew- & Ex. & Bias & CovP & CovP \\
			& \textit{groups} & ator & & abs. & dev. & & ness & Kurto- & (SE) & (95) & (80) \\
			& & & & error& & & & sis & & in \% & in \% \\
			\midrule
			& (1) & (2) & (3) & (4) & (5) & (6) & (7) & (8) & (9) & (10) & (11) \\
			\midrule
			& \multicolumn{11}{c}{N = 2'500} \\
			\midrule
			\textbf{ATE}      & \textit{1} & \textit{mcf}  & 0.225 & 0.225 & 0.063 & 0.234 & 0.11  & 0.11  & 0.001  & 6  & 1  \\
\bottomrule
		\end{tabular}
	}
	\vspace{-8pt}
	\captionsetup{justification=justified,singlelinecheck=false}
	\caption*{\footnotesize \textsf{\textbf{Note:} Table to be continued.}}
\end{table}

\begin{table}[h!]
	\ContinuedFloat
	\sffamily
	\small
	\caption{Linear IATE, medium selectivity}
	\resizebox{\textwidth}{!}{
		\begin{tabular}{lccccccccccc}
			\toprule
			\multicolumn{3}{c}{} & \multicolumn{6}{c}{\textbf{Estimation of effects}} & \multicolumn{3}{c}{\textbf{Estimation of std.}} \\
			\multicolumn{3}{c}{} & \multicolumn{6}{c}{} & \multicolumn{3}{c}{\textbf{errors}} \\
			\midrule
			& \textit{\# of} & Estim- & Bias & Mean & Std. & RMSE& Skew- & Ex. & Bias & CovP & CovP \\
			& \textit{groups} & ator & & abs. & dev. & & ness & Kurto- & (SE) & (95) & (80) \\
			& & & & error& & & & sis & & in \% & in \% \\
			\midrule
			& (1) & (2) & (3) & (4) & (5) & (6) & (7) & (8) & (9) & (10) & (11) \\
			\midrule
			& \multicolumn{11}{c}{N = 10'000} \\
			\midrule
			\textbf{ATE}      & \textit{1} & \textit{mcf}  & 0.191 & 0.191 & 0.029 & 0.192 & 0.14  & -0.10 & 0.003  & 0  & 0  \\
\bottomrule
		\end{tabular}
	}
	\vspace{-8pt}
	\captionsetup{justification=justified,singlelinecheck=false}
	\caption*{\footnotesize \textsf{\textbf{Note:} For GATE and IATE the results are averaged over all effects. \emph{CovP(95, 80)} denotes the (average) probability that the true value is part of the 95\%/80\% confidence interval. 1'000 / 500 replications used for 2'500 / 10'000 obs.}}
\end{table}

\begin{table}[h!]
	\sffamily
	\small
	\caption{Linear IATE, strong selectivity}
	\resizebox{\textwidth}{!}{
		\begin{tabular}{lccccccccccc}
			\toprule
			\multicolumn{3}{c}{} & \multicolumn{6}{c}{\textbf{Estimation of effects}} & \multicolumn{3}{c}{\textbf{Estimation of std.}} \\
			\multicolumn{3}{c}{} & \multicolumn{6}{c}{} & \multicolumn{3}{c}{\textbf{errors}} \\
			\midrule
			& \textit{\# of} & Estim- & Bias & Mean & Std. & RMSE& Skew- & Ex. & Bias & CovP & CovP \\
			& \textit{groups} & ator & & abs. & dev. & & ness & Kurto- & (SE) & (95) & (80) \\
			& & & & error& & & & sis & & in \% & in \% \\
			\midrule
			& (1) & (2) & (3) & (4) & (5) & (6) & (7) & (8) & (9) & (10) & (11) \\
			\midrule
			& \multicolumn{11}{c}{N = 2'500} \\
			\midrule
			\textbf{ATE}      & \textit{1} & \textit{mcf}  & 0.528 & 0.528 & 0.065 & 0.532 & 0.03  & 0.06  & -0.001 & 0  & 0  \\
\bottomrule
		\end{tabular}
	}
	\vspace{-8pt}
	\captionsetup{justification=justified,singlelinecheck=false}
	\caption*{\footnotesize \textsf{\textbf{Note:} Table to be continued.}}
\end{table}

\begin{table}[h!]
	\ContinuedFloat
	\sffamily
	\small
	\caption{Linear IATE, strong selectivity}
	\resizebox{\textwidth}{!}{
		\begin{tabular}{lccccccccccc}
			\toprule
			\multicolumn{3}{c}{} & \multicolumn{6}{c}{\textbf{Estimation of effects}} & \multicolumn{3}{c}{\textbf{Estimation of std.}} \\
			\multicolumn{3}{c}{} & \multicolumn{6}{c}{} & \multicolumn{3}{c}{\textbf{errors}} \\
			\midrule
			& \textit{\# of} & Estim- & Bias & Mean & Std. & RMSE& Skew- & Ex. & Bias & CovP & CovP \\
			& \textit{groups} & ator & & abs. & dev. & & ness & Kurto- & (SE) & (95) & (80) \\
			& & & & error& & & & sis & & in \% & in \% \\
			\midrule
			& (1) & (2) & (3) & (4) & (5) & (6) & (7) & (8) & (9) & (10) & (11) \\
			\midrule
			& \multicolumn{11}{c}{N = 10'000} \\
			\midrule
			\textbf{ATE}      & \textit{1} & \textit{mcf}  & 0.469 & 0.469 & 0.031 & 0.470 & -0.08 & 0.68  & 0.001  & 0  & 0  \\
\bottomrule
		\end{tabular}
	}
	\vspace{-8pt}
	\captionsetup{justification=justified,singlelinecheck=false}
	\caption*{\footnotesize \textsf{\textbf{Note:} For GATE and IATE the results are averaged over all effects. \emph{CovP(95, 80)} denotes the (average) probability that the true value is part of the 95\%/80\% confidence interval. 1'000 / 500 replications used for 2'500 / 10'000 obs.}}
\end{table}

\begin{table}[h!]
	\sffamily
	\small
	\caption{Nonlinear IATE, no selectivity}
	\resizebox{\textwidth}{!}{
		\begin{tabular}{lccccccccccc}
			\toprule
			\multicolumn{3}{c}{} & \multicolumn{6}{c}{\textbf{Estimation of effects}} & \multicolumn{3}{c}{\textbf{Estimation of std.}} \\
			\multicolumn{3}{c}{} & \multicolumn{6}{c}{} & \multicolumn{3}{c}{\textbf{errors}} \\
			\midrule
			& \textit{\# of} & Estim- & Bias & Mean & Std. & RMSE& Skew- & Ex. & Bias & CovP & CovP \\
			& \textit{groups} & ator & & abs. & dev. & & ness & Kurto- & (SE) & (95) & (80) \\
			& & & & error& & & & sis & & in \% & in \% \\
			\midrule
			& (1) & (2) & (3) & (4) & (5) & (6) & (7) & (8) & (9) & (10) & (11) \\
			\midrule
			& \multicolumn{11}{c}{N = 2'500} \\
			\midrule
			\textbf{ATE}      & \textit{1} & \textit{mcf}  & 0.000  & 0.048 & 0.060 & 0.060 & 0.03  & 0.00  & 0.006  & 97 & 85 \\
\bottomrule
		\end{tabular}
	}
	\vspace{-8pt}
	\captionsetup{justification=justified,singlelinecheck=false}
	\caption*{\footnotesize \textsf{\textbf{Note:} Table to be continued.}}
\end{table}

\begin{table}[h!]
	\ContinuedFloat
	\sffamily
	\small
	\caption{Nonlinear IATE, no selectivity}
	\resizebox{\textwidth}{!}{
		\begin{tabular}{lccccccccccc}
			\toprule
			\multicolumn{3}{c}{} & \multicolumn{6}{c}{\textbf{Estimation of effects}} & \multicolumn{3}{c}{\textbf{Estimation of std.}} \\
			\multicolumn{3}{c}{} & \multicolumn{6}{c}{} & \multicolumn{3}{c}{\textbf{errors}} \\
			\midrule
			& \textit{\# of} & Estim- & Bias & Mean & Std. & RMSE& Skew- & Ex. & Bias & CovP & CovP \\
			& \textit{groups} & ator & & abs. & dev. & & ness & Kurto- & (SE) & (95) & (80) \\
			& & & & error& & & & sis & & in \% & in \% \\
			\midrule
			& (1) & (2) & (3) & (4) & (5) & (6) & (7) & (8) & (9) & (10) & (11) \\
			\midrule
			& \multicolumn{11}{c}{N = 10'000} \\
			\midrule
			\textbf{ATE}      & \textit{1} & \textit{mcf}  & 0.000  & 0.022 & 0.030 & 0.030 & 0.25  & 0.03  & 0.003  & 98 & 86 \\
\bottomrule
		\end{tabular}
	}
	\vspace{-8pt}
	\captionsetup{justification=justified,singlelinecheck=false}
	\caption*{\footnotesize \textsf{\textbf{Note:} For GATE and IATE the results are averaged over all effects. \emph{CovP(95, 80)} denotes the (average) probability that the true value is part of the 95\%/80\% confidence interval. 1'000 / 500 replications used for 2'500 / 10'000 obs.}}
\end{table}

\begin{table}[h!]
	\sffamily
	\small
	\caption{Nonlinear IATE, medium selectivity}
	\resizebox{\textwidth}{!}{
		\begin{tabular}{lccccccccccc}
			\toprule
			\multicolumn{3}{c}{} & \multicolumn{6}{c}{\textbf{Estimation of effects}} & \multicolumn{3}{c}{\textbf{Estimation of std.}} \\
			\multicolumn{3}{c}{} & \multicolumn{6}{c}{} & \multicolumn{3}{c}{\textbf{errors}} \\
			\midrule
			& \textit{\# of} & Estim- & Bias & Mean & Std. & RMSE& Skew- & Ex. & Bias & CovP & CovP \\
			& \textit{groups} & ator & & abs. & dev. & & ness & Kurto- & (SE) & (95) & (80) \\
			& & & & error& & & & sis & & in \% & in \% \\
			\midrule
			& (1) & (2) & (3) & (4) & (5) & (6) & (7) & (8) & (9) & (10) & (11) \\
			\midrule
			& \multicolumn{11}{c}{N = 2'500} \\
			\midrule
			\textbf{ATE}      & \textit{1} & \textit{mcf}  & 0.219 & 0.219 & 0.060 & 0.227 & -0.03 & -0.14 & 0.004  & 6  & 1  \\
\bottomrule
		\end{tabular}
	}
	\vspace{-8pt}
	\captionsetup{justification=justified,singlelinecheck=false}
	\caption*{\footnotesize \textsf{\textbf{Note:} Table to be continued.}}
\end{table}

\begin{table}[h!]
	\ContinuedFloat
	\sffamily
	\small
	\caption{Nonlinear IATE, medium selectivity}
	\resizebox{\textwidth}{!}{
		\begin{tabular}{lccccccccccc}
			\toprule
			\multicolumn{3}{c}{} & \multicolumn{6}{c}{\textbf{Estimation of effects}} & \multicolumn{3}{c}{\textbf{Estimation of std.}} \\
			\multicolumn{3}{c}{} & \multicolumn{6}{c}{} & \multicolumn{3}{c}{\textbf{errors}} \\
			\midrule
			& \textit{\# of} & Estim- & Bias & Mean & Std. & RMSE& Skew- & Ex. & Bias & CovP & CovP \\
			& \textit{groups} & ator & & abs. & dev. & & ness & Kurto- & (SE) & (95) & (80) \\
			& & & & error& & & & sis & & in \% & in \% \\
			\midrule
			& (1) & (2) & (3) & (4) & (5) & (6) & (7) & (8) & (9) & (10) & (11) \\
			\midrule
			& \multicolumn{11}{c}{N = 10'000} \\
			\midrule
			\textbf{ATE}      & \textit{1} & \textit{mcf}  & 0.181 & 0.181 & 0.030 & 0.184 & 0.22  & -0.47 & 0.002  & 0  & 0  \\
\bottomrule
		\end{tabular}
	}
	\vspace{-8pt}
	\captionsetup{justification=justified,singlelinecheck=false}
	\caption*{\footnotesize \textsf{\textbf{Note:} For GATE and IATE the results are averaged over all effects. \emph{CovP(95, 80)} denotes the (average) probability that the true value is part of the 95\%/80\% confidence interval. 1'000 / 500 replications used for 2'500 / 10'000 obs.}}
\end{table}

\begin{table}[h!]
	\sffamily
	\small
	\caption{Nonlinear IATE, strong selectivity}
	\resizebox{\textwidth}{!}{
		\begin{tabular}{lccccccccccc}
			\toprule
			\multicolumn{3}{c}{} & \multicolumn{6}{c}{\textbf{Estimation of effects}} & \multicolumn{3}{c}{\textbf{Estimation of std.}} \\
			\multicolumn{3}{c}{} & \multicolumn{6}{c}{} & \multicolumn{3}{c}{\textbf{errors}} \\
			\midrule
			& \textit{\# of} & Estim- & Bias & Mean & Std. & RMSE& Skew- & Ex. & Bias & CovP & CovP \\
			& \textit{groups} & ator & & abs. & dev. & & ness & Kurto- & (SE) & (95) & (80) \\
			& & & & error& & & & sis & & in \% & in \% \\
			\midrule
			& (1) & (2) & (3) & (4) & (5) & (6) & (7) & (8) & (9) & (10) & (11) \\
			\midrule
			& \multicolumn{11}{c}{N = 2'500} \\
			\midrule
			\textbf{ATE}      & \textit{1} & \textit{mcf}  & 0.517 & 0.517 & 0.062 & 0.521 & -0.07 & -0.05 & 0.002  & 0  & 0  \\
\bottomrule
		\end{tabular}
	}
	\vspace{-8pt}
	\captionsetup{justification=justified,singlelinecheck=false}
	\caption*{\footnotesize \textsf{\textbf{Note:} Table to be continued.}}
\end{table}

\begin{table}[h!]
	\ContinuedFloat
	\sffamily
	\small
	\caption{Nonlinear IATE, strong selectivity}
	\resizebox{\textwidth}{!}{
		\begin{tabular}{lccccccccccc}
			\toprule
			\multicolumn{3}{c}{} & \multicolumn{6}{c}{\textbf{Estimation of effects}} & \multicolumn{3}{c}{\textbf{Estimation of std.}} \\
			\multicolumn{3}{c}{} & \multicolumn{6}{c}{} & \multicolumn{3}{c}{\textbf{errors}} \\
			\midrule
			& \textit{\# of} & Estim- & Bias & Mean & Std. & RMSE& Skew- & Ex. & Bias & CovP & CovP \\
			& \textit{groups} & ator & & abs. & dev. & & ness & Kurto- & (SE) & (95) & (80) \\
			& & & & error& & & & sis & & in \% & in \% \\
			\midrule
			& (1) & (2) & (3) & (4) & (5) & (6) & (7) & (8) & (9) & (10) & (11) \\
			\midrule
			& \multicolumn{11}{c}{N = 10'000} \\
			\midrule
			\textbf{ATE}      & \textit{1} & \textit{mcf}  & 0.455 & 0.455 & 0.032 & 0.456 & -0.02 & -0.16 & 0.001  & 0  & 0  \\
\bottomrule
		\end{tabular}
	}
	\vspace{-8pt}
	\captionsetup{justification=justified,singlelinecheck=false}
	\caption*{\footnotesize \textsf{\textbf{Note:} For GATE and IATE the results are averaged over all effects. \emph{CovP(95, 80)} denotes the (average) probability that the true value is part of the 95\%/80\% confidence interval. 1'000 / 500 replications used for 2'500 / 10'000 obs.}}
\end{table}

\begin{table}[h!]
	\sffamily
	\small
	\caption{Quadratic IATE, no selectivity}
	\resizebox{\textwidth}{!}{
		\begin{tabular}{lccccccccccc}
			\toprule
			\multicolumn{3}{c}{} & \multicolumn{6}{c}{\textbf{Estimation of effects}} & \multicolumn{3}{c}{\textbf{Estimation of std.}} \\
			\multicolumn{3}{c}{} & \multicolumn{6}{c}{} & \multicolumn{3}{c}{\textbf{errors}} \\
			\midrule
			& \textit{\# of} & Estim- & Bias & Mean & Std. & RMSE& Skew- & Ex. & Bias & CovP & CovP \\
			& \textit{groups} & ator & & abs. & dev. & & ness & Kurto- & (SE) & (95) & (80) \\
			& & & & error& & & & sis & & in \% & in \% \\
			\midrule
			& (1) & (2) & (3) & (4) & (5) & (6) & (7) & (8) & (9) & (10) & (11) \\
			\midrule
			& \multicolumn{11}{c}{N = 2'500} \\
			\midrule
			\textbf{ATE}      & \textit{1} & \textit{mcf}  & -0.003 & 0.053 & 0.065 & 0.065 & 0.01  & -0.22 & 0.001  & 96 & 80 \\
\bottomrule
		\end{tabular}
	}
	\vspace{-8pt}
	\captionsetup{justification=justified,singlelinecheck=false}
	\caption*{\footnotesize \textsf{\textbf{Note:} Table to be continued.}}
\end{table}

\begin{table}[h!]
	\ContinuedFloat
	\sffamily
	\small
	\caption{Quadratic IATE, no selectivity}
	\resizebox{\textwidth}{!}{
		\begin{tabular}{lccccccccccc}
			\toprule
			\multicolumn{3}{c}{} & \multicolumn{6}{c}{\textbf{Estimation of effects}} & \multicolumn{3}{c}{\textbf{Estimation of std.}} \\
			\multicolumn{3}{c}{} & \multicolumn{6}{c}{} & \multicolumn{3}{c}{\textbf{errors}} \\
			\midrule
			& \textit{\# of} & Estim- & Bias & Mean & Std. & RMSE& Skew- & Ex. & Bias & CovP & CovP \\
			& \textit{groups} & ator & & abs. & dev. & & ness & Kurto- & (SE) & (95) & (80) \\
			& & & & error& & & & sis & & in \% & in \% \\
			\midrule
			& (1) & (2) & (3) & (4) & (5) & (6) & (7) & (8) & (9) & (10) & (11) \\
			\midrule
			& \multicolumn{11}{c}{N = 10'000} \\
			\midrule
			\textbf{ATE}      & \textit{1} & \textit{mcf}  & 0.001  & 0.024 & 0.030 & 0.030 & 0.22  & -0.02  & 0.004  & 97 & 84 \\
\bottomrule
		\end{tabular}
	}
	\vspace{-8pt}
	\captionsetup{justification=justified,singlelinecheck=false}
	\caption*{\footnotesize \textsf{\textbf{Note:} For GATE and IATE the results are averaged over all effects. \emph{CovP(95, 80)} denotes the (average) probability that the true value is part of the 95\%/80\% confidence interval. 1'000 / 500 replications used for 2'500 / 10'000 obs.}}
\end{table}

\begin{table}[h!]
	\sffamily
	\small
	\caption{Quadratic IATE, medium selectivity}
	\resizebox{\textwidth}{!}{
		\begin{tabular}{lccccccccccc}
			\toprule
			\multicolumn{3}{c}{} & \multicolumn{6}{c}{\textbf{Estimation of effects}} & \multicolumn{3}{c}{\textbf{Estimation of std.}} \\
			\multicolumn{3}{c}{} & \multicolumn{6}{c}{} & \multicolumn{3}{c}{\textbf{errors}} \\
			\midrule
			& \textit{\# of} & Estim- & Bias & Mean & Std. & RMSE& Skew- & Ex. & Bias & CovP & CovP \\
			& \textit{groups} & ator & & abs. & dev. & & ness & Kurto- & (SE) & (95) & (80) \\
			& & & & error& & & & sis & & in \% & in \% \\
			\midrule
			& (1) & (2) & (3) & (4) & (5) & (6) & (7) & (8) & (9) & (10) & (11) \\
			\midrule
			& \multicolumn{11}{c}{N = 2'500} \\
			\midrule
			\textbf{ATE}      & \textit{1} & \textit{mcf}  & 0.123  & 0.124 & 0.064 & 0.139 & 0.10  & 0.08  & 0.001  & 52 & 27 \\
\bottomrule
		\end{tabular}
	}
	\vspace{-8pt}
	\captionsetup{justification=justified,singlelinecheck=false}
	\caption*{\footnotesize \textsf{\textbf{Note:} Table to be continued.}}
\end{table}

\begin{table}[h!]
	\ContinuedFloat
	\sffamily
	\small
	\caption{Quadratic IATE, medium selectivity}
	\resizebox{\textwidth}{!}{
		\begin{tabular}{lccccccccccc}
			\toprule
			\multicolumn{3}{c}{} & \multicolumn{6}{c}{\textbf{Estimation of effects}} & \multicolumn{3}{c}{\textbf{Estimation of std.}} \\
			\multicolumn{3}{c}{} & \multicolumn{6}{c}{} & \multicolumn{3}{c}{\textbf{errors}} \\
			\midrule
			& \textit{\# of} & Estim- & Bias & Mean & Std. & RMSE& Skew- & Ex. & Bias & CovP & CovP \\
			& \textit{groups} & ator & & abs. & dev. & & ness & Kurto- & (SE) & (95) & (80) \\
			& & & & error& & & & sis & & in \% & in \% \\
			\midrule
			& (1) & (2) & (3) & (4) & (5) & (6) & (7) & (8) & (9) & (10) & (11) \\
			\midrule
			& \multicolumn{11}{c}{N = 10'000} \\
			\midrule
			\textbf{ATE}      & \textit{1} & \textit{mcf}  & 0.092  & 0.092 & 0.029 & 0.097 & 0.25  & 0.05  & 0.003  & 18 & 3  \\
\bottomrule
		\end{tabular}
	}
	\vspace{-8pt}
	\captionsetup{justification=justified,singlelinecheck=false}
	\caption*{\footnotesize \textsf{\textbf{Note:} For GATE and IATE the results are averaged over all effects. \emph{CovP(95, 80)} denotes the (average) probability that the true value is part of the 95\%/80\% confidence interval. 1'000 / 500 replications used for 2'500 / 10'000 obs.}}
\end{table}

\begin{table}[h!]
	\sffamily
	\small
	\caption{Quadratic IATE, strong selectivity}
	\resizebox{\textwidth}{!}{
		\begin{tabular}{lccccccccccc}
			\toprule
			\multicolumn{3}{c}{} & \multicolumn{6}{c}{\textbf{Estimation of effects}} & \multicolumn{3}{c}{\textbf{Estimation of std.}} \\
			\multicolumn{3}{c}{} & \multicolumn{6}{c}{} & \multicolumn{3}{c}{\textbf{errors}} \\
			\midrule
			& \textit{\# of} & Estim- & Bias & Mean & Std. & RMSE& Skew- & Ex. & Bias & CovP & CovP \\
			& \textit{groups} & ator & & abs. & dev. & & ness & Kurto- & (SE) & (95) & (80) \\
			& & & & error& & & & sis & & in \% & in \% \\
			\midrule
			& (1) & (2) & (3) & (4) & (5) & (6) & (7) & (8) & (9) & (10) & (11) \\
			\midrule
			& \multicolumn{11}{c}{N = 2'500} \\
			\midrule
			\textbf{ATE}      & \textit{1} & \textit{mcf}  & 0.261  & 0.261 & 0.066 & 0.269 & -0.03 & 0.06  & -0.001 & 2  & 0  \\
\bottomrule
		\end{tabular}
	}
	\vspace{-8pt}
	\captionsetup{justification=justified,singlelinecheck=false}
	\caption*{\footnotesize \textsf{\textbf{Note:} Table to be continued.}}
\end{table}

\begin{table}[h!]
	\ContinuedFloat
	\sffamily
	\small
	\caption{Quadratic IATE, strong selectivity}
	\resizebox{\textwidth}{!}{
		\begin{tabular}{lccccccccccc}
			\toprule
			\multicolumn{3}{c}{} & \multicolumn{6}{c}{\textbf{Estimation of effects}} & \multicolumn{3}{c}{\textbf{Estimation of std.}} \\
			\multicolumn{3}{c}{} & \multicolumn{6}{c}{} & \multicolumn{3}{c}{\textbf{errors}} \\
			\midrule
			& \textit{\# of} & Estim- & Bias & Mean & Std. & RMSE& Skew- & Ex. & Bias & CovP & CovP \\
			& \textit{groups} & ator & & abs. & dev. & & ness & Kurto- & (SE) & (95) & (80) \\
			& & & & error& & & & sis & & in \% & in \% \\
			\midrule
			& (1) & (2) & (3) & (4) & (5) & (6) & (7) & (8) & (9) & (10) & (11) \\
			\midrule
			& \multicolumn{11}{c}{N = 10'000} \\
			\midrule
			\textbf{ATE}      & \textit{1} & \textit{mcf}  & 0.198  & 0.198 & 0.032 & 0.201 & -0.04 & 0.57  & 0.001  & 0  & 0  \\
\bottomrule
		\end{tabular}
	}
	\vspace{-8pt}
	\captionsetup{justification=justified,singlelinecheck=false}
	\caption*{\footnotesize \textsf{\textbf{Note:} For GATE and IATE the results are averaged over all effects. \emph{CovP(95, 80)} denotes the (average) probability that the true value is part of the 95\%/80\% confidence interval. 1'000 / 500 replications used for 2'500 / 10'000 obs.}}
\end{table}

\begin{table}[h!]
	\sffamily
	\small
	\caption{Step-function IATE, no selectivity}
	\resizebox{\textwidth}{!}{
		\begin{tabular}{lccccccccccc}
			\toprule
			\multicolumn{3}{c}{} & \multicolumn{6}{c}{\textbf{Estimation of effects}} & \multicolumn{3}{c}{\textbf{Estimation of std.}} \\
			\multicolumn{3}{c}{} & \multicolumn{6}{c}{} & \multicolumn{3}{c}{\textbf{errors}} \\
			\midrule
			& \textit{\# of} & Estim- & Bias & Mean & Std. & RMSE& Skew- & Ex. & Bias & CovP & CovP \\
			& \textit{groups} & ator & & abs. & dev. & & ness & Kurto- & (SE) & (95) & (80) \\
			& & & & error& & & & sis & & in \% & in \% \\
			\midrule
			& (1) & (2) & (3) & (4) & (5) & (6) & (7) & (8) & (9) & (10) & (11) \\
			\midrule
			& \multicolumn{11}{c}{N = 2'500} \\
			\midrule
			\textbf{ATE}      & \textit{1} & \textit{mcf}  & -0.003 & 0.048 & 0.061 & 0.061 & 0.005 & 0.00  & -0.10  & 97 & 83 \\
\bottomrule
		\end{tabular}
	}
	\vspace{-8pt}
	\captionsetup{justification=justified,singlelinecheck=false}
	\caption*{\footnotesize \textsf{\textbf{Note:} Table to be continued.}}
\end{table}

\begin{table}[h!]
	\ContinuedFloat
	\sffamily
	\small
	\caption{Step-function IATE, no selectivity}
	\resizebox{\textwidth}{!}{
		\begin{tabular}{lccccccccccc}
			\toprule
			\multicolumn{3}{c}{} & \multicolumn{6}{c}{\textbf{Estimation of effects}} & \multicolumn{3}{c}{\textbf{Estimation of std.}} \\
			\multicolumn{3}{c}{} & \multicolumn{6}{c}{} & \multicolumn{3}{c}{\textbf{errors}} \\
			\midrule
			& \textit{\# of} & Estim- & Bias & Mean & Std. & RMSE& Skew- & Ex. & Bias & CovP & CovP \\
			& \textit{groups} & ator & & abs. & dev. & & ness & Kurto- & (SE) & (95) & (80) \\
			& & & & error& & & & sis & & in \% & in \% \\
			\midrule
			& (1) & (2) & (3) & (4) & (5) & (6) & (7) & (8) & (9) & (10) & (11) \\
			\midrule
			& \multicolumn{11}{c}{N = 10'000} \\
			\midrule
			\textbf{ATE}      & \textit{1} & \textit{mcf}  & 0.001  & 0.022 & 0.028 & 0.028 & 0.13  & -0.15 & 0.005  & 98 & 84 \\
\bottomrule
		\end{tabular}
	}
	\vspace{-8pt}
	\captionsetup{justification=justified,singlelinecheck=false}
	\caption*{\footnotesize \textsf{\textbf{Note:} For GATE and IATE the results are averaged over all effects. \emph{CovP(95, 80)} denotes the (average) probability that the true value is part of the 95\%/80\% confidence interval. 1'000 / 500 replications used for 2'500 / 10'000 obs.}}
\end{table}

\begin{table}[h!]
	\sffamily
	\small
	\caption{Step-function IATE, medium selectivity}
	\resizebox{\textwidth}{!}{
		\begin{tabular}{lccccccccccc}
			\toprule
			\multicolumn{3}{c}{} & \multicolumn{6}{c}{\textbf{Estimation of effects}} & \multicolumn{3}{c}{\textbf{Estimation of std.}} \\
			\multicolumn{3}{c}{} & \multicolumn{6}{c}{} & \multicolumn{3}{c}{\textbf{errors}} \\
			\midrule
			& \textit{\# of} & Estim- & Bias & Mean & Std. & RMSE& Skew- & Ex. & Bias & CovP & CovP \\
			& \textit{groups} & ator & & abs. & dev. & & ness & Kurto- & (SE) & (95) & (80) \\
			& & & & error& & & & sis & & in \% & in \% \\
			\midrule
			& (1) & (2) & (3) & (4) & (5) & (6) & (7) & (8) & (9) & (10) & (11) \\
			\midrule
			& \multicolumn{11}{c}{N = 2'500} \\
			\midrule
			\textbf{ATE}      & \textit{1} & \textit{mcf}  & 0.175 & 0.175 & 0.062 & 0.186 & 0.11  & 0.13  & 0.002  & 21 & 7  \\
\bottomrule
		\end{tabular}
	}
	\vspace{-8pt}
	\captionsetup{justification=justified,singlelinecheck=false}
	\caption*{\footnotesize \textsf{\textbf{Note:} Table to be continued.}}
\end{table}

\begin{table}[h!]
	\ContinuedFloat
	\sffamily
	\small
	\caption{Step-function IATE, medium selectivity}
	\resizebox{\textwidth}{!}{
		\begin{tabular}{lccccccccccc}
			\toprule
			\multicolumn{3}{c}{} & \multicolumn{6}{c}{\textbf{Estimation of effects}} & \multicolumn{3}{c}{\textbf{Estimation of std.}} \\
			\multicolumn{3}{c}{} & \multicolumn{6}{c}{} & \multicolumn{3}{c}{\textbf{errors}} \\
			\midrule
			& \textit{\# of} & Estim- & Bias & Mean & Std. & RMSE& Skew- & Ex. & Bias & CovP & CovP \\
			& \textit{groups} & ator & & abs. & dev. & & ness & Kurto- & (SE) & (95) & (80) \\
			& & & & error& & & & sis & & in \% & in \% \\
			\midrule
			& (1) & (2) & (3) & (4) & (5) & (6) & (7) & (8) & (9) & (10) & (11) \\
			\midrule
			& \multicolumn{11}{c}{N = 10'000} \\
			\midrule
			\textbf{ATE}      & \textit{1} & \textit{mcf}  & 0.144 & 0.144 & 0.029 & 0.147 & 0.16  & -0.02 & 0.003  & 0  & 0  \\
\bottomrule
		\end{tabular}
	}
	\vspace{-8pt}
	\captionsetup{justification=justified,singlelinecheck=false}
	\caption*{\footnotesize \textsf{\textbf{Note:} For GATE and IATE the results are averaged over all effects. \emph{CovP(95, 80)} denotes the (average) probability that the true value is part of the 95\%/80\% confidence interval. 1'000 / 500 replications used for 2'500 / 10'000 obs.}}
\end{table}

\begin{table}[h!]
	\sffamily
	\small
	\caption{Step-function IATE, strong selectivity}
	\label{tab:c15}
	\resizebox{\textwidth}{!}{
		\begin{tabular}{lccccccccccc}
			\toprule
			\multicolumn{3}{c}{} & \multicolumn{6}{c}{\textbf{Estimation of effects}} & \multicolumn{3}{c}{\textbf{Estimation of std.}} \\
			\multicolumn{3}{c}{} & \multicolumn{6}{c}{} & \multicolumn{3}{c}{\textbf{errors}} \\
			\midrule
			& \textit{\# of} & Estim- & Bias & Mean & Std. & RMSE& Skew- & Ex. & Bias & CovP & CovP \\
			& \textit{groups} & ator & & abs. & dev. & & ness & Kurto- & (SE) & (95) & (80) \\
			& & & & error& & & & sis & & in \% & in \% \\
			\midrule
			& (1) & (2) & (3) & (4) & (5) & (6) & (7) & (8) & (9) & (10) & (11) \\
			\midrule
			& \multicolumn{11}{c}{N = 2'500} \\
			\midrule
			\textbf{ATE}      & \textit{1} & \textit{mcf}  & 0.437 & 0.437 & -0.001 & 0.441 & 0.03  & -0.02 & -0.001 & 0  & 0  \\
\bottomrule
		\end{tabular}
	}
	\vspace{-8pt}
	\captionsetup{justification=justified,singlelinecheck=false}
	\caption*{\footnotesize \textsf{\textbf{Note:} Table to be continued.}}
\end{table}

\begin{table}[h!]
	\ContinuedFloat
	\sffamily
	\small
	\caption{Step-function IATE, strong selectivity}
	\resizebox{\textwidth}{!}{
		\begin{tabular}{lccccccccccc}
			\toprule
			\multicolumn{3}{c}{} & \multicolumn{6}{c}{\textbf{Estimation of effects}} & \multicolumn{3}{c}{\textbf{Estimation of std.}} \\
			\multicolumn{3}{c}{} & \multicolumn{6}{c}{} & \multicolumn{3}{c}{\textbf{errors}} \\
			\midrule
			& \textit{\# of} & Estim- & Bias & Mean & Std. & RMSE& Skew- & Ex. & Bias & CovP & CovP \\
			& \textit{groups} & ator & & abs. & dev. & & ness & Kurto- & (SE) & (95) & (80) \\
			& & & & error& & & & sis & & in \% & in \% \\
			\midrule
			& (1) & (2) & (3) & (4) & (5) & (6) & (7) & (8) & (9) & (10) & (11) \\
			\midrule
			& \multicolumn{11}{c}{N = 10'000} \\
			\midrule
			\textbf{ATE}      & \textit{1} & \textit{mcf}  & 0.375  & 0.375 & 0.031 & 0.377 & -0.06 & 0.70  & 0.001  & 0  & 0  \\
\bottomrule
		\end{tabular}
	}
	\vspace{-8pt}
	\captionsetup{justification=justified,singlelinecheck=false}
	\caption*{\footnotesize \textsf{\textbf{Note:} For GATE and IATE the results are averaged over all effects. \emph{CovP(95, 80)} denotes the (average) probability that the true value is part of the 95\%/80\% confidence interval. 1'000 / 500 replications used for 2'500 / 10'000 obs.}}
\end{table}

\clearpage

\subsection{Extensions to the base specification}\label{appendix-c2}

So far, we investigate the performance of the estimations when we vary
the IATE and the degree of selectivity, as these dimensions are key
parameters in any unconfoundedness setting. However, there are many
other dimensions for which it is interesting to see if the conclusions
concerning estimator behaviour change. Since it is infeasible to vary
all of them simultaneously, we chose the setting with medium
selectivity, the step function specification of IATE heterogeneity, as
well as the other parameters from the previous specification
($k=10$, $p=20$,
$R^{2}(y^{0}) = 10\%$,
$X^{U}$ and $X^{N}$, 2 treatment
groups, a treatment share of 50\%, 5 GATEs, $N=2'500$ and
$N=10'000$). Then, we vary these parameters only individually,
instead of varying them jointly. Therefore, the following tables will
only show the dimension that deviates from this baseline specification.

\begin{table}[h!]
	\sffamily
	\small
	\caption{Fewer covariates ($p$ = 10, $k= p / 2$)}
	\resizebox{\textwidth}{!}{
\begin{tabular}{lccccccccccc}
			\toprule
			\multicolumn{3}{c}{} & \multicolumn{6}{c}{\textbf{Estimation of effects}} & \multicolumn{3}{c}{\textbf{Estimation of std.}} \\
			\multicolumn{3}{c}{} & \multicolumn{6}{c}{} & \multicolumn{3}{c}{\textbf{errors}} \\
			\midrule
			& \textit{\# of} & Estim- & Bias & Mean & Std. & RMSE& Skew- & Ex. & Bias & CovP & CovP \\
			& \textit{groups} & ator & & abs. & dev. & & ness & Kurto- & (SE) & (95) & (80) \\
			& & & & error& & & & sis & & in \% & in \% \\
			\midrule
			& (1) & (2) & (3) & (4) & (5) & (6) & (7) & (8) & (9) & (10) & (11) \\
			\midrule
			& \multicolumn{11}{c}{N = 2'500} \\
			\midrule
			\textbf{ATE}      & \textit{1} & \textit{mcf}  & 0.134  & 0.135 & 0.060 & 0.147 & -0.09 & -0.04 & 0.005  & 44 & 21 \\
\bottomrule
		\end{tabular}
	}
	\vspace{-8pt}
	\captionsetup{justification=justified,singlelinecheck=false}
	\caption*{\footnotesize \textsf{\textbf{Note:} Table to be continued.}}
\end{table}

\begin{table}[h!]
	\ContinuedFloat
	\sffamily
	\small
	\caption{Fewer covariates ($p$ = 10, $k= p / 2$)}
	\resizebox{\textwidth}{!}{
		\begin{tabular}{lccccccccccc}
			\toprule
			\multicolumn{3}{c}{} & \multicolumn{6}{c}{\textbf{Estimation of effects}} & \multicolumn{3}{c}{\textbf{Estimation of std.}} \\
			\multicolumn{3}{c}{} & \multicolumn{6}{c}{} & \multicolumn{3}{c}{\textbf{errors}} \\
			\midrule
			& \textit{\# of} & Estim- & Bias & Mean & Std. & RMSE& Skew- & Ex. & Bias & CovP & CovP \\
			& \textit{groups} & ator & & abs. & dev. & & ness & Kurto- & (SE) & (95) & (80) \\
			& & & & error& & & & sis & & in \% & in \% \\
			\midrule
			& (1) & (2) & (3) & (4) & (5) & (6) & (7) & (8) & (9) & (10) & (11) \\
			\midrule
			& \multicolumn{11}{c}{N = 10'000} \\
			\midrule
			\textbf{ATE}      & \textit{1} & \textit{mcf}  & 0.092  & 0.092 & 0.030 & 0.097 & 0.09  & -0.35 & 0.002  & 18 & 5  \\
\bottomrule
		\end{tabular}
	}
	\vspace{-8pt}
	\captionsetup{justification=justified,singlelinecheck=false}
	\caption*{\footnotesize \textsf{\textbf{Note:} For GATE and IATE the results are averaged over all effects. \emph{CovP(95, 80)} denotes the (average) probability that the true value is part of the 95\%/80\% confidence interval. 1'000 / 500 replications used for 2'500 / 10'000 obs.}}
\end{table}

\begin{table}[h!]
	\sffamily
	\small
	\caption{More covariates ($p = 50$, $k= p / 2$)}
	\label{tab:c17}
	\resizebox{\textwidth}{!}{
		\begin{tabular}{lccccccccccc}
			\toprule
			\multicolumn{3}{c}{} & \multicolumn{6}{c}{\textbf{Estimation of effects}} & \multicolumn{3}{c}{\textbf{Estimation of std.}} \\
			\multicolumn{3}{c}{} & \multicolumn{6}{c}{} & \multicolumn{3}{c}{\textbf{errors}} \\
			\midrule
			& \textit{\# of} & Estim- & Bias & Mean & Std. & RMSE& Skew- & Ex. & Bias & CovP & CovP \\
			& \textit{groups} & ator & & abs. & dev. & & ness & Kurto- & (SE) & (95) & (80) \\
			& & & & error& & & & sis & & in \% & in \% \\
			\midrule
			& (1) & (2) & (3) & (4) & (5) & (6) & (7) & (8) & (9) & (10) & (11) \\
			\midrule
			& \multicolumn{11}{c}{N = 2'500} \\
			\midrule
			\textbf{ATE}      & \textit{1} & \textit{mcf}  & 0.223 & 0.223 & 0.061 & 0.231 & -0.02 & 0.05  & 0.002  & 5  & 1  \\
\bottomrule
		\end{tabular}
	}
	\vspace{-8pt}
	\captionsetup{justification=justified,singlelinecheck=false}
	\caption*{\footnotesize \textsf{\textbf{Note:} Table to be continued.}}
\end{table}

\begin{table}[h!]
	\ContinuedFloat
	\sffamily
	\small
	\caption{More covariates ($p = 50$, $k= p / 2$)}
	\resizebox{\textwidth}{!}{
		\begin{tabular}{lccccccccccc}
			\toprule
			\multicolumn{3}{c}{} & \multicolumn{6}{c}{\textbf{Estimation of effects}} & \multicolumn{3}{c}{\textbf{Estimation of std.}} \\
			\multicolumn{3}{c}{} & \multicolumn{6}{c}{} & \multicolumn{3}{c}{\textbf{errors}} \\
			\midrule
			& \textit{\# of} & Estim- & Bias & Mean & Std. & RMSE& Skew- & Ex. & Bias & CovP & CovP \\
			& \textit{groups} & ator & & abs. & dev. & & ness & Kurto- & (SE) & (95) & (80) \\
			& & & & error& & & & sis & & in \% & in \% \\
			\midrule
			& (1) & (2) & (3) & (4) & (5) & (6) & (7) & (8) & (9) & (10) & (11) \\
			\midrule
			& \multicolumn{11}{c}{N = 10'000} \\
			\midrule
			\textbf{ATE}      & \textit{1} & \textit{mcf}  & 0.201 & 0.201 & 0.031 & 0.204 & 0.01  & -0.07 & 0.001  & 0  & 0  \\
\bottomrule
		\end{tabular}
	}
	\vspace{-8pt}
	\captionsetup{justification=justified,singlelinecheck=false}
	\caption*{\footnotesize \textsf{\textbf{Note:} For GATE and IATE the results are averaged over all effects. \emph{CovP(95, 80)} denotes the (average) probability that the true value is part of the 95\%/80\% confidence interval. 1'000 / 500 replications used for 2'500 / 10'000 obs.}}
\end{table}

\begin{table}[h!]
	\sffamily
	\small
	\caption{Sparser model ($p = 20$, $k= 4$)}
	\resizebox{\textwidth}{!}{
		\begin{tabular}{lccccccccccc}
			\toprule
			\multicolumn{3}{c}{} & \multicolumn{6}{c}{\textbf{Estimation of effects}} & \multicolumn{3}{c}{\textbf{Estimation of std.}} \\
			\multicolumn{3}{c}{} & \multicolumn{6}{c}{} & \multicolumn{3}{c}{\textbf{errors}} \\
			\midrule
			& \textit{\# of} & Estim- & Bias & Mean & Std. & RMSE& Skew- & Ex. & Bias & CovP & CovP \\
			& \textit{groups} & ator & & abs. & dev. & & ness & Kurto- & (SE) & (95) & (80) \\
			& & & & error& & & & sis & & in \% & in \% \\
			\midrule
			& (1) & (2) & (3) & (4) & (5) & (6) & (7) & (8) & (9) & (10) & (11) \\
			\midrule
			& \multicolumn{11}{c}{N = 2'500} \\
			\midrule
			\textbf{ATE}      & \textit{1} & \textit{mcf}  & 0.132  & 0.133 & 0.061 & 0.146 & 0.04  & 0.01  & 0.003  & 47 & 22 \\
\bottomrule
		\end{tabular}
	}
	\vspace{-8pt}
	\captionsetup{justification=justified,singlelinecheck=false}
	\caption*{\footnotesize \textsf{\textbf{Note:} Table to be continued.}}
\end{table}

\begin{table}[h!]
	\ContinuedFloat
	\sffamily
	\small
	\caption{Sparser model ($p = 20$, $k= 4$)}
	\resizebox{\textwidth}{!}{
		\begin{tabular}{lccccccccccc}
			\toprule
			\multicolumn{3}{c}{} & \multicolumn{6}{c}{\textbf{Estimation of effects}} & \multicolumn{3}{c}{\textbf{Estimation of std.}} \\
			\multicolumn{3}{c}{} & \multicolumn{6}{c}{} & \multicolumn{3}{c}{\textbf{errors}} \\
			\midrule
			& \textit{\# of} & Estim- & Bias & Mean & Std. & RMSE& Skew- & Ex. & Bias & CovP & CovP \\
			& \textit{groups} & ator & & abs. & dev. & & ness & Kurto- & (SE) & (95) & (80) \\
			& & & & error& & & & sis & & in \% & in \% \\
			\midrule
			& (1) & (2) & (3) & (4) & (5) & (6) & (7) & (8) & (9) & (10) & (11) \\
			\midrule
			& \multicolumn{11}{c}{N = 10'000} \\
			\midrule
			\textbf{ATE}      & \textit{1} & \textit{mcf}  & 0.097  & 0.097 & 0.029 & 0.102 & 0.16  & -0.31 & 0.003  & 14 & 2  \\
\bottomrule
		\end{tabular}
	}
	\vspace{-8pt}
	\captionsetup{justification=justified,singlelinecheck=false}
	\caption*{\footnotesize \textsf{\textbf{Note:} For GATE and IATE the results are averaged over all effects. \emph{CovP(95, 80)} denotes the (average) probability that the true value is part of the 95\%/80\% confidence interval. 1'000 / 500 replications used for 2'500 / 10'000 obs.}}
\end{table}

\begin{table}[h!]
	\sffamily
	\small
	\caption{Less sparse model ($p = 20$, $k= 16$)}
	\resizebox{\textwidth}{!}{
		\begin{tabular}{lccccccccccc}
			\toprule
			\multicolumn{3}{c}{} & \multicolumn{6}{c}{\textbf{Estimation of effects}} & \multicolumn{3}{c}{\textbf{Estimation of std.}} \\
			\multicolumn{3}{c}{} & \multicolumn{6}{c}{} & \multicolumn{3}{c}{\textbf{errors}} \\
			\midrule
			& \textit{\# of} & Estim- & Bias & Mean & Std. & RMSE& Skew- & Ex. & Bias & CovP & CovP \\
			& \textit{groups} & ator & & abs. & dev. & & ness & Kurto- & (SE) & (95) & (80) \\
			& & & & error& & & & sis & & in \% & in \% \\
			\midrule
			& (1) & (2) & (3) & (4) & (5) & (6) & (7) & (8) & (9) & (10) & (11) \\
			\midrule
			& \multicolumn{11}{c}{N = 2'500} \\
			\midrule
			\textbf{ATE}      & \textit{1} & \textit{mcf}  & 0.192 & 0.192 & 0.061 & 0.201 & 0.08  & 0.23  & 0.002  & 14 & 4  \\
\bottomrule
		\end{tabular}
	}
	\vspace{-8pt}
	\captionsetup{justification=justified,singlelinecheck=false}
	\caption*{\footnotesize \textsf{\textbf{Note:} Table to be continued.}}
\end{table}

\begin{table}[h!]
	\ContinuedFloat
	\sffamily
	\small
	\caption{Less sparse model ($p = 20$, $k= 16$)}
	\resizebox{\textwidth}{!}{
		\begin{tabular}{lccccccccccc}
			\toprule
			\multicolumn{3}{c}{} & \multicolumn{6}{c}{\textbf{Estimation of effects}} & \multicolumn{3}{c}{\textbf{Estimation of std.}} \\
			\multicolumn{3}{c}{} & \multicolumn{6}{c}{} & \multicolumn{3}{c}{\textbf{errors}} \\
			\midrule
			& \textit{\# of} & Estim- & Bias & Mean & Std. & RMSE& Skew- & Ex. & Bias & CovP & CovP \\
			& \textit{groups} & ator & & abs. & dev. & & ness & Kurto- & (SE) & (95) & (80) \\
			& & & & error& & & & sis & & in \% & in \% \\
			\midrule
			& (1) & (2) & (3) & (4) & (5) & (6) & (7) & (8) & (9) & (10) & (11) \\
			\midrule
			& \multicolumn{11}{c}{N = 10'000} \\
			\midrule
			\textbf{ATE}      & \textit{1} & \textit{mcf}  & 0.168 & 0.168 & 0.029 & 0.170 & 0.06  & 0.07  & 0.003  & 0  & 0  \\
\bottomrule
		\end{tabular}
	}
	\vspace{-8pt}
	\captionsetup{justification=justified,singlelinecheck=false}
	\caption*{\footnotesize \textsf{\textbf{Note:} For GATE and IATE the results are averaged over all effects. \emph{CovP(95, 80)} denotes the (average) probability that the true value is part of the 95\%/80\% confidence interval. 1'000 / 500 replications used for 2'500 / 10'000 obs.}}
\end{table}

\begin{table}[h!]
	\sffamily
	\small
	\caption{Covariates irrelevant for $Y^0$ ($R^2(y^0) = 0\%$)}
	\resizebox{\textwidth}{!}{
		\begin{tabular}{lccccccccccc}
			\toprule
			\multicolumn{3}{c}{} & \multicolumn{6}{c}{\textbf{Estimation of effects}} & \multicolumn{3}{c}{\textbf{Estimation of std.}} \\
			\multicolumn{3}{c}{} & \multicolumn{6}{c}{} & \multicolumn{3}{c}{\textbf{errors}} \\
			\midrule
			& \textit{\# of} & Estim- & Bias & Mean & Std. & RMSE& Skew- & Ex. & Bias & CovP & CovP \\
			& \textit{groups} & ator & & abs. & dev. & & ness & Kurto- & (SE) & (95) & (80) \\
			& & & & error& & & & sis & & in \% & in \% \\
			\midrule
			& (1) & (2) & (3) & (4) & (5) & (6) & (7) & (8) & (9) & (10) & (11) \\
			\midrule
			& \multicolumn{11}{c}{N = 2'500} \\
			\midrule
			\textbf{ATE}      & \textit{1} & \textit{mcf}  & 0.024  & 0.050 & 0.059 & 0.064 & 0.10  & 0.07  & 0.00   & 93 & 76 \\
\bottomrule
		\end{tabular}
	}
	\vspace{-8pt}
	\captionsetup{justification=justified,singlelinecheck=false}
	\caption*{\footnotesize \textsf{\textbf{Note:} Table to be continued.}}
\end{table}

\begin{table}[h!]
	\ContinuedFloat
	\sffamily
	\small
	\caption{Covariates irrelevant for $Y^0$ ($R^2(y^0) = 0\%$)}
	\resizebox{\textwidth}{!}{
		\begin{tabular}{lccccccccccc}
			\toprule
			\multicolumn{3}{c}{} & \multicolumn{6}{c}{\textbf{Estimation of effects}} & \multicolumn{3}{c}{\textbf{Estimation of std.}} \\
			\multicolumn{3}{c}{} & \multicolumn{6}{c}{} & \multicolumn{3}{c}{\textbf{errors}} \\
			\midrule
			& \textit{\# of} & Estim- & Bias & Mean & Std. & RMSE& Skew- & Ex. & Bias & CovP & CovP \\
			& \textit{groups} & ator & & abs. & dev. & & ness & Kurto- & (SE) & (95) & (80) \\
			& & & & error& & & & sis & & in \% & in \% \\
			\midrule
			& (1) & (2) & (3) & (4) & (5) & (6) & (7) & (8) & (9) & (10) & (11) \\
			\midrule
			& \multicolumn{11}{c}{N = 10'000} \\
			\midrule
			\textbf{ATE}      & \textit{1} & \textit{mcf}  & 0.016  & 0.026 & 0.027 & 0.032 & 0.27  & 0.04  & 0.003  & 94 & 79 \\
\bottomrule
		\end{tabular}
	}
	\vspace{-8pt}
	\captionsetup{justification=justified,singlelinecheck=false}
	\caption*{\footnotesize \textsf{\textbf{Note:} For GATE and IATE the results are averaged over all effects. \emph{CovP(95, 80)} denotes the (average) probability that the true value is part of the 95\%/80\% confidence interval. 1'000 / 500 replications used for 2'500 / 10'000 obs.}}
\end{table}

\begin{table}[h!]
	\sffamily
	\small
	\caption{Covariates very important for $Y^0$ ($R^2(y^0) = 45\%$)}
	\label{tab:c21}
	\resizebox{\textwidth}{!}{
		\begin{tabular}{lccccccccccc}
			\toprule
			\multicolumn{3}{c}{} & \multicolumn{6}{c}{\textbf{Estimation of effects}} & \multicolumn{3}{c}{\textbf{Estimation of std.}} \\
			\multicolumn{3}{c}{} & \multicolumn{6}{c}{} & \multicolumn{3}{c}{\textbf{errors}} \\
			\midrule
			& \textit{\# of} & Estim- & Bias & Mean & Std. & RMSE& Skew- & Ex. & Bias & CovP & CovP \\
			& \textit{groups} & ator & & abs. & dev. & & ness & Kurto- & (SE) & (95) & (80) \\
			& & & & error& & & & sis & & in \% & in \% \\
			\midrule
			& (1) & (2) & (3) & (4) & (5) & (6) & (7) & (8) & (9) & (10) & (11) \\
			\midrule
			& \multicolumn{11}{c}{N = 2'500} \\
			\midrule
			\textbf{ATE}      & \textit{1} & \textit{mcf}  & 0.421 & 0.421 & 0.074 & 0.427 & 0.09  & 0.17  & 0.016  & 0  & 0  \\
\bottomrule
		\end{tabular}
	}
	\vspace{-8pt}
	\captionsetup{justification=justified,singlelinecheck=false}
	\caption*{\footnotesize \textsf{\textbf{Note:} Table to be continued.}}
\end{table}

\begin{table}[h!]
	\ContinuedFloat
	\sffamily
	\small
	\caption{Covariates very important for $Y^0$ ($R^2(y^0) = 45\%$)}
	\resizebox{\textwidth}{!}{
		\begin{tabular}{lccccccccccc}
			\toprule
			\multicolumn{3}{c}{} & \multicolumn{6}{c}{\textbf{Estimation of effects}} & \multicolumn{3}{c}{\textbf{Estimation of std.}} \\
			\multicolumn{3}{c}{} & \multicolumn{6}{c}{} & \multicolumn{3}{c}{\textbf{errors}} \\
			\midrule
			& \textit{\# of} & Estim- & Bias & Mean & Std. & RMSE& Skew- & Ex. & Bias & CovP & CovP \\
			& \textit{groups} & ator & & abs. & dev. & & ness & Kurto- & (SE) & (95) & (80) \\
			& & & & error& & & & sis & & in \% & in \% \\
			\midrule
			& (1) & (2) & (3) & (4) & (5) & (6) & (7) & (8) & (9) & (10) & (11) \\
			\midrule
			& \multicolumn{11}{c}{N = 10'000} \\
			\midrule
			\textbf{ATE}      & \textit{1} & \textit{mcf}  & 0.346 & 0.346 & 0.034 & 0.347 & -0.09 & -0.12 & 0.011  & 0  & 0  \\
\bottomrule
		\end{tabular}
	}
	\vspace{-8pt}
	\captionsetup{justification=justified,singlelinecheck=false}
	\caption*{\footnotesize \textsf{\textbf{Note:} For GATE and IATE the results are averaged over all effects. \emph{CovP(95, 80)} denotes the (average) probability that the true value is part of the 95\%/80\% confidence interval. 1'000 / 500 replications used for 2'500 / 10'000 obs.}}
\end{table}

\begin{table}[h!]
	\sffamily
	\small
	\caption{Uniformly distributed covariates ($X^U$)}
	\resizebox{\textwidth}{!}{
		\begin{tabular}{lccccccccccc}
			\toprule
			\multicolumn{3}{c}{} & \multicolumn{6}{c}{\textbf{Estimation of effects}} & \multicolumn{3}{c}{\textbf{Estimation of std.}} \\
			\multicolumn{3}{c}{} & \multicolumn{6}{c}{} & \multicolumn{3}{c}{\textbf{errors}} \\
			\midrule
			& \textit{\# of} & Estim- & Bias & Mean & Std. & RMSE& Skew- & Ex. & Bias & CovP & CovP \\
			& \textit{groups} & ator & & abs. & dev. & & ness & Kurto- & (SE) & (95) & (80) \\
			& & & & error& & & & sis & & in \% & in \% \\
			\midrule
			& (1) & (2) & (3) & (4) & (5) & (6) & (7) & (8) & (9) & (10) & (11) \\
			\midrule
			& \multicolumn{11}{c}{N = 2'500} \\
			\midrule
			\textbf{ATE}      & \textit{1} & \textit{mcf}  & 0.179 & 0.179 & 0.058 & 0.188 & 0.115 & -0.24 & 0.006  & 19 & 5  \\
\bottomrule
		\end{tabular}
	}
	\vspace{-8pt}
	\captionsetup{justification=justified,singlelinecheck=false}
	\caption*{\footnotesize \textsf{\textbf{Note:} Table to be continued.}}
\end{table}

\begin{table}[h!]
	\ContinuedFloat
	\sffamily
	\small
	\caption{Uniformly distributed covariates ($X^U$)}
	\resizebox{\textwidth}{!}{
		\begin{tabular}{lccccccccccc}
			\toprule
			\multicolumn{3}{c}{} & \multicolumn{6}{c}{\textbf{Estimation of effects}} & \multicolumn{3}{c}{\textbf{Estimation of std.}} \\
			\multicolumn{3}{c}{} & \multicolumn{6}{c}{} & \multicolumn{3}{c}{\textbf{errors}} \\
			\midrule
			& \textit{\# of} & Estim- & Bias & Mean & Std. & RMSE& Skew- & Ex. & Bias & CovP & CovP \\
			& \textit{groups} & ator & & abs. & dev. & & ness & Kurto- & (SE) & (95) & (80) \\
			& & & & error& & & & sis & & in \% & in \% \\
			\midrule
			& (1) & (2) & (3) & (4) & (5) & (6) & (7) & (8) & (9) & (10) & (11) \\
			\midrule
			& \multicolumn{11}{c}{N = 10'000} \\
			\midrule
			\textbf{ATE}      & \textit{1} & \textit{mcf}  & 0.145 & 0.145 & 0.028 & 0.148 & 0.01  & -0.58 & 0.004  & 0  & 0  \\
\bottomrule
		\end{tabular}
	}
	\vspace{-8pt}
	\captionsetup{justification=justified,singlelinecheck=false}
	\caption*{\footnotesize \textsf{\textbf{Note:} For GATE and IATE the results are averaged over all effects. \emph{CovP(95, 80)} denotes the (average) probability that the true value is part of the 95\%/80\% confidence interval. 1'000 / 500 replications used for 2'500 / 10'000 obs.}}
\end{table}

\begin{table}[h!]
	\sffamily
	\small
	\caption{Normally distributed covariates ($X^N$)}
	\resizebox{\textwidth}{!}{
		\begin{tabular}{lccccccccccc}
			\toprule
			\multicolumn{3}{c}{} & \multicolumn{6}{c}{\textbf{Estimation of effects}} & \multicolumn{3}{c}{\textbf{Estimation of std.}} \\
			\multicolumn{3}{c}{} & \multicolumn{6}{c}{} & \multicolumn{3}{c}{\textbf{errors}} \\
			\midrule
			& \textit{\# of} & Estim- & Bias & Mean & Std. & RMSE& Skew- & Ex. & Bias & CovP & CovP \\
			& \textit{groups} & ator & & abs. & dev. & & ness & Kurto- & (SE) & (95) & (80) \\
			& & & & error& & & & sis & & in \% & in \% \\
			\midrule
			& (1) & (2) & (3) & (4) & (5) & (6) & (7) & (8) & (9) & (10) & (11) \\
			\midrule
			& \multicolumn{11}{c}{N = 2'500} \\
			\midrule
			\textbf{ATE}      & \textit{1} & \textit{mcf}  & 0.191 & 0.191 & 0.062 & 0.201 & -0.06 & 0.14  & 0.002  & 15 & 4  \\
\bottomrule
		\end{tabular}
	}
	\vspace{-8pt}
	\captionsetup{justification=justified,singlelinecheck=false}
	\caption*{\footnotesize \textsf{\textbf{Note:} Table to be continued.}}
\end{table}

\begin{table}[h!]
	\ContinuedFloat
	\sffamily
	\small
	\caption{Normally distributed covariates ($X^N$)}
	\resizebox{\textwidth}{!}{
		\begin{tabular}{lccccccccccc}
			\toprule
			\multicolumn{3}{c}{} & \multicolumn{6}{c}{\textbf{Estimation of effects}} & \multicolumn{3}{c}{\textbf{Estimation of std.}} \\
			\multicolumn{3}{c}{} & \multicolumn{6}{c}{} & \multicolumn{3}{c}{\textbf{errors}} \\
			\midrule
			& \textit{\# of} & Estim- & Bias & Mean & Std. & RMSE& Skew- & Ex. & Bias & CovP & CovP \\
			& \textit{groups} & ator & & abs. & dev. & & ness & Kurto- & (SE) & (95) & (80) \\
			& & & & error& & & & sis & & in \% & in \% \\
			\midrule
			& (1) & (2) & (3) & (4) & (5) & (6) & (7) & (8) & (9) & (10) & (11) \\
			\midrule
			& \multicolumn{11}{c}{N = 10'000} \\
			\midrule
			\textbf{ATE}      & \textit{1} & \textit{mcf}  & 0.156 & 0.156 & 0.030 & 0.159 & -0.15 & -0.03 & 0.002  & 0  & 0  \\
\bottomrule
		\end{tabular}
	}
	\vspace{-8pt}
	\captionsetup{justification=justified,singlelinecheck=false}
	\caption*{\footnotesize \textsf{\textbf{Note:} For GATE and IATE the results are averaged over all effects. \emph{CovP(95, 80)} denotes the (average) probability that the true value is part of the 95\%/80\% confidence interval. 1'000 / 500 replications used for 2'500 / 10'000 obs.}}
\end{table}

\begin{table}[h!]
	\sffamily
	\small
	\caption{Normally, uniformly distributed, and indicator covariates ($X^D, X^N, X^U$)}
	\resizebox{\textwidth}{!}{
		\begin{tabular}{lccccccccccc}
			\toprule
			\multicolumn{3}{c}{} & \multicolumn{6}{c}{\textbf{Estimation of effects}} & \multicolumn{3}{c}{\textbf{Estimation of std.}} \\
			\multicolumn{3}{c}{} & \multicolumn{6}{c}{} & \multicolumn{3}{c}{\textbf{errors}} \\
			\midrule
			& \textit{\# of} & Estim- & Bias & Mean & Std. & RMSE& Skew- & Ex. & Bias & CovP & CovP \\
			& \textit{groups} & ator & & abs. & dev. & & ness & Kurto- & (SE) & (95) & (80) \\
			& & & & error& & & & sis & & in \% & in \% \\
			\midrule
			& (1) & (2) & (3) & (4) & (5) & (6) & (7) & (8) & (9) & (10) & (11) \\
			\midrule
			& \multicolumn{11}{c}{N = 2'500} \\
			\midrule
			\textbf{ATE}      & \textit{1} & \textit{mcf}  & 0.160 & 0.161 & 0.062 & 0.172 & 0.21  & 0.30  & 0.003  & 30 & 10 \\
\bottomrule
		\end{tabular}
	}
	\vspace{-8pt}
	\captionsetup{justification=justified,singlelinecheck=false}
	\caption*{\footnotesize \textsf{\textbf{Note:} Table to be continued.}}
\end{table}

\begin{table}[h!]
	\ContinuedFloat
	\sffamily
	\small
	\caption{Normally, uniformly distributed, and indicator covariates ($X^D, \allowbreak X^N, \allowbreak X^U$)}
	\resizebox{\textwidth}{!}{
		\begin{tabular}{lccccccccccc}
			\toprule
			\multicolumn{3}{c}{} & \multicolumn{6}{c}{\textbf{Estimation of effects}} & \multicolumn{3}{c}{\textbf{Estimation of std.}} \\
			\multicolumn{3}{c}{} & \multicolumn{6}{c}{} & \multicolumn{3}{c}{\textbf{errors}} \\
			\midrule
			& \textit{\# of} & Estim- & Bias & Mean & Std. & RMSE& Skew- & Ex. & Bias & CovP & CovP \\
			& \textit{groups} & ator & & abs. & dev. & & ness & Kurto- & (SE) & (95) & (80) \\
			& & & & error& & & & sis & & in \% & in \% \\
			\midrule
			& (1) & (2) & (3) & (4) & (5) & (6) & (7) & (8) & (9) & (10) & (11) \\
			\midrule
			& \multicolumn{11}{c}{N = 10'000} \\
			\midrule
			\textbf{ATE}      & \textit{1} & \textit{mcf}  & 0.120 & 0.120 & 0.028 & 0.124 & 0.32  & 0.41  & 0.004  & 2  & 0  \\
\bottomrule
		\end{tabular}
	}
	\vspace{-8pt}
	\captionsetup{justification=justified,singlelinecheck=false}
	\caption*{\footnotesize \textsf{\textbf{Note:} For GATE and IATE the results are averaged over all effects. \emph{CovP(95, 80)} denotes the (average) probability that the true value is part of the 95\%/80\% confidence interval. 1'000 / 500 replications used for 2'500 / 10'000 obs.}}
\end{table}

\begin{table}[h!]
	\sffamily
	\small
	\caption{Four treatments}
	\resizebox{\textwidth}{!}{
		\begin{tabular}{lccccccccccc}
			\toprule
			\multicolumn{3}{c}{} & \multicolumn{6}{c}{\textbf{Estimation of effects}} & \multicolumn{3}{c}{\textbf{Estimation of std.}} \\
			\multicolumn{3}{c}{} & \multicolumn{6}{c}{} & \multicolumn{3}{c}{\textbf{errors}} \\
			\midrule
			& \textit{\# of} & Estim- & Bias & Mean & Std. & RMSE& Skew- & Ex. & Bias & CovP & CovP \\
			& \textit{groups} & ator & & abs. & dev. & & ness & Kurto- & (SE) & (95) & (80) \\
			& & & & error& & & & sis & & in \% & in \% \\
			\midrule
			& (1) & (2) & (3) & (4) & (5) & (6) & (7) & (8) & (9) & (10) & (11) \\
			\midrule
			& \multicolumn{11}{c}{N = 2'500} \\
			\midrule
			\textbf{ATE}      & \textit{1} & \textit{mcf}  & 0.086  & 0.099 & 0.087 & 0.122 & 0.205 & -0.06 & 0.008  & 87 & 67 \\
\bottomrule
		\end{tabular}
	}
	\vspace{-8pt}
	\captionsetup{justification=justified,singlelinecheck=false}
	\caption*{\footnotesize \textsf{\textbf{Note:} Table to be continued.}}
\end{table}

\begin{table}[h!]
	\ContinuedFloat
	\sffamily
	\small
	\caption{Four treatments}
	\resizebox{\textwidth}{!}{
		\begin{tabular}{lccccccccccc}
			\toprule
			\multicolumn{3}{c}{} & \multicolumn{6}{c}{\textbf{Estimation of effects}} & \multicolumn{3}{c}{\textbf{Estimation of std.}} \\
			\multicolumn{3}{c}{} & \multicolumn{6}{c}{} & \multicolumn{3}{c}{\textbf{errors}} \\
			\midrule
			& \textit{\# of} & Estim- & Bias & Mean & Std. & RMSE& Skew- & Ex. & Bias & CovP & CovP \\
			& \textit{groups} & ator & & abs. & dev. & & ness & Kurto- & (SE) & (95) & (80) \\
			& & & & error& & & & sis & & in \% & in \% \\
			\midrule
			& (1) & (2) & (3) & (4) & (5) & (6) & (7) & (8) & (9) & (10) & (11) \\
			\midrule
			& \multicolumn{11}{c}{N = 10'000} \\
			\midrule
			\textbf{ATE}      & \textit{1} & \textit{mcf}  & 0.075  & 0.077 & 0.044 & 0.087 & 0.00  & 0.65  & 0.005  & 69 & 38 \\
\bottomrule
		\end{tabular}
	}
	\vspace{-8pt}
	\captionsetup{justification=justified,singlelinecheck=false}
	\caption*{\footnotesize \textsf{\textbf{Note:} For GATE and IATE the results are averaged over all effects. \emph{CovP(95, 80)} denotes the (average) probability that the true value is part of the 95\%/80\% confidence interval. 1'000 / 500 replications used for 2'500 / 10'000 obs. Results are shown for the comparison of treatments 1 to 0. \\ For the \emph{grf}, the results differ substantially for the different treatment comparisons (which all have the same effect size), with an RMSE of the ATE/GATE/IATE in the range of 0.066-0.154 / 0.136-0.196 / 0.236-0.266 for $N=2'500$ and 0.045-0.110 / 0.076-0.126 / 0.111-0.147 for $N=10'000$. \\ For the centred \emph{grf}, the RMSE of the ATE/GATE/IATE is in the range of 0.062-0.083 / 0.134-0.145 / 0.228-0.239 for $N=2'500$ and 0.030-0.042 / 0.066-0.073 / 0.113-0.122 for $N=10'000$.}}
\end{table}

\begin{table}[h!]
	\sffamily
	\small
	\caption{Larger sample}
	\resizebox{\textwidth}{!}{
\begin{tabular}{lccccccccccc}
			\toprule
			\multicolumn{3}{c}{} & \multicolumn{6}{c}{\textbf{Estimation of effects}} & \multicolumn{3}{c}{\textbf{Estimation of std.}} \\
			\multicolumn{3}{c}{} & \multicolumn{6}{c}{} & \multicolumn{3}{c}{\textbf{errors}} \\
			\midrule
			& \textit{\# of} & Estim- & Bias & Mean & Std. & RMSE& Skew- & Ex. & Bias & CovP & CovP \\
			& \textit{groups} & ator & & abs. & dev. & & ness & Kurto- & (SE) & (95) & (80) \\
			& & & & error& & & & sis & & in \% & in \% \\
			\midrule
			& (1) & (2) & (3) & (4) & (5) & (6) & (7) & (8) & (9) & (10) & (11) \\
			\midrule
			& \multicolumn{11}{c}{N = 40'000} \\
			\midrule
			\textbf{ATE}      & \textit{1} & \textit{mcf}  & 0.111 & 0.111 & 0.012 & 0.112 & 0.12  & -0.92 & 0.004  & 0  & 0  \\
\bottomrule
		\end{tabular}
	}
	\vspace{-8pt}
	\captionsetup{justification=justified,singlelinecheck=false}
	\caption*{\footnotesize \textsf{\textbf{Note:} For GATE and IATE the results are averaged over all effects. \emph{CovP(95, 80)} denotes the (average) probability that the true value is part of the 95\%/80\% confidence interval. 62 replications.}}
\end{table}

\begin{table}[h!]
	\sffamily
	\small
	\caption{Only 25\% treated}
	\resizebox{\textwidth}{!}{
		\begin{tabular}{lccccccccccc}
			\toprule
			\multicolumn{3}{c}{} & \multicolumn{6}{c}{\textbf{Estimation of effects}} & \multicolumn{3}{c}{\textbf{Estimation of std.}} \\
			\multicolumn{3}{c}{} & \multicolumn{6}{c}{} & \multicolumn{3}{c}{\textbf{errors}} \\
			\midrule
			& \textit{\# of} & Estim- & Bias & Mean & Std. & RMSE& Skew- & Ex. & Bias & CovP & CovP \\
			& \textit{groups} & ator & & abs. & dev. & & ness & Kurto- & (SE) & (95) & (80) \\
			& & & & error& & & & sis & & in \% & in \% \\
			\midrule
			& (1) & (2) & (3) & (4) & (5) & (6) & (7) & (8) & (9) & (10) & (11) \\
			\midrule
			& \multicolumn{11}{c}{N = 2'500} \\
			\midrule
			\textbf{ATE}      & \textit{1} & \textit{mcf}  & 0.247 & 0.247 & 0.073 & 0.257 & -0.04 & -0.12 & 0.001  & 8  & 2  \\
\bottomrule
		\end{tabular}
	}
	\vspace{-8pt}
	\captionsetup{justification=justified,singlelinecheck=false}
	\caption*{\footnotesize \textsf{\textbf{Note:} Table to be continued.}}
\end{table}

\begin{table}[h!]
	\ContinuedFloat
	\sffamily
	\small
	\caption{Only 25\% treated}
	\resizebox{\textwidth}{!}{
		\begin{tabular}{lccccccccccc}
			\toprule
			\multicolumn{3}{c}{} & \multicolumn{6}{c}{\textbf{Estimation of effects}} & \multicolumn{3}{c}{\textbf{Estimation of std.}} \\
			\multicolumn{3}{c}{} & \multicolumn{6}{c}{} & \multicolumn{3}{c}{\textbf{errors}} \\
			\midrule
			& \textit{\# of} & Estim- & Bias & Mean & Std. & RMSE& Skew- & Ex. & Bias & CovP & CovP \\
			& \textit{groups} & ator & & abs. & dev. & & ness & Kurto- & (SE) & (95) & (80) \\
			& & & & error& & & & sis & & in \% & in \% \\
			\midrule
			& (1) & (2) & (3) & (4) & (5) & (6) & (7) & (8) & (9) & (10) & (11) \\
			\midrule
			& \multicolumn{11}{c}{N = 10'000} \\
			\midrule
			\textbf{ATE}      & \textit{1} & \textit{mcf}  & 0.189 & 0.189 & 0.035 & 0.192 & 0.06  & -0.10 & 0.002  & 0  & 0  \\
\bottomrule
		\end{tabular}
	}
	\vspace{-8pt}
	\captionsetup{justification=justified,singlelinecheck=false}
	\caption*{\footnotesize \textsf{\textbf{Note:} For GATE and IATE the results are averaged over all effects. \emph{CovP(95, 80)} denotes the (average) probability that the true value is part of the 95\%/80\% confidence interval. 1'000 / 500 replications used for 2'500 / 10'000 obs.}}
\end{table}

\begin{table}[h!]
	\sffamily
	\small
	\caption{More GATE groups, no selectivity}
	\label{tab:c28}
	\resizebox{\textwidth}{!}{
		\begin{tabular}{lccccccccccc}
			\toprule
			\multicolumn{3}{c}{} & \multicolumn{6}{c}{\textbf{Estimation of effects}} & \multicolumn{3}{c}{\textbf{Estimation of std.}} \\
			\multicolumn{3}{c}{} & \multicolumn{6}{c}{} & \multicolumn{3}{c}{\textbf{errors}} \\
			\midrule
			& \textit{\# of} & Estim- & Bias & Mean & Std. & RMSE& Skew- & Ex. & Bias & CovP & CovP \\
			& \textit{groups} & ator & & abs. & dev. & & ness & Kurto- & (SE) & (95) & (80) \\
			& & & & error& & & & sis & & in \% & in \% \\
			\midrule
			& (1) & (2) & (3) & (4) & (5) & (6) & (7) & (8) & (9) & (10) & (11) \\
			\midrule
			& \multicolumn{11}{c}{N = 2'500} \\
			\midrule
			\textbf{GATE} & \textit{5}  & \textit{mcf}  & -0.004 & 0.092 & 0.087 & 0.115 & 0.00  & 0.01  & 0.002  & 86 & 67 \\
\bottomrule
		\end{tabular}
	}
	\vspace{-8pt}
	\captionsetup{justification=justified,singlelinecheck=false}
	\caption*{\footnotesize \textsf{\textbf{Note:} Table to be continued.}}
\end{table}

\begin{table}[h!]
	\ContinuedFloat
	\sffamily
	\small
	\caption{More GATE groups, no selectivity}
	\resizebox{\textwidth}{!}{
		\begin{tabular}{lccccccccccc}
			\toprule
			\multicolumn{3}{c}{} & \multicolumn{6}{c}{\textbf{Estimation of effects}} & \multicolumn{3}{c}{\textbf{Estimation of std.}} \\
			\multicolumn{3}{c}{} & \multicolumn{6}{c}{} & \multicolumn{3}{c}{\textbf{errors}} \\
			\midrule
			& \textit{\# of} & Estim- & Bias & Mean & Std. & RMSE& Skew- & Ex. & Bias & CovP & CovP \\
			& \textit{groups} & ator & & abs. & dev. & & ness & Kurto- & (SE) & (95) & (80) \\
			& & & & error& & & & sis & & in \% & in \% \\
			\midrule
			& (1) & (2) & (3) & (4) & (5) & (6) & (7) & (8) & (9) & (10) & (11) \\
			\midrule
			& \multicolumn{11}{c}{N = 10'000} \\
			\midrule
			\textbf{GATE} & \textit{5}  & \textit{mcf}  & 0.001  & 0.049 & 0.045 & 0.063 & 0.07  & -0.21 & 0.002  & 87 & 66 \\
\bottomrule
		\end{tabular}
	}
	\vspace{-8pt}
	\captionsetup{justification=justified,singlelinecheck=false}
	\caption*{\footnotesize \textsf{\textbf{Note:} For GATE, the results are averaged over all effects. \emph{CovP(95, 80)} denotes the (average) probability that the true value is part of the 95\%/80\% confidence interval. 1'000 / 500 replications used for 2'500 / 10'000 obs.}}
\end{table}

\begin{table}[h!]
	\ContinuedFloat
	\sffamily
	\small
	\caption{More GATE groups, no selectivity - Distribution of GATE minus true value of the centred mcf}
	\begin{tabular}{c}
		\toprule
		N = 2'500 \\
		\midrule
		\includegraphics[width = \textwidth]{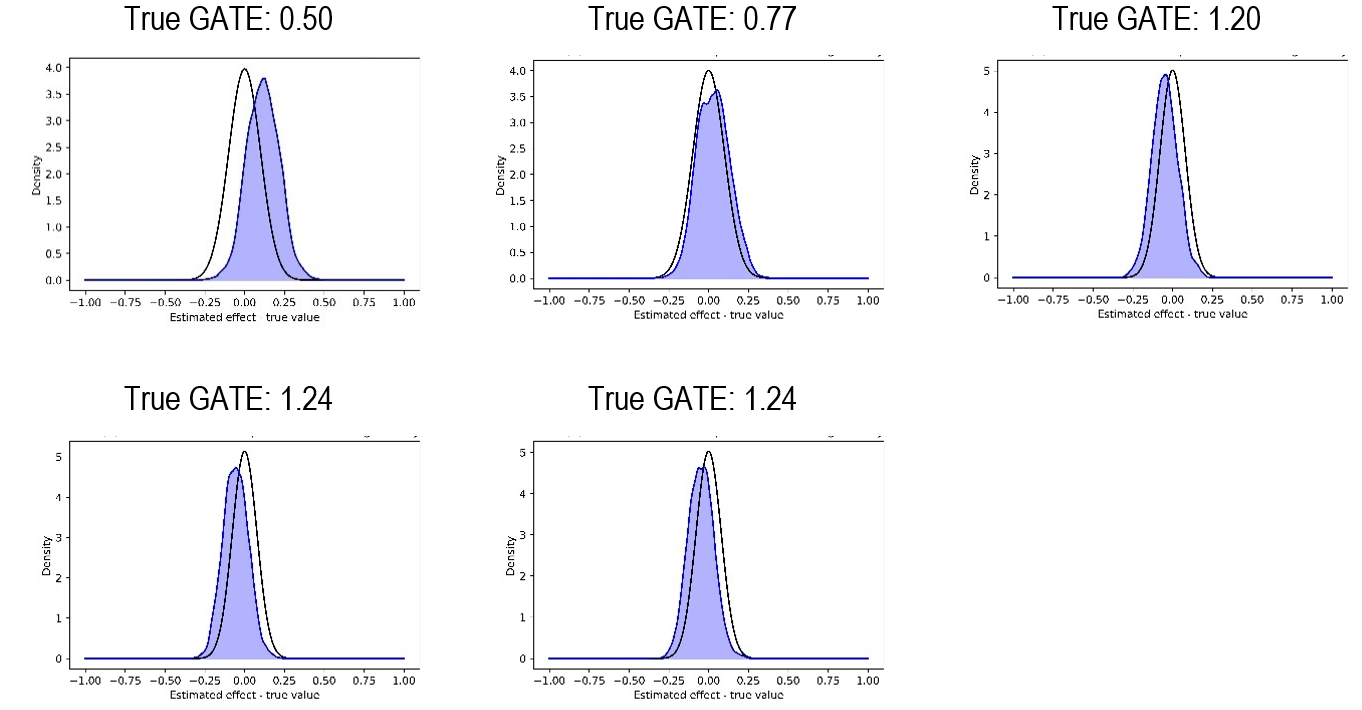} \\
		\midrule
		N = 10'000 \\
		\midrule
		\includegraphics[width = \textwidth]{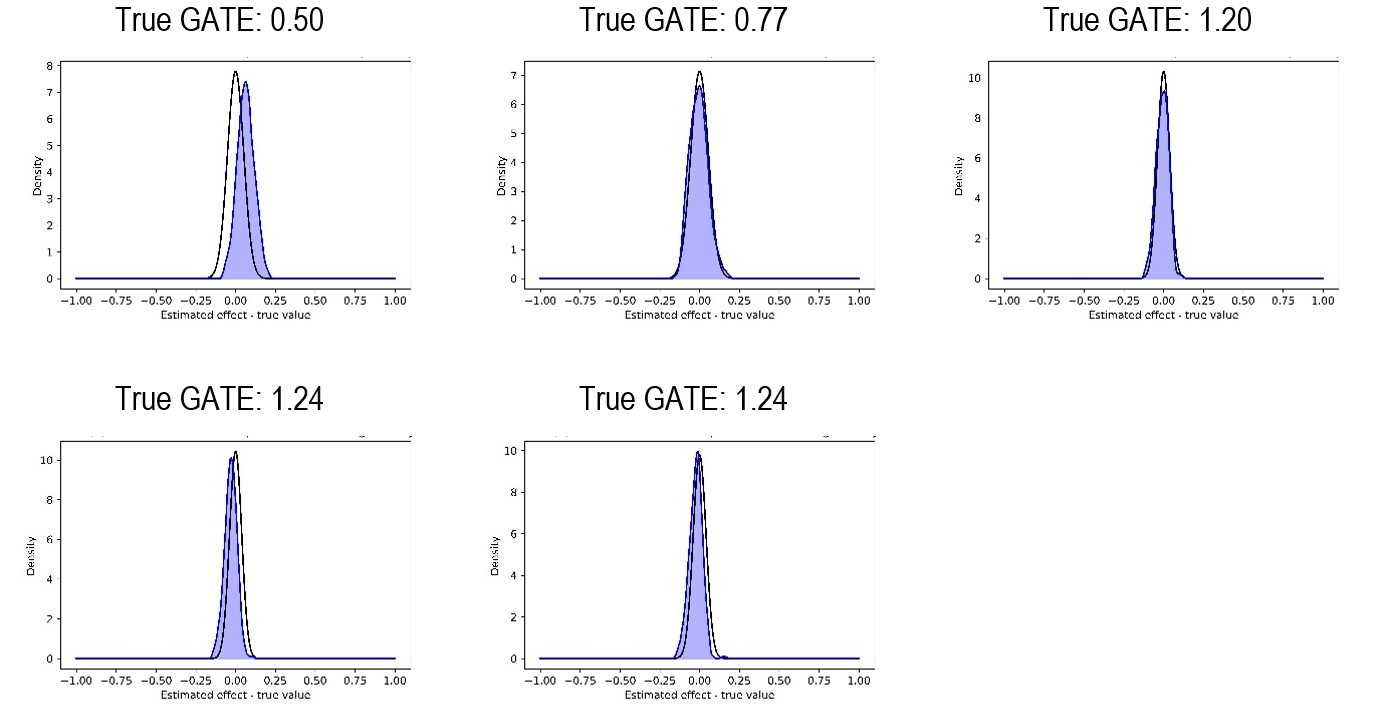} \\
		\bottomrule
	\end{tabular}
\end{table}

\begin{table}[h!]
	\sffamily
	\small
	\caption{More GATE groups, medium selectivity}
	\label{tab:c29}
	\resizebox{\textwidth}{!}{
		\begin{tabular}{lccccccccccc}
			\toprule
			\multicolumn{3}{c}{} & \multicolumn{6}{c}{\textbf{Estimation of effects}} & \multicolumn{3}{c}{\textbf{Estimation of std.}} \\
			\multicolumn{3}{c}{} & \multicolumn{6}{c}{} & \multicolumn{3}{c}{\textbf{errors}} \\
			\midrule
			& \textit{\# of} & Estim- & Bias & Mean & Std. & RMSE& Skew- & Ex. & Bias & CovP & CovP \\
			& \textit{groups} & ator & & abs. & dev. & & ness & Kurto- & (SE) & (95) & (80) \\
			& & & & error& & & & sis & & in \% & in \% \\
			\midrule
			& (1) & (2) & (3) & (4) & (5) & (6) & (7) & (8) & (9) & (10) & (11) \\
			\midrule
			& \multicolumn{11}{c}{N = 2'500} \\
			\midrule
						\textbf{GATE} & \textit{5}  & \textit{mcf}  & 0.175 & 0.178 & 0.090 & 0.211 & 0.06  & -0.04 & 0.008  & 53 & 32 \\
\bottomrule
		\end{tabular}
	}
	\vspace{-8pt}
	\captionsetup{justification=justified,singlelinecheck=false}
	\caption*{\footnotesize \textsf{\textbf{Note:} Table to be continued.}}
\end{table}

\begin{table}[h!]
	\ContinuedFloat
	\sffamily
	\small
	\caption{More GATE groups, medium selectivity}
	\resizebox{\textwidth}{!}{
		\begin{tabular}{lccccccccccc}
			\toprule
			\multicolumn{3}{c}{} & \multicolumn{6}{c}{\textbf{Estimation of effects}} & \multicolumn{3}{c}{\textbf{Estimation of std.}} \\
			\multicolumn{3}{c}{} & \multicolumn{6}{c}{} & \multicolumn{3}{c}{\textbf{errors}} \\
			\midrule
			& \textit{\# of} & Estim- & Bias & Mean & Std. & RMSE& Skew- & Ex. & Bias & CovP & CovP \\
			& \textit{groups} & ator & & abs. & dev. & & ness & Kurto- & (SE) & (95) & (80) \\
			& & & & error& & & & sis & & in \% & in \% \\
			\midrule
			& (1) & (2) & (3) & (4) & (5) & (6) & (7) & (8) & (9) & (10) & (11) \\
			\midrule
			& \multicolumn{11}{c}{N = 10'000} \\
			\midrule
			\textbf{GATE} & \textit{5}  & \textit{mcf}  & 0.144 & 0.144 & 0.048 & 0.158 & 0.16  & -0.03 & -0.001 & 21 & 7  \\
\bottomrule
		\end{tabular}
	}
	\vspace{-8pt}
	\captionsetup{justification=justified,singlelinecheck=false}
	\caption*{\footnotesize \textsf{\textbf{Note:} For GATE, the results are averaged over all effects. \emph{CovP(95, 80)} denotes the (average) probability that the true value is part of the 95\%/80\% confidence interval. 1'000 / 500 replications used for 2'500 / 10'000 obs.}}
\end{table}

\begin{table}[h!]
	\ContinuedFloat
	\sffamily
	\small
	\caption{More GATE groups, medium selectivity - Distribution of GATE minus true value of the centred mcf}
	\begin{tabular}{c}
		\toprule
		N = 2'500 \\
		\midrule
		\includegraphics[width = \textwidth]{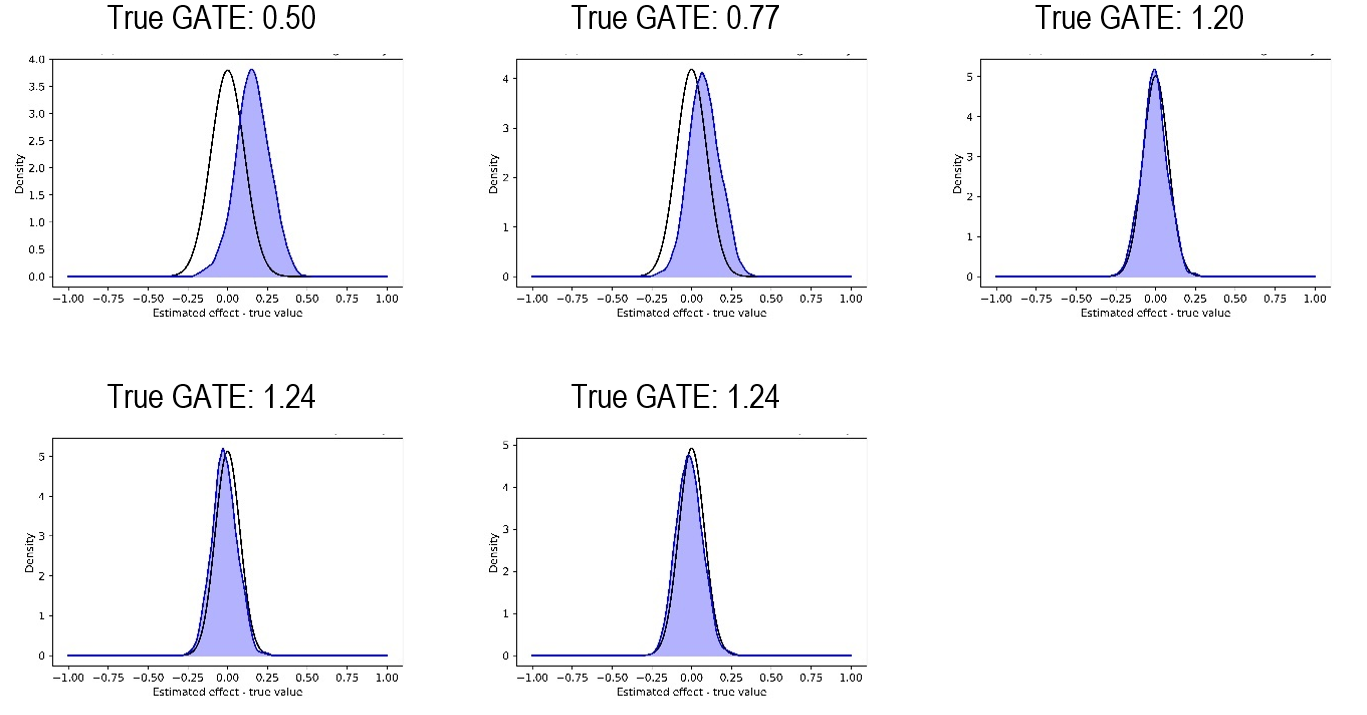} \\
		\midrule
		N = 10'000 \\
		\midrule
		\includegraphics[width = \textwidth]{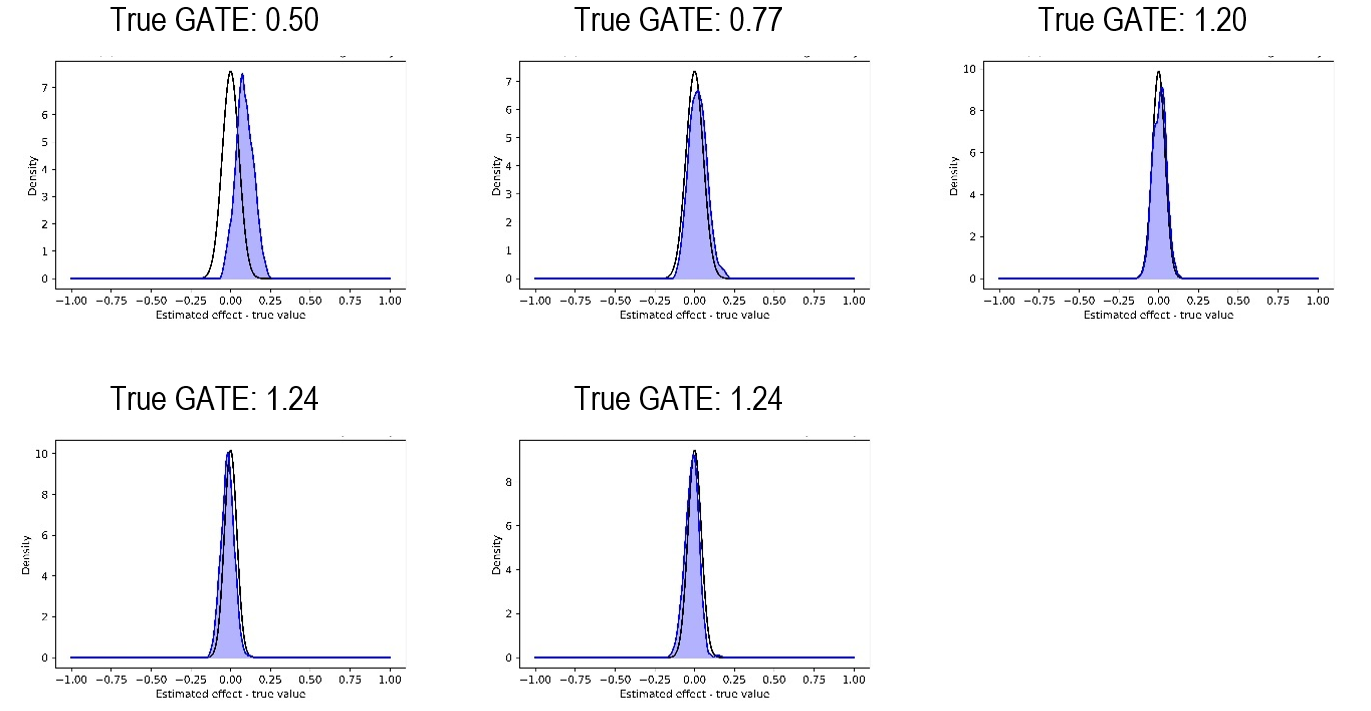} \\
		\bottomrule
	\end{tabular}
\end{table}

\begin{table}[h!]
	\sffamily
	\small
	\caption{More GATE groups, strong selectivity}
	\label{tab:c30}
	\resizebox{\textwidth}{!}{
		\begin{tabular}{lccccccccccc}
			\toprule
			\multicolumn{3}{c}{} & \multicolumn{6}{c}{\textbf{Estimation of effects}} & \multicolumn{3}{c}{\textbf{Estimation of std.}} \\
			\multicolumn{3}{c}{} & \multicolumn{6}{c}{} & \multicolumn{3}{c}{\textbf{errors}} \\
			\midrule
			& \textit{\# of} & Estim- & Bias & Mean & Std. & RMSE& Skew- & Ex. & Bias & CovP & CovP \\
			& \textit{groups} & ator & & abs. & dev. & & ness & Kurto- & (SE) & (95) & (80) \\
			& & & & error& & & & sis & & in \% & in \% \\
			\midrule
			& (1) & (2) & (3) & (4) & (5) & (6) & (7) & (8) & (9) & (10) & (11) \\
			\midrule
			& \multicolumn{11}{c}{N = 2'500} \\
			\midrule
			\textbf{GATE} & \textit{5}  & \textit{mcf}  & 0.436 & 0.436 & 0.010 & 0.455 & 0.00  & 0.06  & 0.001  & 1  & 0  \\
\bottomrule
		\end{tabular}
	}
	\vspace{-8pt}
	\captionsetup{justification=justified,singlelinecheck=false}
	\caption*{\footnotesize \textsf{\textbf{Note:} Table to be continued.}}
\end{table}

\begin{table}[h!]
	\ContinuedFloat
	\sffamily
	\small
	\caption{More GATE groups, strong selectivity}
	\resizebox{\textwidth}{!}{
		\begin{tabular}{lccccccccccc}
			\toprule
			\multicolumn{3}{c}{} & \multicolumn{6}{c}{\textbf{Estimation of effects}} & \multicolumn{3}{c}{\textbf{Estimation of std.}} \\
			\multicolumn{3}{c}{} & \multicolumn{6}{c}{} & \multicolumn{3}{c}{\textbf{errors}} \\
			\midrule
			& \textit{\# of} & Estim- & Bias & Mean & Std. & RMSE& Skew- & Ex. & Bias & CovP & CovP \\
			& \textit{groups} & ator & & abs. & dev. & & ness & Kurto- & (SE) & (95) & (80) \\
			& & & & error& & & & sis & & in \% & in \% \\
			\midrule
			& (1) & (2) & (3) & (4) & (5) & (6) & (7) & (8) & (9) & (10) & (11) \\
			\midrule
			& \multicolumn{11}{c}{N = 10'000} \\
			\midrule
			\textbf{GATE} & \textit{5}  & \textit{mcf}  & 0.375  & 0.375 & 0.053 & 0.384 & -0.10 & -0.10 & -0.002 & 0  & 0  \\
\bottomrule
		\end{tabular}
	}
	\vspace{-8pt}
	\captionsetup{justification=justified,singlelinecheck=false}
	\caption*{\footnotesize \textsf{\textbf{Note:} For GATE, the results are averaged over all effects. \emph{CovP(95, 80)} denotes the (average) probability that the true value is part of the 95\%/80\% confidence interval. 1'000 / 500 replications used for 2'500 / 10'000 obs.}}
\end{table}

\begin{table}[h!]
	\ContinuedFloat
	\sffamily
	\small
	\caption{More GATE groups, strong selectivity - Distribution of GATE minus true value of the centred mcf}
	\begin{tabular}{c}
		\toprule
		N = 2'500 \\
		\midrule
		\includegraphics[width = \textwidth]{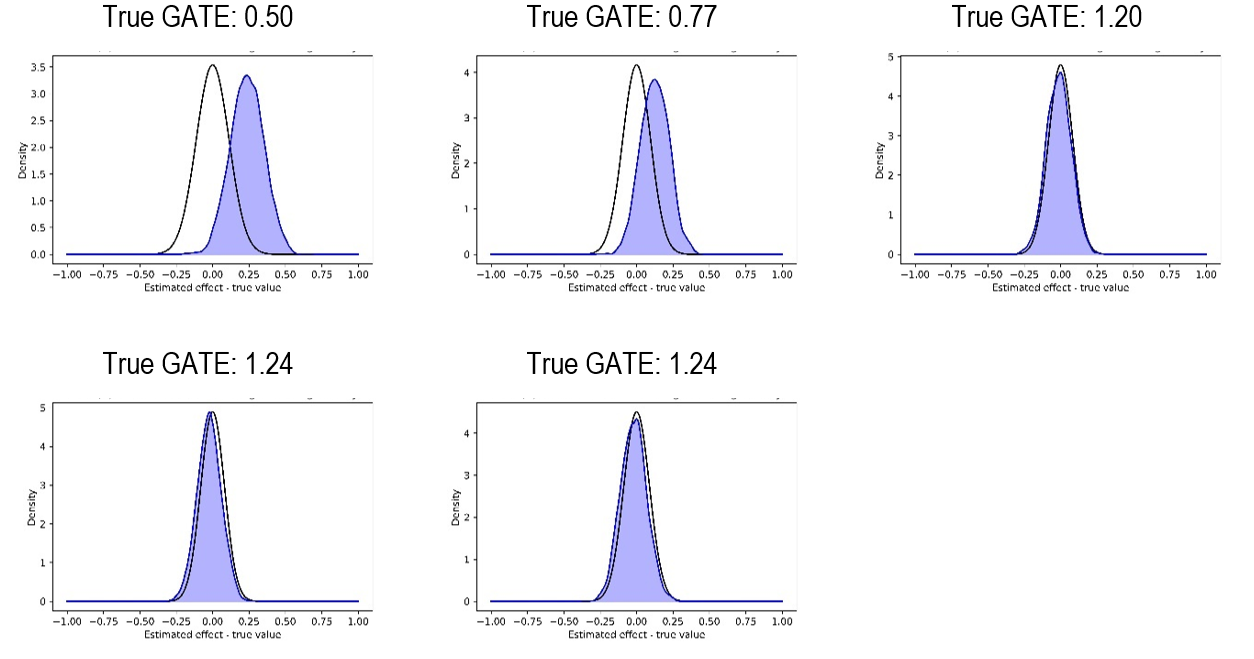} \\
		\midrule
		N = 10'000 \\
		\midrule
		\includegraphics[width = \textwidth]{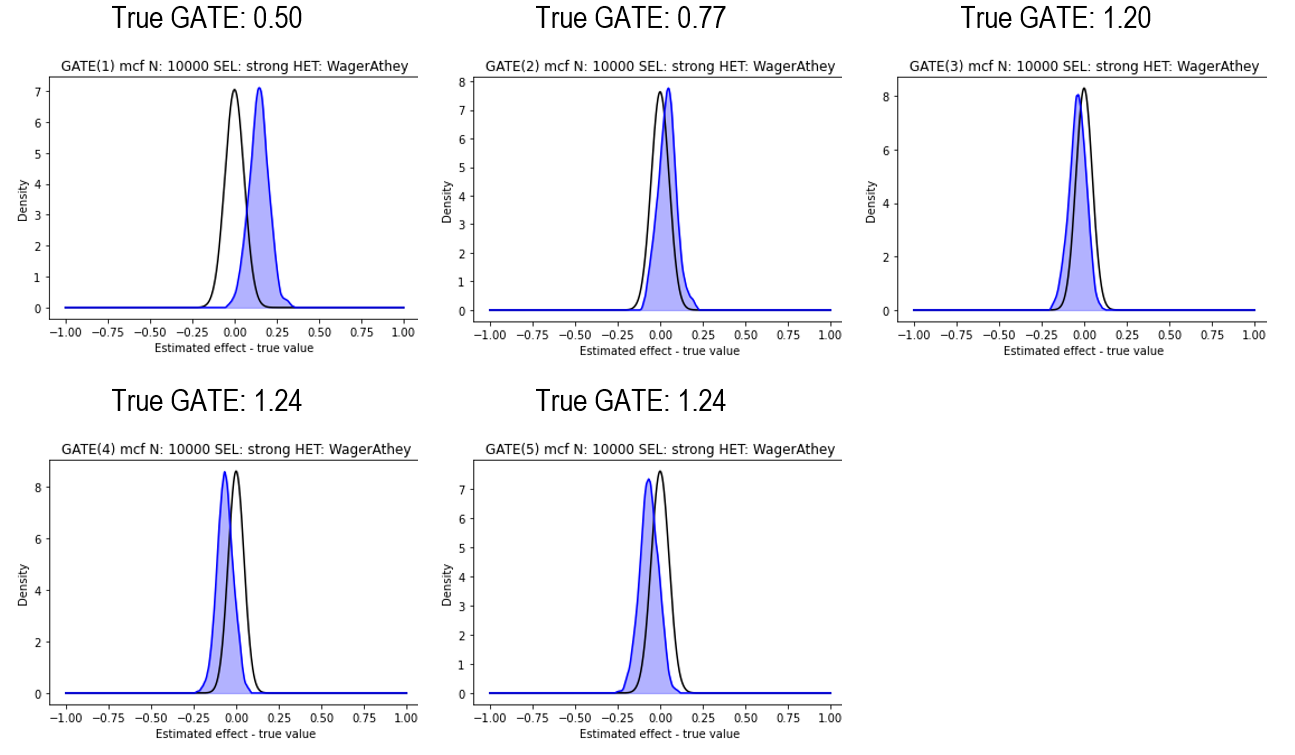} \\
		\bottomrule
	\end{tabular}
\end{table}